\documentstyle[./inc/psfig]{./inc/mn11inch} 

\newcommand{\Wo}{\mbox{${\rm W}_0$}}
\newcommand{\msun}{\mbox{${\rm M}_\odot$}}

\newcommand{\lsun}{\mbox{${\rm L}_\odot$}}
\newcommand{\rsun}{\mbox{${\rm R}_\odot$}}
\newcommand{\kms}{\mbox{${\rm km~s}^{-1}$}}

\newcommand{\kmpskpc}{\mbox{${\rm km~s}^{-1} \,{\rm kpc}^{-1}$}}
\newcommand{\Zytkow}{$\dot{\mbox{Z}}$ytkow}
\newcommand{\TZO}{$\mbox{T}\dot{\mbox{Z}}$\mbox{O}}
\newcommand{\TZ}{$\mbox{T}\dot{\mbox{Z}}$}

\newcommand{\kpc}{\mbox{${\rm kpc}$}}

\newcommand{\kira}{\mbox{${\sf kira}$}}
\newcommand{\Kira}{\mbox{${\sf kira}$}}
\newcommand{\SeBa}{\mbox{${\sf SeBa}$}}
\newcommand{\paperI}{\mbox{paper~{\rm I}}}
\newcommand{\paperII}{\mbox{paper~{\rm II}}}
\newcommand{\paperIII}{\mbox{paper~{\rm III}}}
\newcommand{\nbody}{\mbox{{{\em N}-body}}}
\newcommand{\thm}{\mbox{${t_{\rm hm}}$}}

\newcommand{\thc}{\mbox{${t_{\rm hm}}$}}
\newcommand{\tcrss}{\mbox{${t_{\rm hm}}$}}
\newcommand{\trlx}{\mbox{${t_{\rm rlx}}$}}

\newcommand{\mavc}{\mbox{${\langle m\rangle_{\rm c}}$}}
\newcommand{\mavh}{\mbox{${\langle m\rangle_{\rm h}}$}}
\newcommand{\mavt}{\mbox{${\langle m\rangle_{\rm t}}$}}
\newcommand{\mc}{\mbox{${m_{\rm core}}$}}
\newcommand{\mh}{\mbox{${m_{\rm hm}}$}}
\newcommand{\mt}{\mbox{${m_{\rm tide}}$}}
\newcommand{\mtot}{\mbox{${M_{\rm tot}}$}}

\newcommand{\mm}{\mbox{$\langle m \rangle$}}
\newcommand{\mmean}{\mbox{${\langle m \rangle}$}}

\newcommand{\rcore}{\mbox{${r_{\rm core}}$}}
\newcommand{\rc}{\mbox{${r_{\rm core}}$}}
\newcommand{\rvir}{\mbox{${r_{\rm vir}}$}}
\newcommand{\rv}{\mbox{${r_{\rm vir}}$}}
\newcommand{\rhm}{\mbox{${r_{\rm hm}}$}}

\newcommand{\rt}{\mbox{${r_{\rm tide}}$}}
\newcommand{\rtide}{\mbox{${r_{\rm tide}}$}}
\newcommand{\rJ}{\mbox{${r_{\rm Jacob}}$}}

\newcommand{\Et}{\mbox{${E_{\rm tide}}$}}

\newcommand{\vdisp}{\mbox{$\langle v \rangle$}}
\newcommand{\rlof}{\mbox{RLOF}}

\newcommand{\fwdgs}{\mbox{$f_{\rm {wd \over gs}}$}}

\newcommand{\nbsgwd}{\mbox{$N_{\rm bss:gs:wd}$}}
\newcommand{\fsbt}{\mbox{$f_{\rm s:b:t}$}}

\newcommand{\HRD}{Hertzsprung--Russel diagram}

\newcommand{\pc}{\mbox{${\rm pc}$}}

\def\unit#1{{\mbox{[{\rm #1}]}}}
\def\apgt{\ {\raise-.5ex\hbox{$\buildrel>\over\sim$}}\ }
\def\aplt{\ {\raise-.5ex\hbox{$\buildrel<\over\sim$}}\ }
\def\unit#1{{\mbox{[{\rm #1}]}}}
\def\apgt{\ {\raise-.5ex\hbox{$\buildrel>\over\sim$}}\ }
\def\aplt{\ {\raise-.5ex\hbox{$\buildrel<\over\sim$}}\ }
\title{Star cluster ecology IVa:
       Dissection of an open star cluster---photometry}

\author[Portegies Zwart et al.]
	{Simon F.\ Portegies Zwart,$^1$\thanks{Hubble Fellow}
	Stephen L.\ W.\ McMillan,$^2$
    	Piet Hut,$^3$
	Junichiro Makino,$^4$\\
$^1$	Department of Astronomy,
		 Boston University,
		 725 Commonwealth Ave.,
		 Boston, MA 02215, USA \\
$^2$Department of Physics,
		Drexel University,
                Philadelphia, PA 19104, USA\\
$^3$Institute for Advanced Study,
		Princeton, NJ 08540, USA\\
$^4$Department of Information Science and Graphics,
		College of Arts and Science,
		University of Tokyo, 3-8-1 Komaba,
		Meguro-ku, Tokyo 153, Japan
}

\date{Accepted 1993 December 11. Received 1993 March 17}

\begin{document}
\maketitle

\begin{abstract}

The evolution of star clusters is studied using {\nbody} simulations
in which the evolution of single stars and binaries are taken
self-consistently into account.  Initial conditions are chosen to
represent relatively young Galactic open clusters, such as the
Pleiades, Praesepe and the Hyades.  The calculations include a
realistic mass function, primordial binaries and the external
potential of the parent Galaxy.

Our model clusters are generally significantly flattened in the
Galactic tidal field, and dissolve before deep core collapse occurs.
The binary fraction decreases initially due to the destruction of soft
binaries, but increases later because lower mass single stars escape
more easily than the more massive binaries. At late times, the cluster
core is quite rich in giants and white dwarfs.  There is no evidence
for preferential evaporation of old white dwarfs, on the contrary the
formed white dwarfs are likely to remain in the cluster.  Stars tend
to escape from the cluster through the first and second Lagrange
points, in the direction of and away from the Galactic center.

Mass segregation manifests itself in our models well within an initial
relaxation time.  As expected, giants and white dwarfs are much more
strongly affected by mass segregation than main-sequence stars.
However, the clusters to which we compare our results all appear to be
somewhat more affected by mass segregation than our models would
suggest.

Open clusters are dynamically rather inactive.  However, the combined
effect of stellar mass loss and evaporation of stars from the cluster
potential drives its dissolution on a much shorter timescale than if
these effects are neglected.  The often-used argument that a star
cluster is barely older than its relaxation time and therefore cannot
be dynamically evolved is clearly in error for the majority of star
clusters.

An observation of a blue straggler in an eccentric orbit around an
unevolved star or a blue straggler of more than twice the turn off
mass might indicate past dynamical activity.  We find two distinct
populations of blue stragglers: those formed above the main-sequence
turn off, and those which appear as blue stragglers as the cluster's
turnoff drops below the mass of the rejuvenated star.


\end{abstract}


\section{Introduction}
Star clusters are remarkable laboratories for the study of fundamental
astrophysical processes spanning the range of stellar evolution, from
birth to death.  Open clusters are relatively small (typically $\aplt
10^4$ stars), young (generally $\aplt 1$ Gyr old) systems found
primarily in the Galactic disk.  Roughly 1100 are known to lie within
a few kiloparsecs of the Sun (Lynga 1987a)\nocite{Lynga1987}.  As a
class, they afford us the opportunity to witness star formation and
explore the complex evolution of young stellar systems.  Since stars
are born in star clusters and become part of the general Galactic
population only when the parent cluster dissolves in the Galactic
tidal field, it is of considerable interest to understand how this
process occurs, on what time scale it operates, and what are the
detailed observable consequences of cluster evolution.

At least as a first approximation, many star clusters are quite well
characterized dynamically as pure {\nbody} systems, in that there are
few collisions or close encounters between stars and the clusters
themselves are relatively isolated in space.  The same argument can be
made for the evolution of the stars and binaries without interactions,
which may also quite well explain the properties of open clusters.
This simple picture rapidly becomes inadequate when both effects are
combined---cluster evolution is actuality an intricate mix of
dynamics, stellar evolution, and external (tidal) influences, and the
subtle interplay between stellar dynamics and stellar physics makes
for a formidable modeling problem.  Of the many physical processes
influencing cluster evolution, probably the most important are the
effects of stellar evolution, the tidal field of the Galaxy, and the
evolution and dynamics of binary stars.  We present here a series of
models in which all these effects are taken self-consistently into
account.

The processes just mentioned are tightly coupled, complicating the
evolution and making it difficult to isolate the importance of each
individual effect.  We refer to Meylan \& Heggie
(1997)\nocite{1997A&ARv...8....1M} for a recent review.  Generally,
speaking, we can say mass loss from stellar evolution is of greatest
importance during the first few tens of millions of years of cluster
evolution, and may well result in the disruption of the entire cluster
(cf.~Chernoff \& Weinberg 1990,\nocite{1990ApJ...351..121C} Fukushige
\& Heggie 1995,\nocite{1995MNRAS.276..206F} Takahashi \& Portegies
Zwart 2000).  If the cluster survives this early phase, stellar
evolutionary time scales soon become longer than the time scales for
dynamical evolution, and two-body relaxation and tidal effects become
dominant.  Ultimately, these effects cause the cluster to dissociate
and the stars to become part of the general ``field'' population of
the disk.

There is strong observational evidence that star clusters contain
substantial binary populations, quite possibly as rich as those found
in the field (see Rubenstein 1997).\nocite{1997PASP..109..933R} The
properties and numbers of observed cluster binaries cannot be
explained by internal formation processes, such as three body dynamics
or two body tidal capture (Aarseth \& Lecar
1975).\nocite{1975ARA&A..13....1A} The majority must be primordial,
i.e.~formed together with the single stars at the cluster's birth.
These observations are important to cluster dynamics, as a cluster's
evolution depends strongly on its binary population and even a small
initial binary fraction can play a pivotal role in governing cluster
dynamics (Goodman \& Hut 1989,\nocite{1989Natur.339...40G} McMillan et
al.~1990, 1991a and
1991b\nocite{1990ApJ...362..522M}\nocite{1991ApJ...372..111M}\nocite{1991fesc.book..421M}
Gao et al. 1991, Hut et al.~1992\nocite{hmg+92}, McMillan \& Hut
1994).\nocite{1994ApJ...427..793M} Binaries are also crucial to
cluster stellar evolution.  The possibility of mass transfer between
binary components permits wholly new stellar evolutionary states to
arise; in addition, the presence of binaries will enhance the rate of
stellar collisions and close encounters, through the temporary capture
of single stars and other binaries in three-body resonance encounters
(Verbunt \& Hut 1987; Portegies Zwart et
al.~1998).\nocite{1987IAUS..125..187V}

Until quite recently, dynamical models have tended to exclude
binaries, for the good practical reasons that (1) binaries slow down
the calculations dramatically and tend to induce numerical errors, and
(2) their internal evolution is much more complicated than the
evolution of single stars.  However, it has also long been known that
cluster models lacking adequate treatments of binary systems are of at
best limited validity.  In this paper we begin to address these
limitations by performing calculations of open clusters containing
substantial numbers of primordial binaries, with the goal of studying
the mutual influence of stellar evolution and stellar dynamics in
these systems---the ``ecology'' of star clusters (Heggie
1992).\nocite{1992Natur.359..772H}

The first paper in this series (Portegies Zwart et al.~1997a,
hereafter \paperI)\nocite{pzhv97} quantified the effect of collisions
on stellar evolution and attempted to assess the corresponding changes
in the stellar population.  The stellar number density was held
constant in these calculations, thus excluding the possibility of any
interplay between the dynamical evolution of the cluster and
collisions between stars.  In the second paper in this series
(Portegies Zwart et al.~1997b, hereafter \paperII),\nocite{pzhmv97} the
evolution of a population of primordial binaries was followed in time
by tracking in detail the results of encounters between single stars
and binaries.  The assumption of constant stellar number density was
relaxed in Portegies Zwart et al.~(1999, hereafter
\paperIII),\nocite{pzmmh99} where the dynamical evolution of a star
cluster (without primordial binaries) was followed in detail using
\nbody\, calculations.

This paper continues the process of relaxing the simplifying
assumptions made in Paper I.  As in Paper III, we calculate the
dynamical evolution of our model system by direct (\nbody) integration
of the system, but now including both stars and binaries in the
computational mix.  Binary evolution is incorporated into the \nbody\,
treatment, accounting for changes in binary orbital parameters due to
stellar mass loss, supernovae, tidal forces between the stars, mass
transfer from one star to its companion, general relativistic
corrections, etc.  Encounters between binaries and single stars and
higher order encounters (between binaries and binaries and between
binaries and triples etc.) are fully integrated as are the orbits of
the other stars in the \nbody\, system.  All changes in the stellar
and binary population caused by stellar evolution are fed back into
the dynamical evolution of the parent cluster, allowing stellar
dynamics and the stellar evolution to be studied simultaneously and
self-consistently.

As an initial case study, we concentrate here on open star clusters
near the Sun (between $\sim6$ and 10\,\kpc\, from the Galactic
center), which are less than 1 billion years old, and which initially
contain a few thousand ($\sim2000$--3000) stars, roughly half of them
in binaries.  Well known clusters fitting this general description are
the Pleiades, Praesepe and the Hyades.  Such systems are small enough
that multiple simulations can be performed in order to improve
statistical coverage of their properties, yet they are large enough
and old enough that both stellar evolution and stellar dynamics have
had time to play significant roles in determining their present
structure and appearance.

In an effort to close the gap between theoretical and observational
studies of cluster structure, in this and subsequent papers we will
attempt wherever possible to ``observe'' our model clusters using
techniques similar to those employed by observers.  For this paper we
have chosen to adopt a ``photometric'' approach.  Consequently, we
present little detailed information about the binaries in our
calculations.  Binary properties will be discussed in depth in a
``spectroscopic'' companion paper (Portegies Zwart et.~al, Paper IVb,
in preparation).

Section \ref{sect:init} discusses the choice of initial conditions and
parameters for our model clusters.  The results of the calculations
are presented in \S\ref{sect:results}, followed in section
\ref{sect:discussion}, which compares our results with selected
observations on a cluster-by-cluster basis.  Section\,
\ref{sect:other_work} our results with some previous studies in this
area.  We summarize in \S\ref{sect:summary}.  Two appendices are
included, Appendix \ref{sect:terminology} gives an overview of the
terminology used throughout the paper; Appendix \ref{sect:Starlab}
reviews the ``Starlab'' software package and the implementation and
coupling of its two main constituents \Kira\, (the \nbody\, integrator
\S\ref{sect:Kira}) and \SeBa\, (the stellar/binary evolution
program\S\ref{sect:SeBa}).


\section{Initial conditions}\label{sect:init}
\begin{table*}
\caption[]{Observed and derived parameters for several open star
clusters with which our simulations may be compared.  Subsequent
columns give (3) the distance to the cluster (in pc), (4) the cluster
age (in Myr), (5) the half mass relaxation time (in Myr), (6) the
total mass (in \msun), (7) the tidal radius (in pc), (8) estimate for
the half mass radius (in pc), and (9) the core radius (in pc).  In
cases where the parameters (relaxation time, mass, etc.) are not
accessible in the literature, we calculate it; these entries are
printed {\em in italics}.  In most cases these numbers can be
calculated using Eq.\,\,\ref{Eq:trlx} or Eq.\,\ref{Eq:Rtide}.  Dashes
and question marks indicate that we cannot derive these numbers from
the literature.  The final two columns contain information on the
cluster stellar content.  The column labeled $\fsbt$ indicates the
number of single stars, binaries and triples (separated by colons).
For clusters where the numbers are given directly by observations, the
table gives the observed numbers of each system.  If the binary
fraction is derived by other methods, we give the relative fractions
normalized to the number of single stars.  The last column ($\nbsgwd$)
gives the number of observed blue stragglers, giants and white dwarfs,
separated by colons.\footnote{Data on various clusters is also
available via Mermilliods' {\tt WEBDA} online catalog via {\tt
http://obswww.unige.ch/webda/}.}
}
\bigskip
\begin{tabular}{llrrrrrrrrrcc} \hline
name    &ref.&$d$ &$t$&    \trlx&$M$ &\rt\  &\rhm\   &\rc\ 
                                                      &\fsbt&\nbsgwd \\
   &  &[pc]&\multicolumn{2}{c}{~~~~~[Myr]}&[\msun]&\multicolumn{3}{c}{[pc]}\\
\hline
NGC\,2516 &a  & 373&  110     &{\em 220}&{\em 1000}&{\em 13}& 2.9 & ---
					               &16:6:?&6:4:4\\
Pleiades &b  & 135&  115    &      150&$\sim$1500&   16&  2--4 & 1.4  
                                                       &137:60:2 & 0:3:3\\
NGC\,2287 &c  & 655&  160--200&  ---    &$\apgt${\em 120} & 6.3 & ---  & ---
						       &1:0.6:? &3:8:3\\
Praesepe&d  & 174&  400--900&{\em 370}& 1160     & 12  & {\em 3.5} & 2.8  
                                                       &1:0.3:0.03 &5:5:11\\
Hyades  &e  &  46& 625      &{\em 390}&500--1000& 10.3&{\em 3.7}&2.6
                                                       &1:0.4:0 & 1:4:10\\
NGC\,2660&f  & 2884&900--1200&{\em 315}&$\apgt 400$&{\em 9.6}& 4 & {\em 1.5}
                                                      &1:0.3:? & 18:39:?\\
NGC\,3680&g  & 735&1450      &{\em 28} &$\apgt 100$&     4.3 & 1.2     &{\em 0.6}
                                                      &44:25:0 & 4:17:?\\
\hline
\end{tabular}
\\
References to the literature (second column) are:
(a) Abt \& Levy (1972)\nocite{1972ApJ...172..355A}; 
    Dachs, J \& Kabus (1989)\nocite{1989A&AS...78...25D};
    Hawley et al. (1999).\nocite{1999AJ....117.1341H}
(Note: we interpret the quoted limiting cluster radius as the
half mass radius.)
(b) Pinfield et al.\, (1998);\nocite{1998MNRAS.299..955P}
    Raboud \& Mermilliod, (1998);\nocite{1998A&A...329..101R}
    Bouvier et al (1998);\nocite{1997A&A...323..139B}
(c) Harris et al. (1993);\nocite{1993AJ....106.1533H}
    Ianna et al (1987);\nocite{1987AJ.....93..347I}
    Cox (1954).\nocite{1954ApJ...119..188C}
(d) Andrievsky (1998);\nocite{1998A&A...334..139A} 
    Jones \& Stauffer (1991);\nocite{1991AJ....102.1080J} 
    Mermilliod \& Mayor (1999);\nocite{1999astro.ph.11405M}
    Mermilliod et al. (1990);\nocite{1990A&A...235..114M}
    Hodgkin et al. (1999).\nocite{1999MNRAS.310...87H}	
(Note: we interpret the quoted central radius for the cluster as the
half mass radius.)
(e) Perryman et al.\, (1998 and references therein)
    \nocite{1998A&A...331...81P} 
    Reid \& Hawley (1999);\nocite{1999AJ....117..343R}
%
(f) Frandsen et al. (1989);\nocite{1989A&A...215..287F} 
    Hartwick \& Hesser (1971);\nocite{1971PASP...83...53H} 
    Sandrelli et al. (1999).\nocite{1999MNRAS.309..739S}
(g) Hawley et al. 1999 \nocite{1999AJ....117.1341H}; 
    Nordstr\"om et al. (1997);\nocite{1997A&A...322..460N} 
    Nordstr\"om et al.\,(1996),\nocite{1996A&AS..118..407N}
Data on numbers of white dwars was taken from Anthony-Twarog
(1984)\nocite{1984AJ.....89..267A} for Praesepe, from Koester \&
Reimers (1987)\nocite{1981A&A....99L...8K} for NGC 2287 and from von
Hippel (1998)\nocite{1998AJ....115.1536V} for the other clusters.
\label{Tab:observed} 
\end{table*}

In order to begin a simulation, a number of critical cluster
parameters must be chosen: the mass, virial radius, and tidal radius
(or its equivalent), the initial mass function, and the initial
distributions of binary spatial density and orbital properties.  Table
\ref{Tab:observed} presents an overview of observed parameters for
some star clusters with similar overall masses, stellar membership and
half-mass radii.  The ages of these clusters vary widely, from about
110 Myr (Pleiades) to over 1 Gyr (NGC\,3680).  Their core and half mass
radii suggest that they may be described approximately by King models
(King 1966)\nocite{1966AJ.....71...64K} with dimensionless depths in
the range $\Wo\sim4-6$, where larger $\Wo$ corresponds to higher
central concentration.

We define our mass scale by arbitrarily adopting a ``Hyades-like''
model, in which the mass of the system at an age of 625 Myr is
$\sim1000\msun$.  We then estimate the initial mass of a cluster by
applying a number of corrections.  We adopt the initial mass function
for the solar neighborhood described by Scalo (1986).\nocite{scalo86}
Table \ref{Tab:mass} shows how the Scalo mass function evolves in
time.  Approximately 20\% of the initial mass is lost by purely
stellar-evolutionary processes.  (Binary evolution complicates matters
which we neglect in these rough numbers.)

\begin{table}
\caption{Number $N_x$ of stars of type $x$ and total mass $M$,
from the Scalo (1986) initial mass
function, evolved with \SeBa. Calculation was performed with a
population of 100\,k stars, but the numbers are normalized to 1\,k
stars.}
\medskip
\begin{tabular}{lr|rrrrrr} \hline
time [Myr]    &      0&   100&   200&   400&   600&  800  \\ \hline
ms            &   1024&1014.7&1007.7& 996.7& 988.3& 982.1 \\
gs            &      0&   2.7&   4.3&   6.2&   6.2&   5.8 \\
wd            &      0&   3.0&	 8.5&  17.6&  26.0&  32.7 \\
ns            &      0&   3.6&   3.6&   3.6&   3.6&   3.6 \\
$M$ [\msun] &  624.4& 562.3& 541.5& 517.6& 501.1& 490.6 \\ \hline
\end{tabular}
\label{Tab:mass} 
\end{table}

In addition to stellar evolution, dynamical evolution of the star
cluster and the tidal field of the Galaxy also tend to consume (eject)
cluster stars, at a rate of about 10\% per relaxation time (Spitzer
1987).\nocite{1987degc.book.....S} With a current relaxation time for
Hyades of about 400Myr (see Table \ref{Tab:observed}) we estimate that
the amount of mass lost by dynamical processes up to an age of
625\,Myr is similar to the mass lost to stellar evolution.  Adding
these numbers provides a conservative lower limit to the amount of
mass lost by the star cluster. This limit is conservative because we
neglected the interaction between stellar mass loss and dynamical mass
loss and the extra mass loss induced by interactions with giant
molecular clouds.  We therefore adopt an initial cluster mass of
$M_0\sim1600\,\msun$. This is close to the 1800\,\msun\, for the
initial mass of Hyades derived by Weidemann (1992).  We assume a Scalo
(1986) initial mass function, with minimum and maximum masses of
0.1\,{\msun} and 100\,\msun, respectively, and mean mass $\mm\simeq
0.6\msun$.  Consistent with the above estimates, our simulations are
performed with 1024 single stars and 1024 binaries, for a total of
3096 (3k) stars and a binary fraction of 50\%.

Stars and binaries within our model are initialized as follows.  A
total of 2k single stars are selected from the initial mass function
and placed in an equilibrium configuration in the selected density
distribution (see below).  We then randomly select half the stars and
add a second companion star to them.  The masses of the companions are
randomly selected between 0.1\,{\msun} and the primary mass.  For
low-mass primaries, the mass ratio distribution
peaks at unity, whereas the distribution is flat for more massive
primaries.  Once stellar masses are chosen, other binary parameters
are determined.  Binary eccentricities are selected from a thermal
distribution between 0 and 1.  Orbital separations $a$ are selected
with equal probability in $\log a$ with the lower limit set by the
separation at which the primary fills its Roche lobe or at 1\,\rsun,
whichever is smaller. The upper limit for the initial semi-major axis
is taken at $10^6$\,\rsun\ (about 0.02\,pc, Duquennoy \& Mayor
1991).\nocite{1991A&A...248..485D} When a binary appears to be in
contact at pericenter, new orbital parameters are selected.
Table\,\ref{Tab:Binit} gives an overview of the various distribution
functions from which stars and binaries are initialized.

\begin{table}
\caption[]{Initial conditions for the stellar and binary population.
The first column gives the parameter, the second and third columns
give the symbol and the distribution function, followed by the lower
and upper limits adopted.}
\medskip
\begin{tabular}{lcl|cc} \hline
                  &   &                        & \multicolumn{2}{c}{limits} \\
parameter         &   & function               & lower & upper \\ 
primary mass      &$M$& $P(M) =$ Scalo (1986)  & 0.1\,\msun  & 100\,\msun \\
secondary mass    &$m$& $P(m) = {\rm constant}$& 0.1\,\msun  & $M$ \\
orbital separation&$a$& $P(a) = 1/a$           & \rlof       & $10^6$\,\rsun \\
eccentricity      &$e$& $P(e) = 2e$            & 0           & 1    \\ \hline
\end{tabular}
\label{Tab:Binit} \end{table}

We select initial density profiles from the anisotropic density
distributions described by Heggie \& Ramamani
(1995)\nocite{1995MNRAS.272..317H} with $\Wo=4$ and $\Wo=6$, and refer
to these models as W4 and W6, respectively throughout this paper.  The
Heggie-Ramamani models are derived from King (1966) models, but take
into account the velocity anisotropy and non-spherical shape of the
critical zero-velocity (Jacobi) surface of the cluster in the Galactic
tidal field.  (The classical King models have spherical boundaries.)
Within the half mass radius, the Heggie-Ramamani models are quite
isotropic.

All models are started with a virial radius of $\rvir = 2.5$ pc.  For
our adopted parameters, the initial cluster dynamical time scale is
then $\tcrss \equiv (GM/\rvir^3)^{-1/2} \sim 1.5$ Myr.  Each cluster
is assumed to precisely fill its Jacobi surface at birth.  (Expressed
less precisely, we could say that the limiting radius of the initial
King model is equal to the Roche radius of the cluster in the Galactic
tidal field.)  Given the Oort constants in the solar neighborhood, we
find that the models with $\Wo=6$ are somewhat farther
($\sim$12.1\,kpc) from the Galactic center than is the Sun, while a
model with $\Wo=4$ is slightly closer ($\sim$6.3\,kpc).

For a total cluster mass of 1600\,\msun, the Lagrange points of our
two standard clusters lie, respectively, at 14.5 pc ($\Wo=4$) and 21.6
pc ($\Wo=6$) from the cluster center.  A star is removed from a
simulation when its distance from the cluster's density center exceeds
twice the distance from the center to the first Lagrangian point.



Table\,\ref{Tab:init} reviews the adopted parameters and initial
conditions of our models.  In order to improve statistics, we
performed four calculations (labeled I through IV) for each set of
initial conditions. A fifth run is performed as a pioneering study for
each set of initial conditions but they are terminated at 400\,Myears.

\begin{table*}
\caption[]{Initial conditions and parameters for the selected models.
The columns give the model name, initial mass (\msun), number of
stars, dimensionless central depth of the potential well (\Wo), the
distance from the Galactic center (kpc), the initial relaxation and
half mass crossing times (both in Myr), $r_x$, $r_y$, and $r_z$, the
distances from the cluster center to the Jacobi surface (so $r_x$ is
simply the Jacobi radius, $r_J$; see Appendix A), and the virial, half
mass and core radius (all in parsec).  }
\begin{flushleft}
\begin{tabular}{l|rrrrrrrrrrrrrr} \hline
name &$M$  &$N$&\Wo&$R_{\rm Gal}$
             & \trlx &\thc&\multicolumn{3}{c}{\rJ} &\rv &\rhm &\rc \\
     &[\msun]&&&[kpc]&\multicolumn{2}{c}{[Myr]}
             &&\multicolumn{3}{c}{[pc]}  & \multicolumn{3}{c}{[pc]} \\ \hline
W4   & 1600&3k& 4& 6.3&109&4.07&  14.5& 9.7& 7.2   & 2.5&2.14 & 0.83 \\
W6   & 1600&3k& 6&12.1&102&4.15&  21.6&14.4&10.8   & 2.5&2.00 & 0.59 \\ \hline
\end{tabular}
\end{flushleft}
\label{Tab:init} 
\end{table*}


\section{Results}\label{sect:results}
We now discuss the ``photometric'' properties of our model clusters.
As mentioned above, we defer the discussion of ``spectroscopic''
properties, including details on the various types of binaries found
in our simulations, to Paper IVb.


\subsection{Global properties}
\begin{figure*}
(a)\psfig{figure=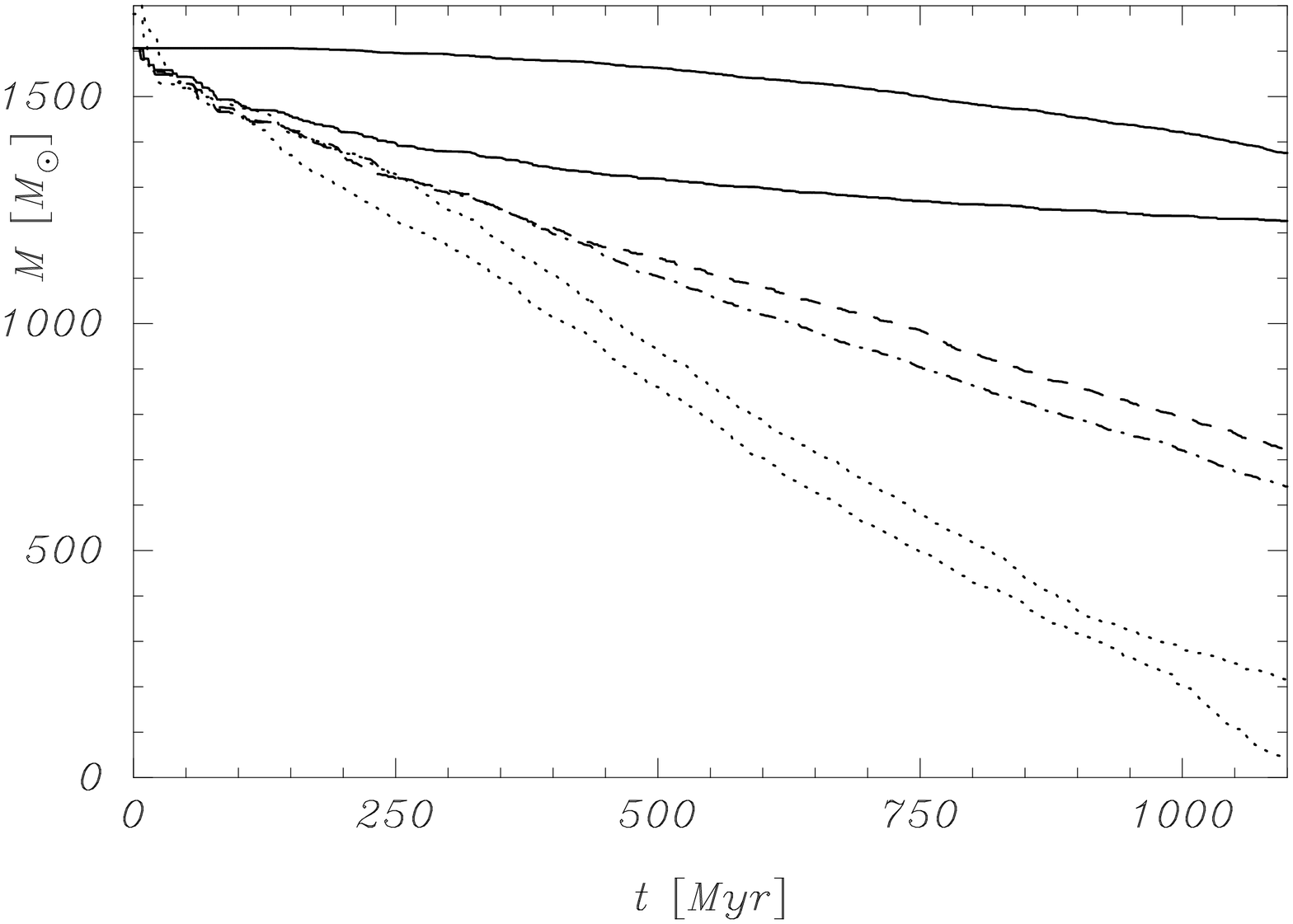,width=7.5cm,angle=-90}
(b)\psfig{figure=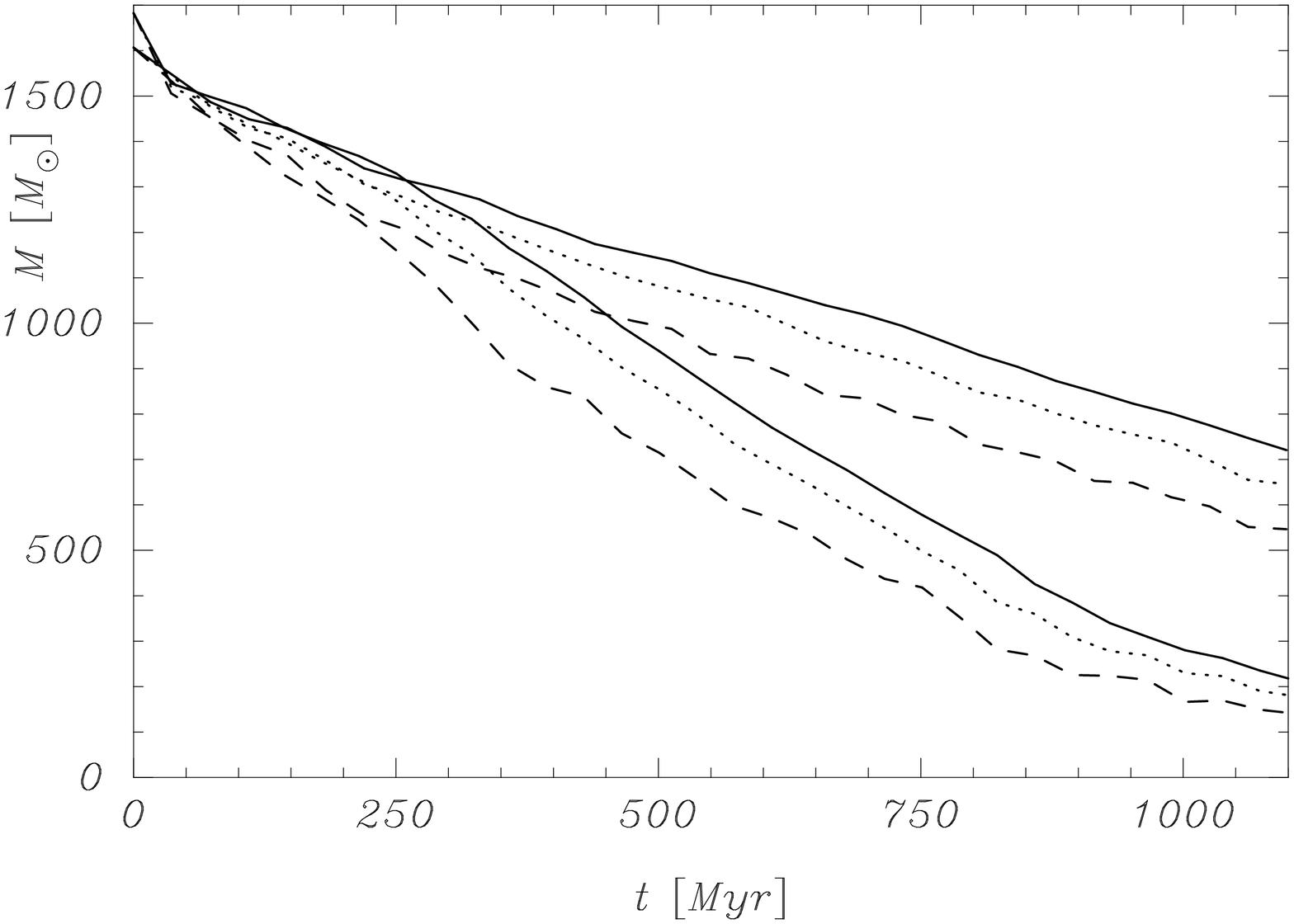,width=7.5cm,angle=-90}

\caption[]{(a) Mass (in \msun) as a function of time (in Myr) for
various models.  The solid lines show the time-dependence of the total
mass with (upper curve) stellar evolution suppressed, and (lower
curve) without dynamical evolution.  The dashed line is the total mass
of model W6-III.  The dot-dashed line is the total mass of a
calculations with the identical mass function as in model W6-III, but
all stars were single and redistributed in an identical density
distribution.  This line therefore gives the difference in a
calculation with or without primordial binaries, but with all the
other effects takes the same.  The dotted lines represent the total
mass of models W4-II (upper) and W4-IV(lower).
(b) Various definitions of the cluster mass (in \msun) as functions of
time for model W6-III (upper set of 3 lines) and model W4-II (lower
set of lines).  Solid lines represent the total mass in the {\nbody}
system, dotted lines the mass within the Jacobi surface, and
dashed lines the total mass seen in projection within a distance $r_J$
of the cluster center.
}
\label{fig:tm_all}
\label{fig:tm_Jacob}
\end{figure*}

%
%
Figure \ref{fig:tm_all}(a) shows the mass of the cluster as a function
of time for several models.  In this figure, the ``total mass'' of a
model is simply taken to be the sum of the masses of all stars
remaining within the {\nbody} system.  This overestimates both the
bound mass of the system and probably also the mass that would be
derived by observers, and therefore provides a firm upper limit to the
``true'' mass.  The dashed line gives the results from model W6-III.
The two dotted lines show the mass evolution of two of the W4 models
(upper line: model W4-III, lower line: W4-IV), illustrating the
run-to-run variations in dynamical evolution (which are mainly the
result of an initial offset between the masses of the two models,
caused by different random seeds).

Table\,\ref{Tab:modelW6III} indicates the differences between the
various line styles in Fig.\,\ref{fig:tm_all}(a). All lines listed are
computed using exaclty the same stellar masses as in model W6-III. 

\begin{table}
\caption{Overfiew of the variation in parameters for model W6-III.
The realization of the initial mass function was identical for all
these models. Other parameters are as in Tab.\,\ref{Tab:init}.  For
each of these calculations some feature of starlab was switched on (+)
or off (-); \kira (the \nbody\, integrator), \SeBa\, (the stellar and
binary evolution model) and wether the calcualtion started with
primordial binaries.  }
\medskip
\begin{tabular}{l|ccc} \hline
            &\multicolumn{2}{c}{Starlab} & Primordial \\
	    & \kira & \SeBa              & binaries   \\
\hline
upper solid & -  &  +   & + \\
lower solid & +  &  -   & + \\
dashes      & +  &  +   & + \\
dash-dots   & +  &  +   & - \\
\hline
\end{tabular}
\label{Tab:modelW6III}
\end{table}

The two solid lines in Figure \ref{fig:tm_all}(a) show (upper line)
the total mass of an {\nbody} system without stellar evolution, but
otherwise with the same initial conditions as run W6-III, and (lower
line) the total mass of stars, excluding stellar dynamics but
including mass loss by stellar evolution (with the same initial
stellar masses as in model W6-III; see Table.\,\ref{Tab:modelW6III}
for an review of the line styles).  Mass loss in the absence of
stellar evolution is not linear with time, as one expects from a equal
mass {\nbody} system; the presence of a mass function causes heavier
stars generally to be lost later, resulting in larger mass loss at
later time.  However, the loss rate of stars is roughly constant, and
the curvature of the upper solid line in Figure \ref{fig:tm_all}(a)
demonstrates the strong effects of mass segregation and cluster
dynamics.

The actual variation of the cluster mass (under the combined effects
of stellar mass loss and dynamical evolution) is larger than the sum
of the two separate effects by about a factor two.  For the selected
initial conditions, the interplay between stellar evolution and
stellar dynamics is especially important during the later stages of
the evolution.  The presence of primordial binaries has little effect
on the evaporation rate of the clusters.  The dash-dotted line gives
the total mass of the same model, but without binaries---all stars in
the initial mass function (including binary secondaries) are
redistributed with the same density profile as model W6-III.  As noted
by McMillan \& Hut (1994), primordial binaries have little effect on
the overall rate at which mass is lost from the cluster.


Figure \ref{fig:tm_Jacob}(b) shows the masses of two model clusters
(W6-III and W4-II), with different criteria for the limiting radius of
the stellar system.  The solid lines give the total mass in the
{\nbody} system (see also Figure \ref{fig:tm_all}[a]), the dotted
lines the total mass within the Jacobi surface, and the dashed lines
the total mass within the Jacobi radius of the cluster center, as seen
by an observer looking along the $y$-axis.  The total mass in stars in
the {\nbody} system overestimates the cluster's mass; the mass within
the zero velocity surface may give a better measure of the mass one
would observe in a real situation.  The cluster mass within the Jacobi
radius, viewed along the $x$-axis, always lies between the solid and
dotted lines; the corresponding mass viewed along the $z$ axis lies
between the dotted and the dashed lines.  These trends are
found in all models studied.

\begin{figure*}
(a)\psfig{figure=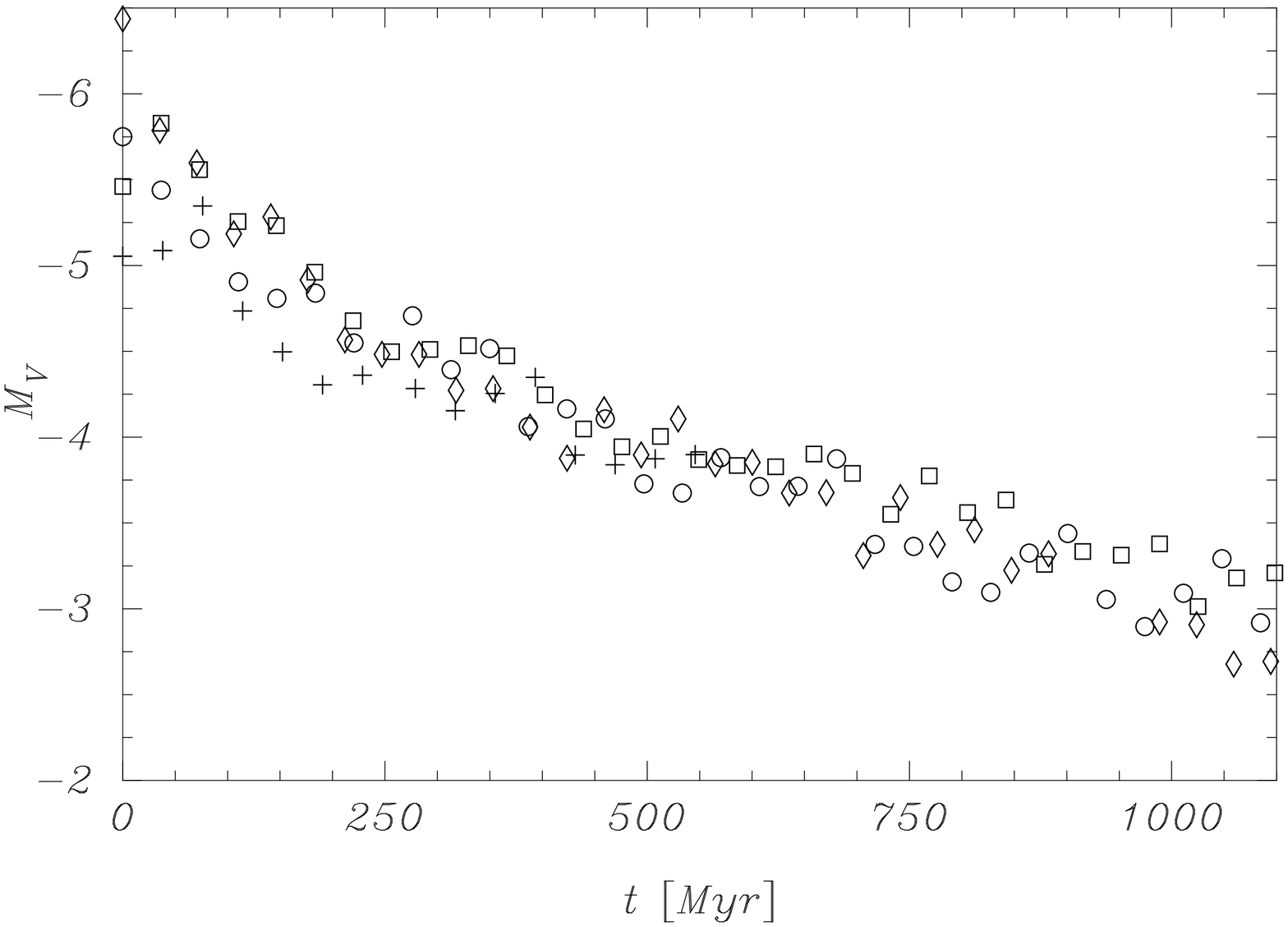,width=7.5cm,angle=-90}
(b)\psfig{figure=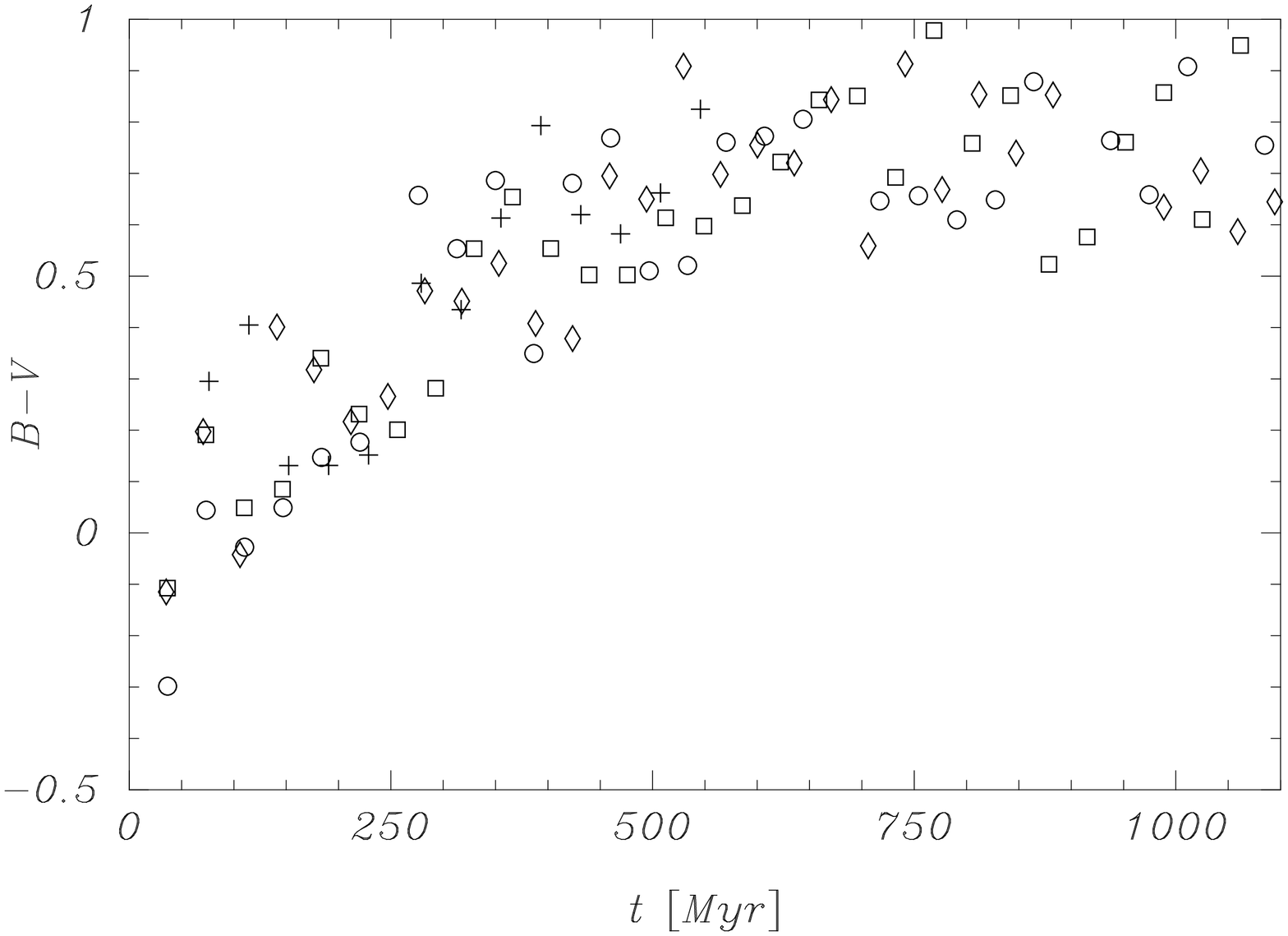,width=7.5cm,angle=-90}
\caption[]{(a) Integrated $M_V$ magnitude as a function of time of the
four W6 models (plus, circle, square, and diamond give the data for
the models I--IV).  The variation between the points at similar times
within a single calculation is comparable to the visible spread in the
data.  (b) Integrated $B-V$ color as a function of time for the W6
models.}
\label{fig:tV_W6all}
\label{fig:tBmV_W6all}
\end{figure*}

Figure \ref{fig:tV_W6all}(a) presents the observational equivalent of
Figure \ref{fig:tm_Jacob}(b), showing the total $M_V$ magnitude of the W6
models at various times.  The spread in $M_V$ is caused by the intrinsic
differences between runs---initial conditions, run-to-run variations
and fluctuations due to the small numbers of giants, which dominate the
total magnitude.  Note that, for technical reasons, the output
intervals are not the same for the four calculations shown here.

Figure \ref{fig:tBmV_W6all}(b) plots the integrated $B-V$ color of the
cluster as a function of time. The initial color of all models is
confined to a small range between $B-V \simeq -0.15$ and $-0.28$, but
rapidly grows to larger values with a larger spread: $B-V \simeq
0.1$--$0.4$ at 50\,Myr and $B-V \simeq 0.5$--$0.8$ at 500\,Myr. As in
Figure \ref{fig:tV_W6all}(a), the intrinsic spread is caused by the
presence of a relatively small number of giant stars.  The increase in
the color index indicates that the cluster gets redder with age, which
is mainly due to the loss of the massive blue stars and the formation
of red giants.  The color variation of the W4 models is similar.

\begin{figure*}
(a)\psfig{figure=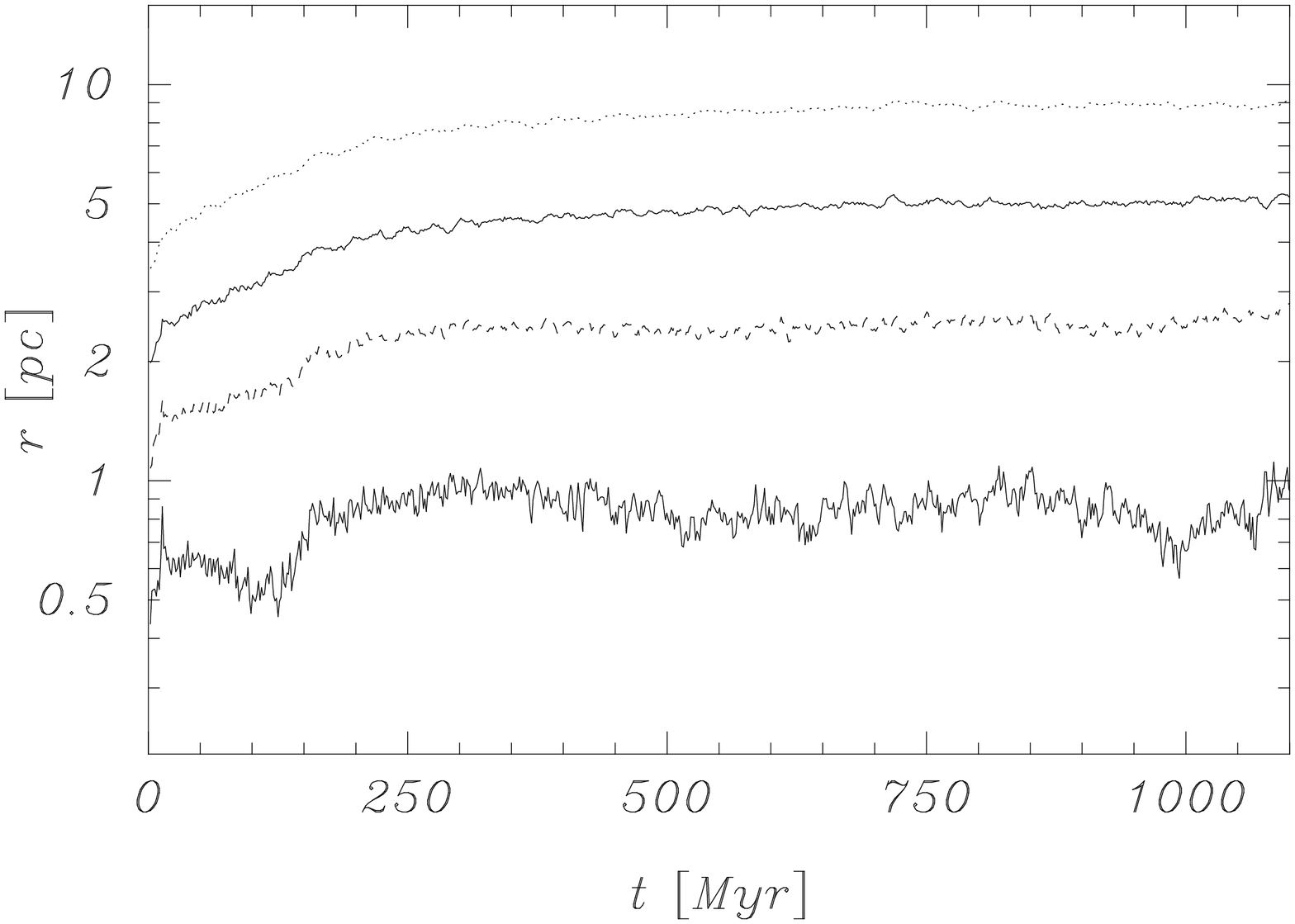,width=7.5cm,angle=-90}
(b)\psfig{figure=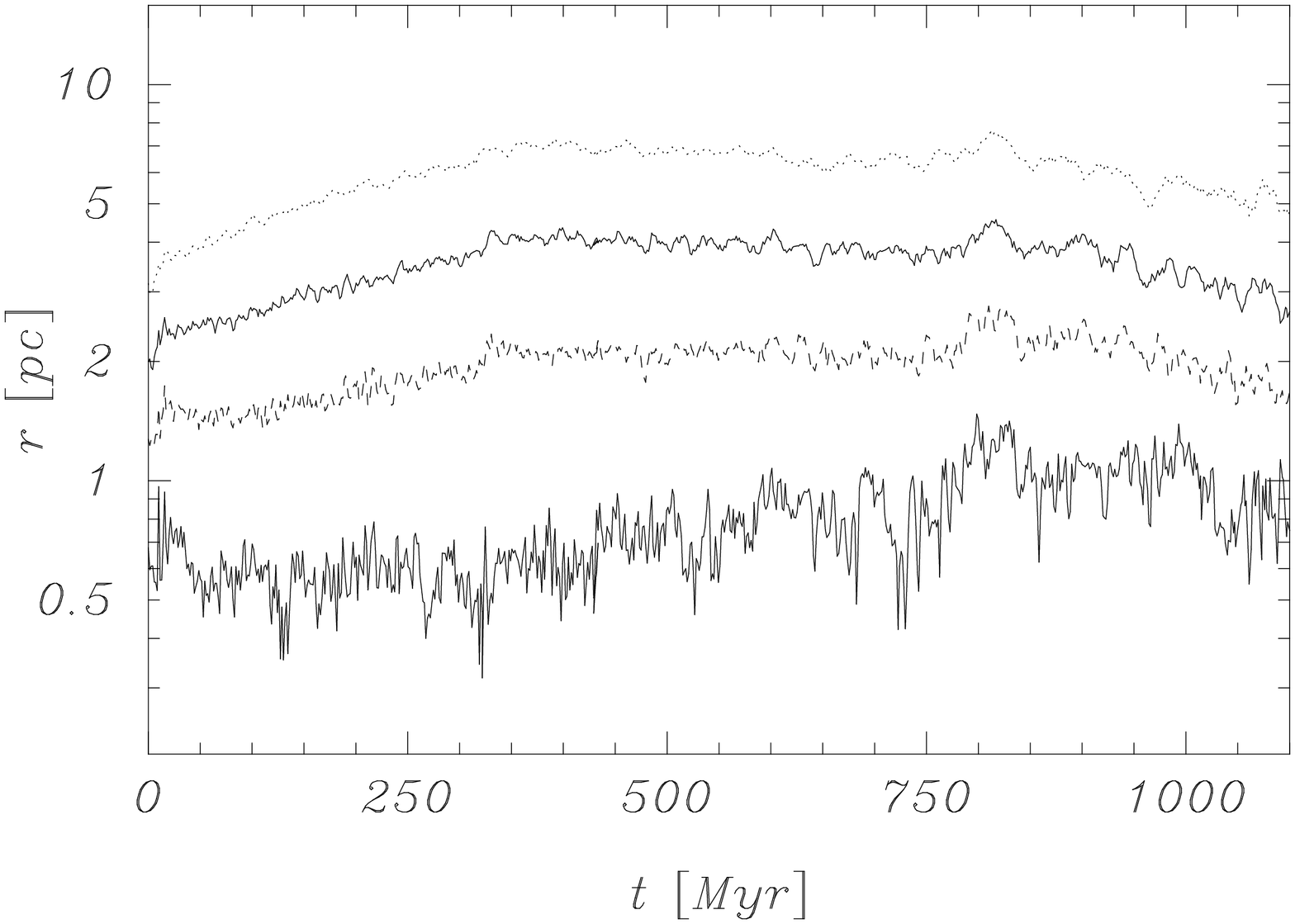,width=7.5cm,angle=-90}
\caption[]{Lagrangian radii (in pc) as functions of time.  The data in
(a) represent the mean of the four W6 runs, those in (b) model W4-II.
From top to bottom, the radii contain the following mass fractions:
75\% (dots), 50\% (upper solid), 25\% (dashes) and 5\% (lower solid).}
\label{fig:lagrad_W6R2TF}
\label{fig:lagrad_W4}
\end{figure*}

Figures \ref{fig:lagrad_W6R2TF} shows how the radii of models W4 and
W6 evolve with time.  All stars in the {\nbody} system are taken into
account in calculating the Lagrangian radii.  The outer radii
therefore expand much more than they would do if only stars within the
Jacobi surface were considered.  Note the absence of any discernible
core collapse in either case.  There is a slight, barely noticeable,
core contraction between $t=100$ and 150\,Myr for the W6 models, and
somewhat earlier for the W4 models, but neither is very deep.  This
shallow core contraction phase demonstrates the importance of stellar
mass loss and, to a lesser extent, binary heating to the dynamical
evolution of these systems; mass segregation for example, which also
plays an important role here (see Fig.\,\ref{fig:mm_W6R2TF}).  A
comparable model with primordial binaries and without stellar
evolution would experience core collapse (McMillan et al.\, 1990,
1991),\nocite{1990ApJ...362..522M} \nocite{1991ApJ...372..111M} even
in the presence of a Galactic tidal field (McMillan \& Hut
1994).\nocite{1994ApJ...427..793M}

In contrast to the W6 radii shown in Figure
\ref{fig:lagrad_W6R2TF}(a), the Lagrangian radii of model W4-II
(Figure \ref{fig:lagrad_W6R2TF}[b]) expand for about 300 Myr, and
subsequently shrink.  The decrease in Lagrangian radii at late times
is an indicator of cluster evaporation.  As the cluster dissolves in
the Galactic tidal field, its tidal radius decreases, accelerating the
dissolution and causing the Lagrangian radii to decrease.  The W6
clusters show the same behavior, but at somewhat later times.  We show
the result of a single W4 model to illustrate the intrinsic
fluctuations within a single {\nbody} run.

%
%
%
%

\begin{figure}
\psfig{figure=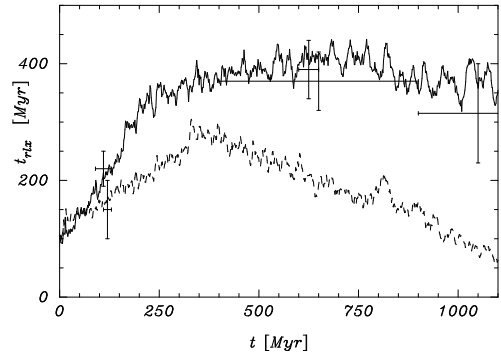,width=7.5cm,angle=-90} 
\caption[]{Half-mass relaxation time as a function of time for the
models W6-III (solid) and for model W4-II (dashes).  The error bars
indicate the computed relaxation times for the observed star clusters
from Table \ref{Tab:observed}. (Confidence intervals are not listed in
the table.)  }
\label{fig:ttrlx_W6W4}
\end{figure}

Finally, Figure \ref{fig:ttrlx_W6W4} shows the half-mass relaxation
time (Eq.\,\ref{Eq:trlx}) as a function of time for models W6-III
(solid line) and W4-II (dashed line).  Note that the relaxation time
peaks around the cluster's ``half-life'' epoch---750 and 400\,Myr for
models W6-III and W4-II, respectively.  Consequently, estimates of the
present-day relaxation time of observed open clusters may provide poor
indicators of the dynamical age of the stellar system.  The often-used
argument that a star cluster is barely older than its relaxation time
and therefore cannot be dynamically evolved is clearly in error for
the majority of star clusters (see also McMillan \& Hut 1994).

\subsection{Mass segregation}
\begin{figure*}
(a)\psfig{figure=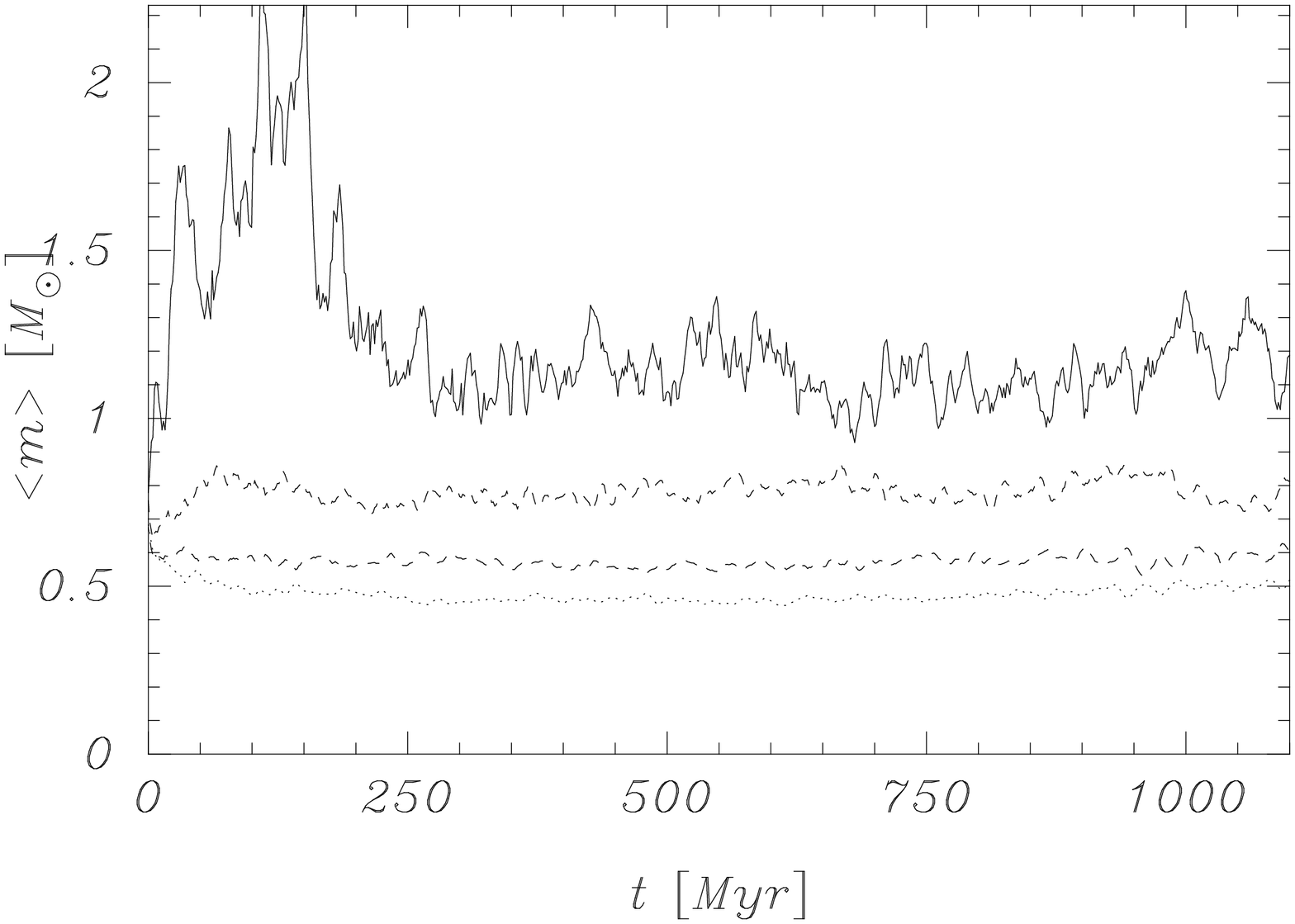,width=7.5cm,angle=-90}
(b)\psfig{figure=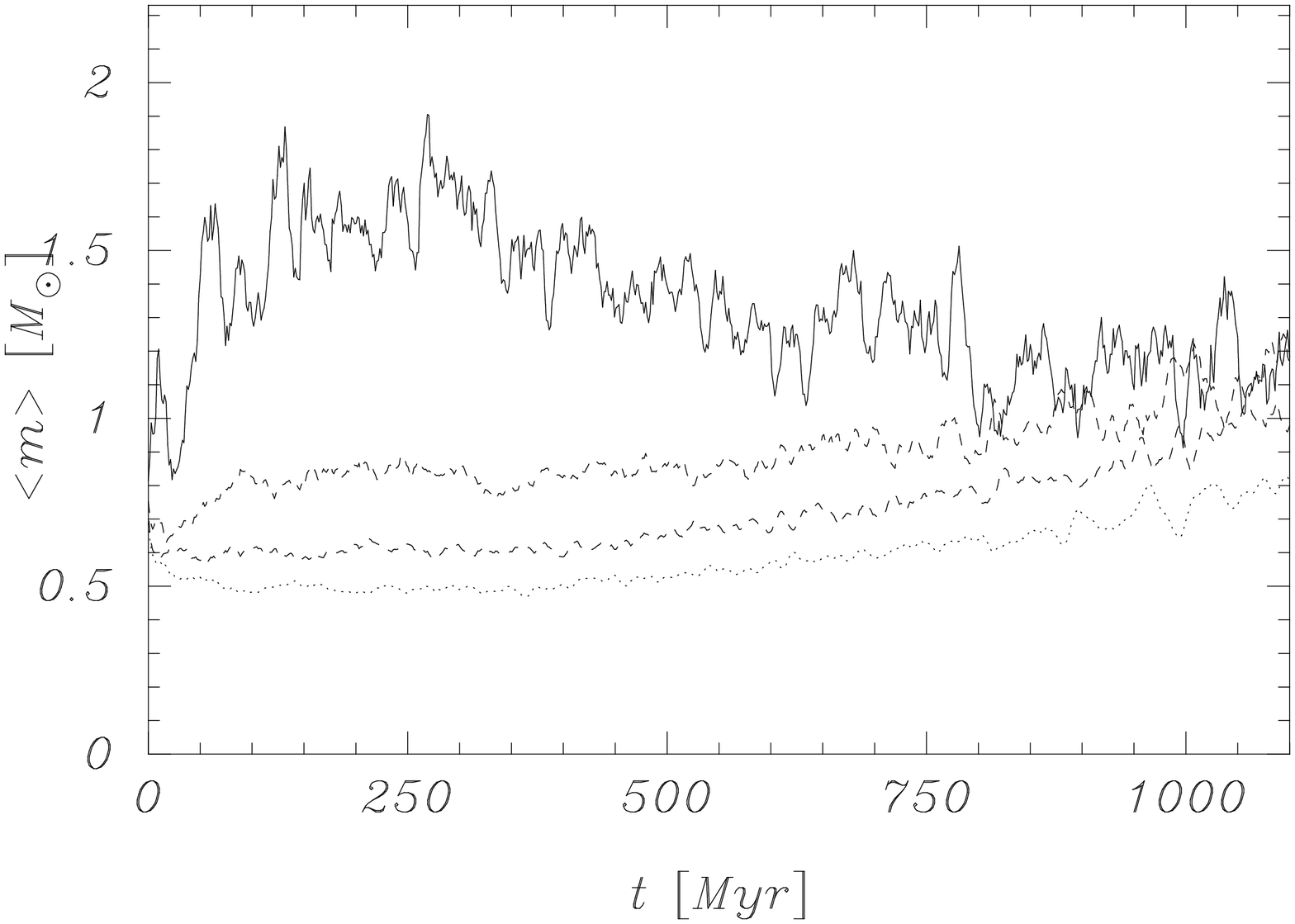,width=7.5cm,angle=-90}
\caption[]{ (a and b) The mean mass \mm\, of model W6-III (a) and
W4-II (b), as functions of time.  The data have been smoothed over
time intervals of 8.8\,Myr.  From top to bottom, the lines represent
the mean mass within the 5\% (solid), 25\% and 50\% (dashes), and 75\%
(dots) Lagrangian radii (see Fig. \ref{fig:lagrad_W6R2TF} for the
corresponding radii.)  }
\label{fig:mm_W6R2TF}
\end{figure*}

%
%
The effect of mass segregation is clearly visible in Figure
\ref{fig:mm_W6R2TF}(a), which shows the mean mass \mm\, of stars
within the 5\%, 25\%, 50\% and 75\% Lagrangian radii as functions of
time, averaged over the four W6 models.  The initial increase in the
mean mass within the inner 5\% Lagrangian radius is the result of mass
segregation.  The value of \mm\, in the cluster center decreases again
after about 100 Myr, when the most massive stars leave the mean
sequence and lose most of their mass on the Asymptotic Giant Branch.
For the remainder of the calculation \mm\, stays more or less constant
in each Lagrangian zone, but with a significantly higher value in the
inner zones. 

Figure\,\ref{fig:mm_W6R2TF}(b) shows the evolution of the mean mass
\mmean, in model W4-II. Mass segregation in this model proceeds on a
longer time scale than in the W6 models, but the cluster dissolves on
a shorter timescale. Near the end of the cluster lifetime the mean
mass in the outer regions increase, caused by the preferential loss of
the lower mass stars by evaporation. The more massive stars have
greater difficulty to climb out of the potential well.

\begin{figure}
\hspace*{1.cm}
\psfig{figure=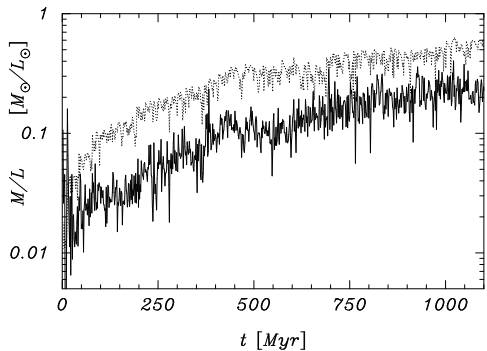,width=7.5cm,angle=-90}
\caption[]{ The mean mass to light ratio for models W6-I to IV, within
the 5\% (solid) and 75\% (dots) Lagrangian radii.}
\label{fig:ML_W6R2TF}
\end{figure}

Figure \ref{fig:ML_W6R2TF} shows the mean mass to light ratio for the
W6 models.  Except for the first few million years, the mass-to light
ratio in the cluster core is always substantially smaller (and
noisier) than that in the halo, as mass segregation causes the most
massive (i.e.~brightest) stars to sink rapidly to the cluster center.
The general increase in mass-to light ratio with time is the result of
stellar evolution, the loss of lower-mass stars by tidal stripping
compensates somewhat.  The occasional dips in the mass-to-light ratio
are caused by individual red giant stars; most of these dips occur in
the core, where mass segregation causes the giants to accumulate.  We
can reduce this ``noise'' considerably by averaging over the four W6
calculations, but the effect remains visible.



\begin{figure*}
(a)\psfig{figure=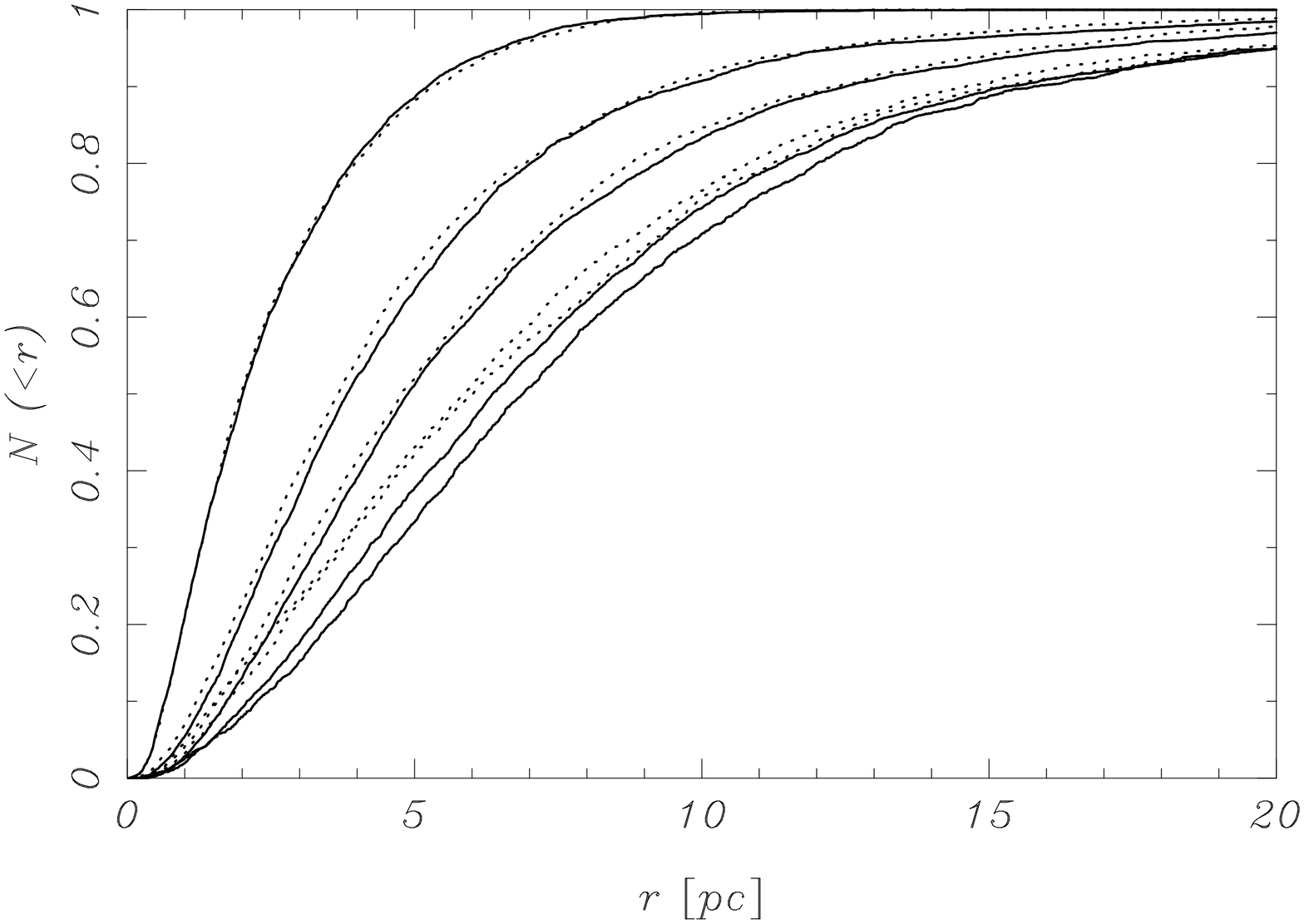,width=7.5cm,angle=-90}
(b)\psfig{figure=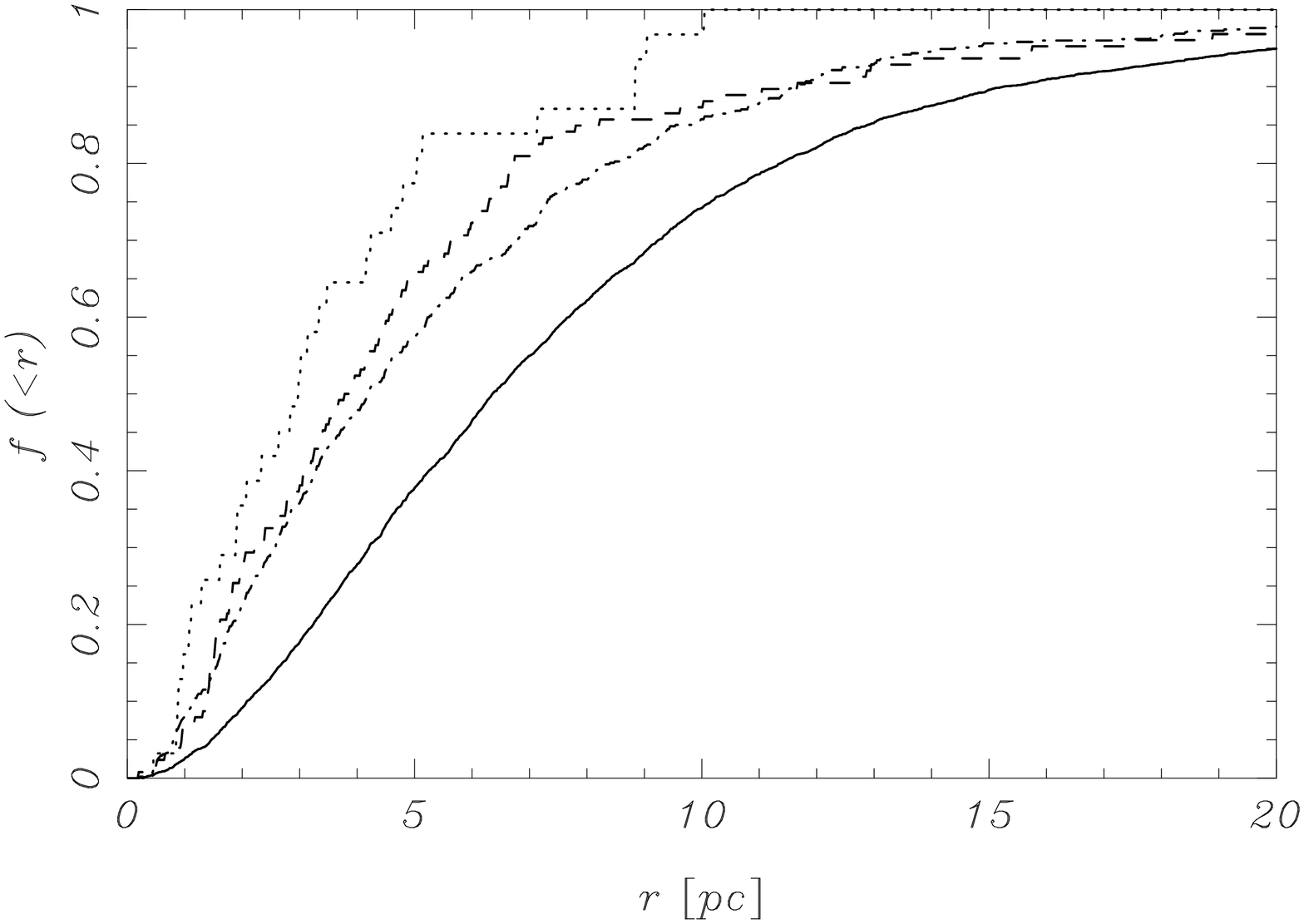,width=7.5cm,angle=-90}
\caption[]{(a) Cumulative distribution of single stars (solid lines)
and binaries (dotted lines), averaged over the four W6 models, at
(upper left to lower right) $t$ = 0, 100, 200, 600, and 800 Myr.  (b)
Cumulative distribution of various stellar populations for the four W6
runs at an age of 600 Myr.  Solid line: the distribution of all stars
(see also the lower solid line in Figure \ref{fig:cumd_W6R2TF}[a]);
dash-dotted line: stars with $L>0.5$\,\lsun; dashed line: white
dwarfs; dotted line: (sub)giants.}
\label{fig:cumd_W6R2TF}
\label{fig:cumd_rG_W6t600}
\end{figure*}

Figure \ref{fig:cumd_W6R2TF}(a) shows the radial stellar distribution
in the W6 models at various epochs.  The cluster expands as it ages.
The binaries (dotted lines) closely follow the distribution of the
single stars for the first 100 Myr, but become more centrally
concentrated at later epochs.  Mass segregation is also clearly
visible if we compare the radial distributions of low mass (faint)
stars with the more massive (bright) stars (Figure
\ref{fig:cumd_rG_W6t600}[b]).  Stars with $L>0.5$ {\lsun} are clearly
more centrally concentrated than the mean cluster star, while giants
(although there are only a few) are even more strongly concentrated in
the cluster center.

\begin{figure*}
(a)\psfig{figure=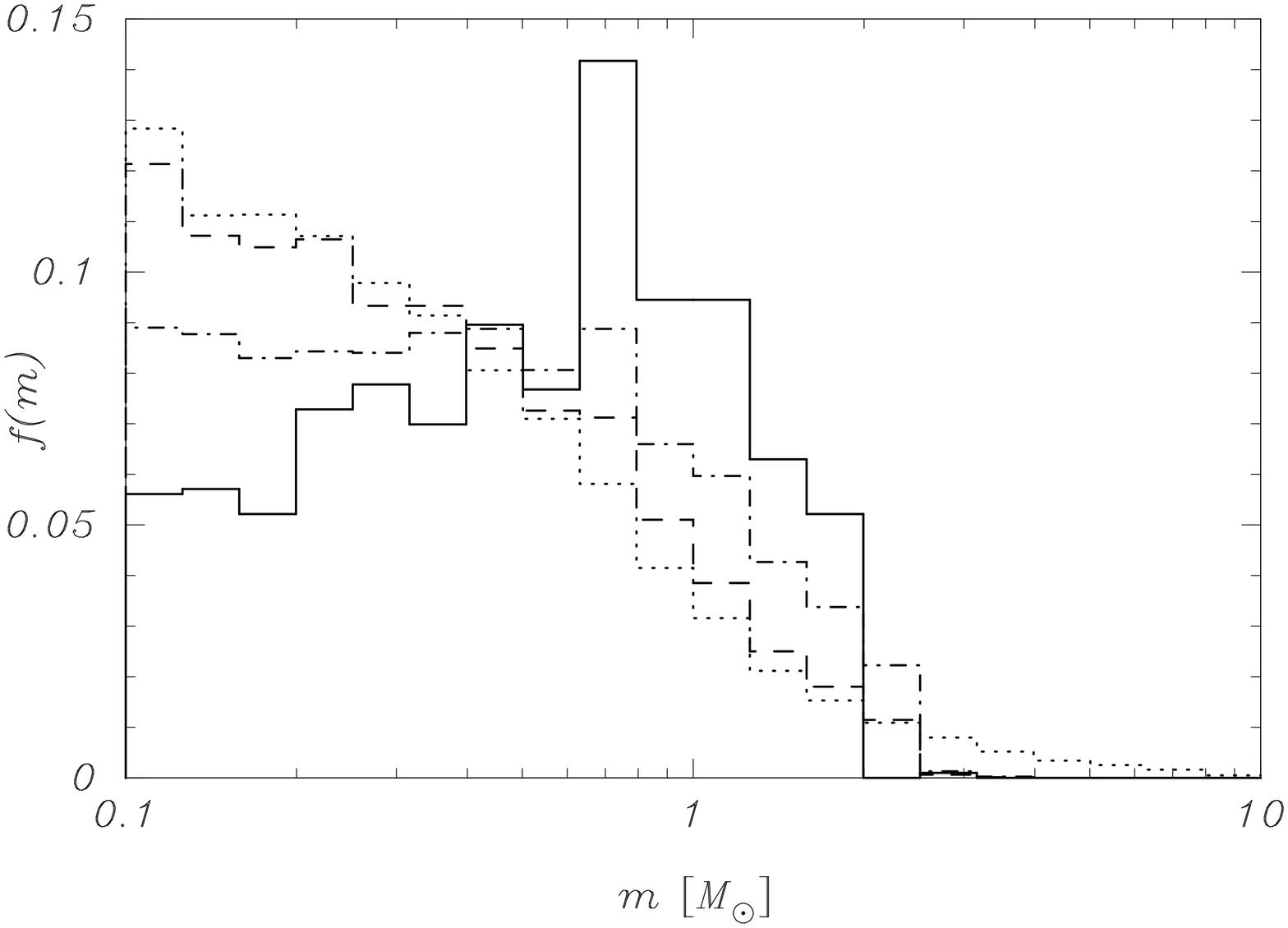,width=7.5cm,angle=-90}
(b)\psfig{figure=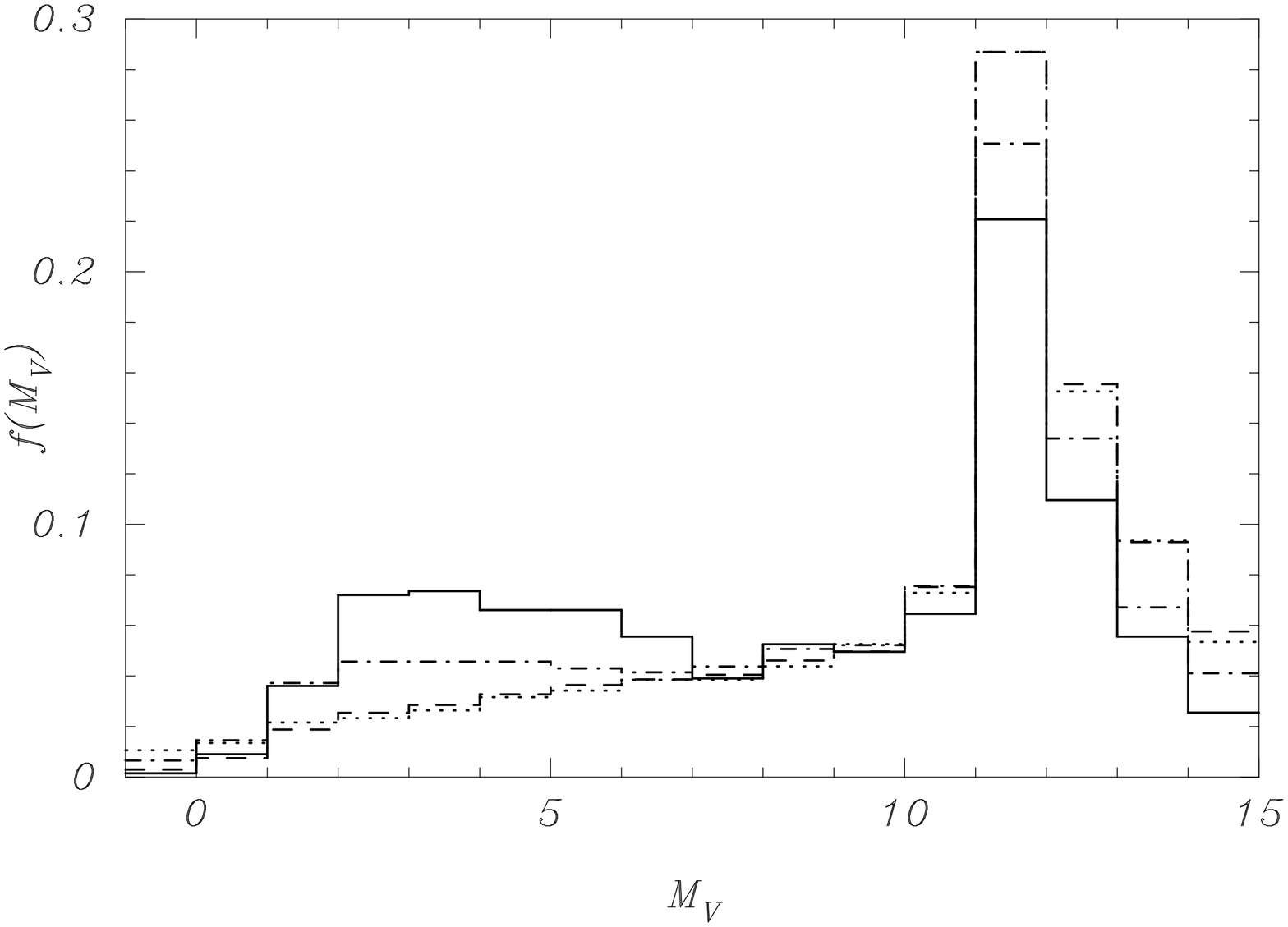,width=7.5cm,angle=-90}
\caption[]{mass function (a) and luminosity function (b) for model W6.
The dotted lines give the $t=0$ distributions and the dashed lines
give the distributions at $t=600$ Myr (averaged over all models).  The
dash-dotted lines are the mass (luminosity) functions for stars within
3.4\,pc (the half-mass radius for W6 models at $t=600$ Myr) of the
density center (for [b] in projection, as viewed along the $x$ axis).
The solid line gives the mass fucntion of the W6 models within 3.4\,pc
(projected again for [b]) at an age of 1100\,Myear.  }
\label{fig:Vt600_W6W4}
\label{fig:Mft600_W6.ps}
\end{figure*}

Mass segregation can also be observed in the cluster's mass and
luminosity functions.  Figure \ref{fig:Mft600_W6.ps}(a) shows global
mass functions for all single stars and binary primaries for the W6
models at birth and at $t=600$ Myr.  The global mass function of the
cluster is affected only slightly by stellar evolution and mass
segregation.  However, the mass function at $t=600$ Myr for stars in
the inner part of the cluster (dot-dashed line) is clearly different
from the global mass function at that time.

The white dwarfs are more centrally concentrated than stars with
luminosity $>0.5$\,\lsun\, and slighly less concentrated that the
giants (see Fig.\,\ref{fig:cumd_rG_W6t600}[b]). This is caused by the
progenitors of the white dwarfs, the giants, being centrally
concentrated while after their envelepes are shed they have masses
comparable to the the mean cluster stars. Segregating outwards takes
more time than sinking inwards and at the same time more white dwarfs
are produced in the cluster center (see the end of this section for
more detailes).

Figure \ref{fig:Vt600_W6W4}(b) shows the luminosity
functions (in $M_V$) for stars (including binaries) in models
W6 at zero age and at 600 Myr.  Note that the luminosity function
at the later time shows a {\em larger} faction of bright stars.  
The reason is two fold: 1) stellar evolution has removed only the most
massive stars by this time, while mass segregation has concentrated
the remaining massive (bright) stars in the cluster core, at the same
time causing the lower mass stars to escape and 2) heaviest stars turn
into giants, which for older (lighter) stars are much brighter than
main-sequence stars, whereas for younger (heavier) stars, the
difference in brightness between giant and main-sequence star is much
smaller.

The W4 models more strongly affected by mass segregation (not shows
but see sect\,\ref{sect:discussion}).  In part, this is caused by the
more rapid evaporation of these models compared to the W6 models.
This is consistent with the findings of Takahashi \& Portegies Zwart
(2000), who noted that clusters which are close to complete disruption
contain a higher fraction of high-mass stars.

%
%
%


\subsection{Hertzsprung--Russel diagrams}\label{sec:HRD}
\begin{figure*}
(a)\psfig{figure=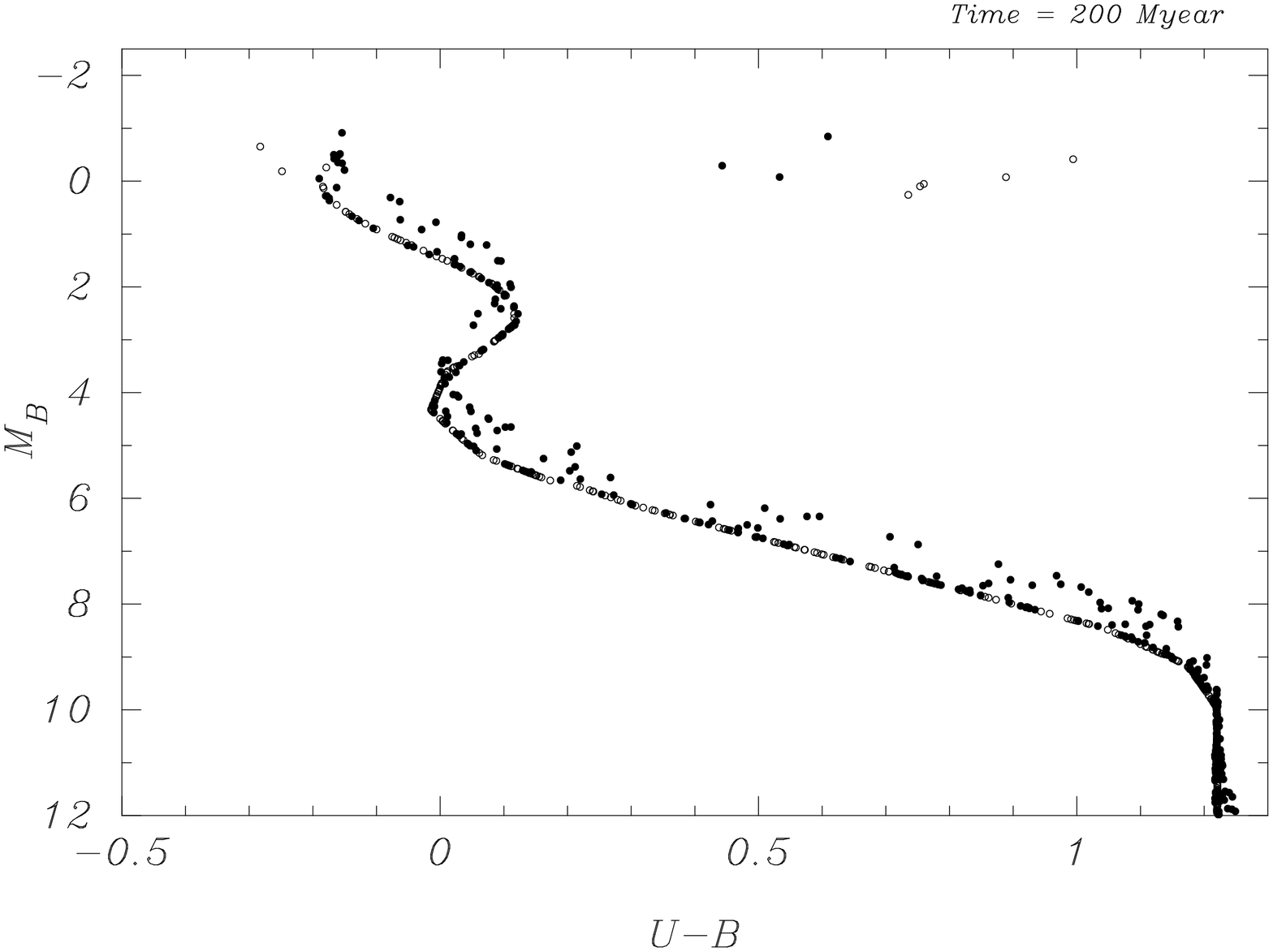,width=7.5cm,angle=-90}
(b)\psfig{figure=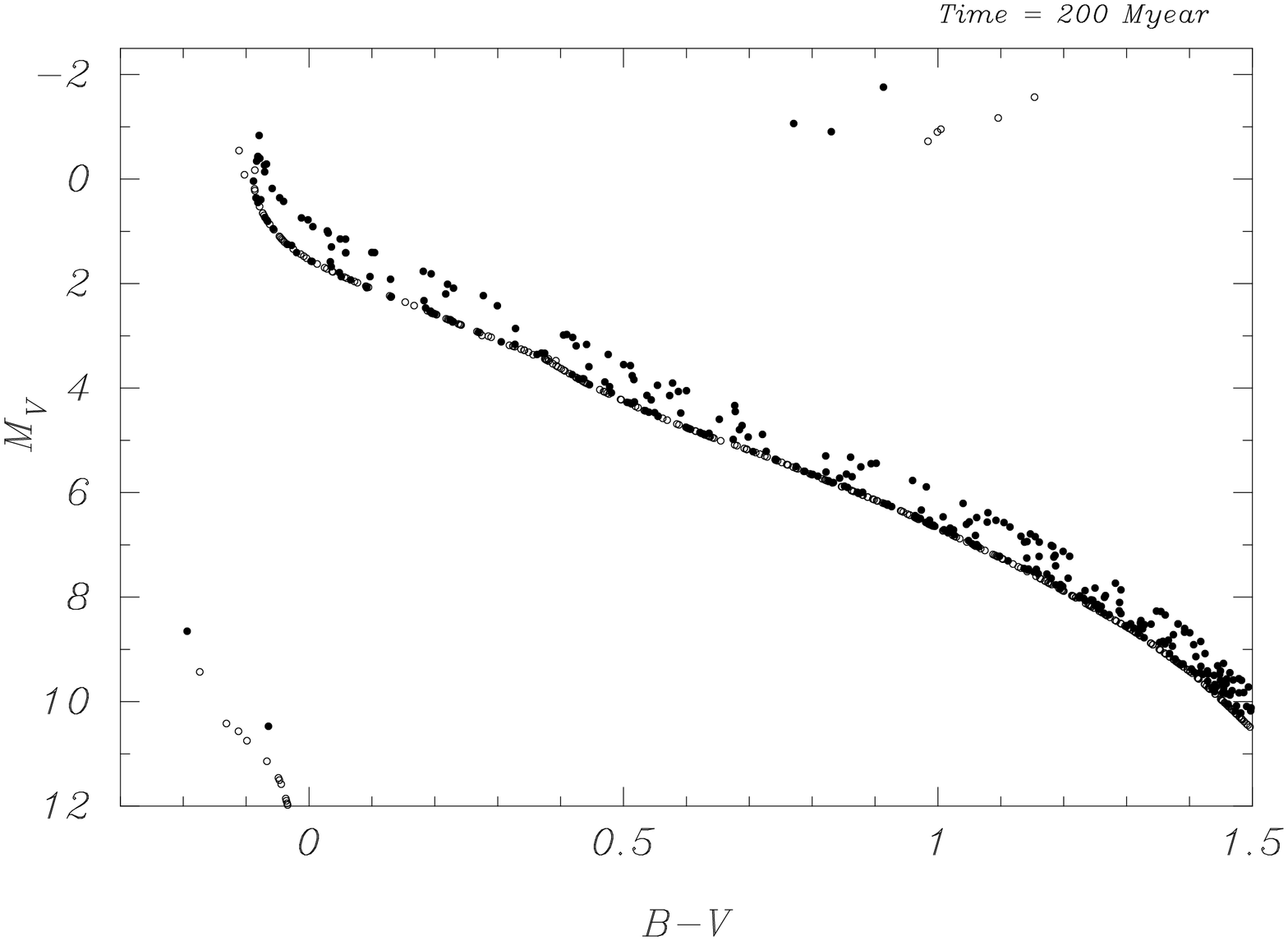,width=7.5cm,angle=-90}
(a)\psfig{figure=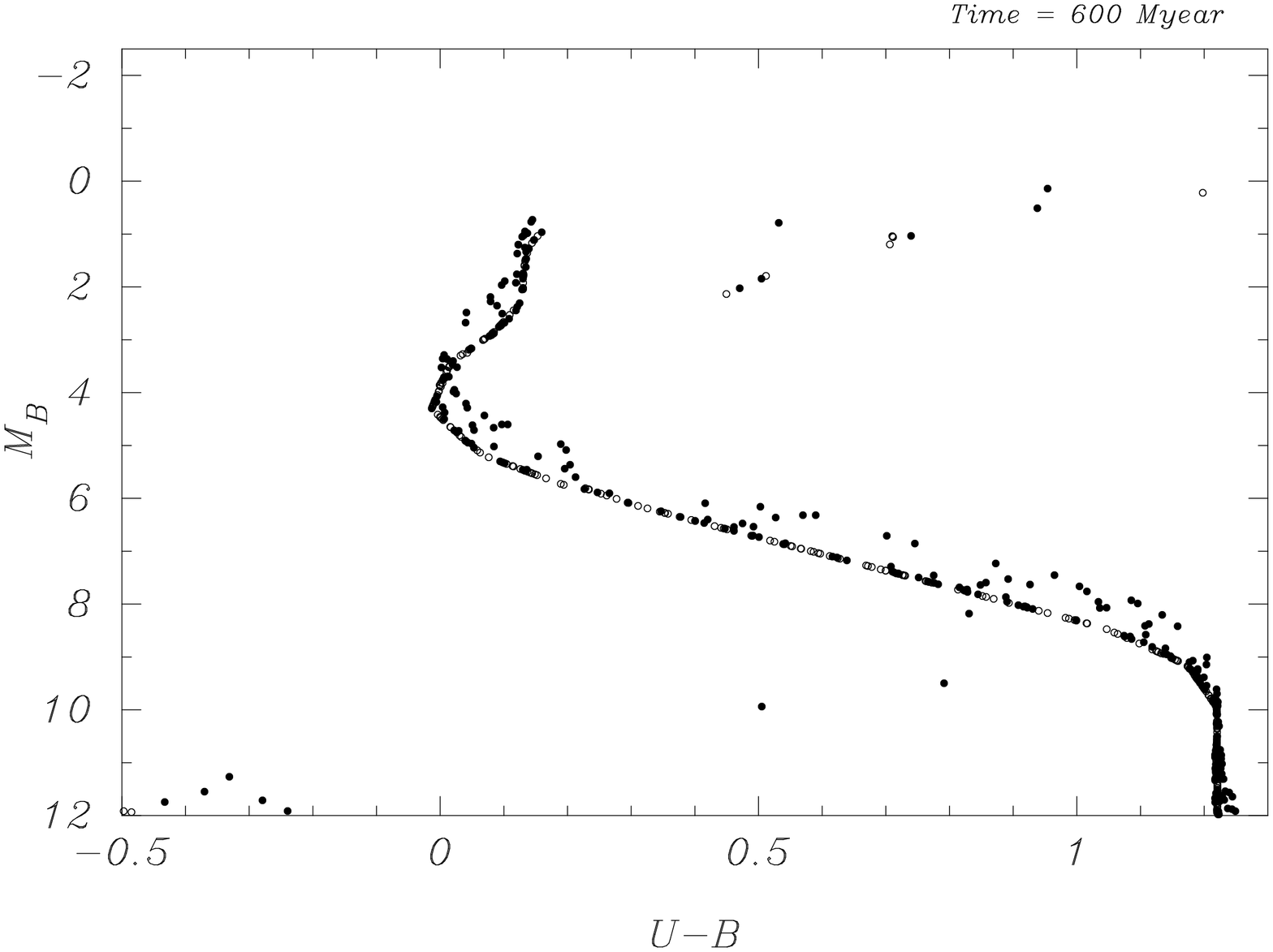,width=7.5cm,angle=-90}
(b)\psfig{figure=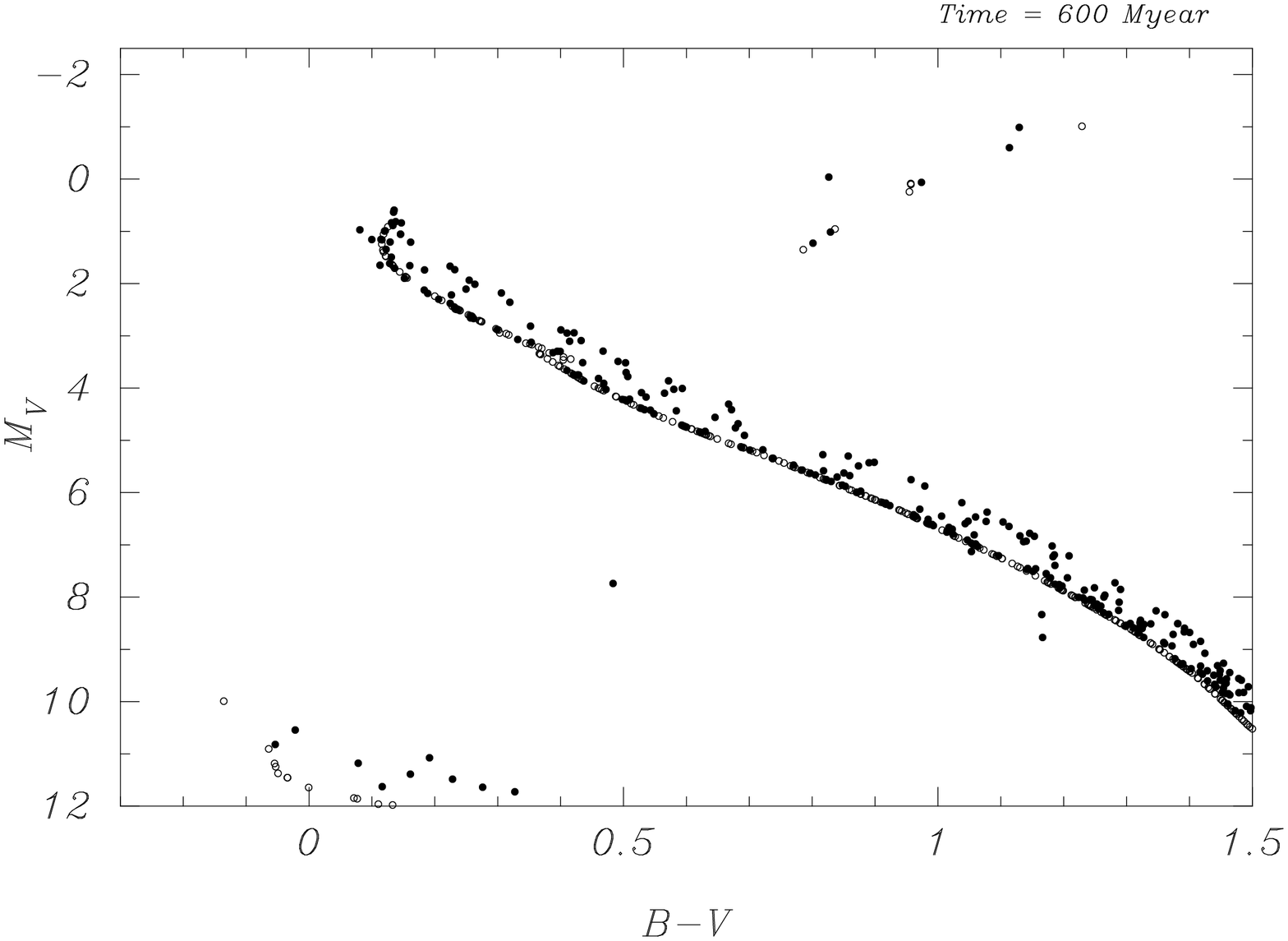,width=7.5cm,angle=-90}
(a)\psfig{figure=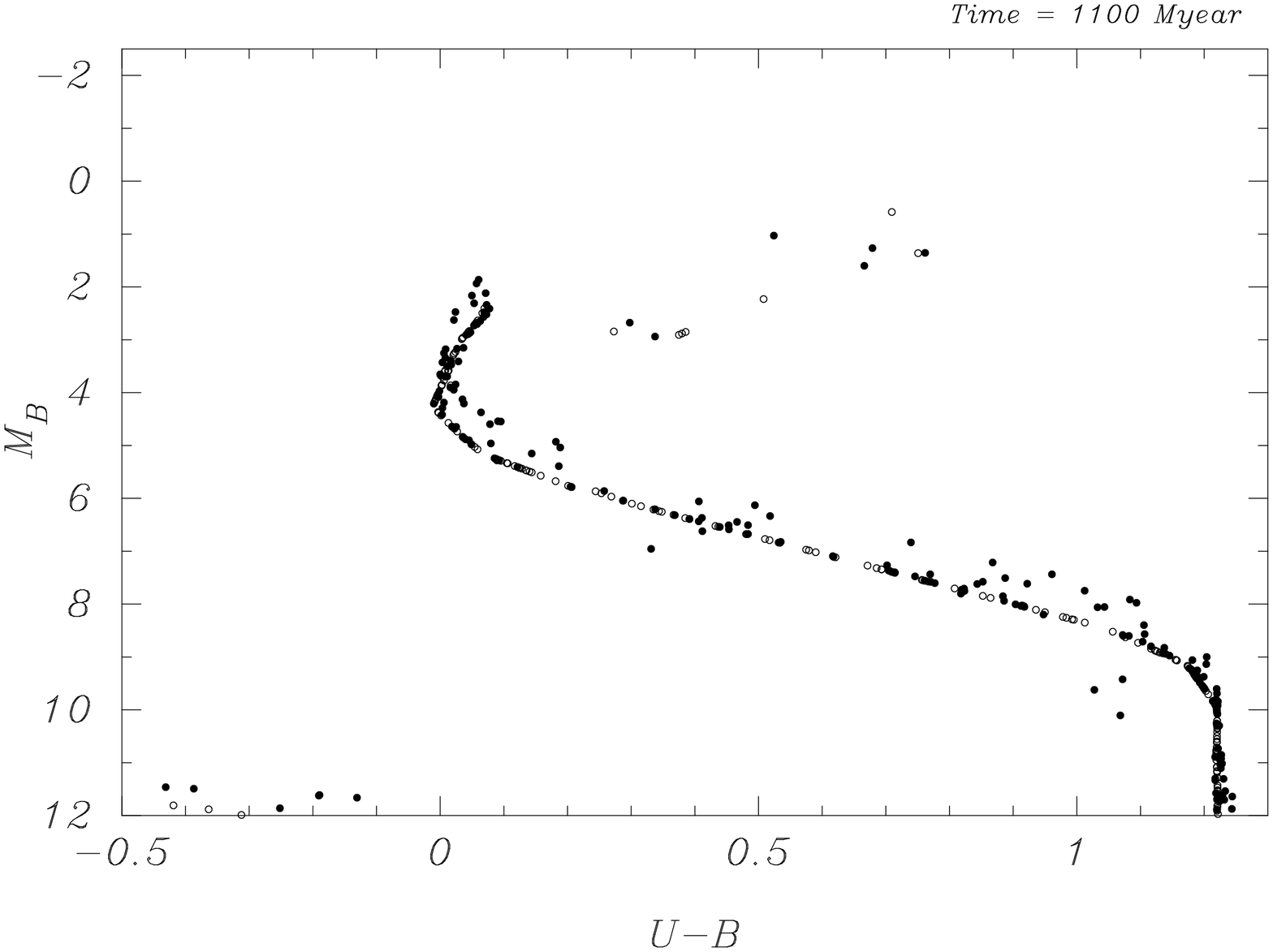,width=7.5cm,angle=-90}
(b)\psfig{figure=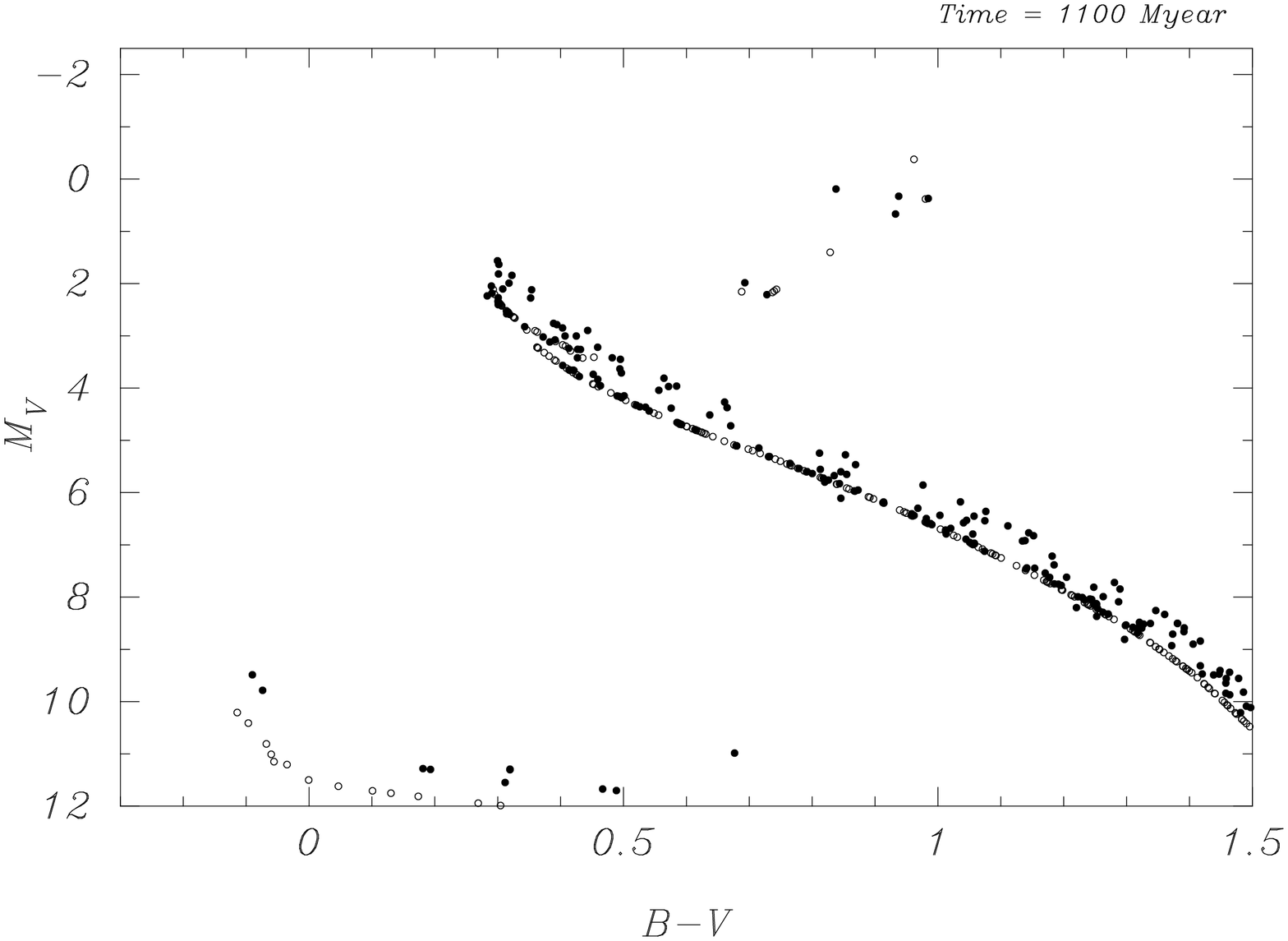,width=7.5cm,angle=-90}
\caption[]{{\HRD} of all stars in model W6-III at ages of 200 Myr
$U-B$ in the left panels and $B-V$ in the right panels (upper panels;
1983 objects), 600 Myr (middle; 1603 objects), and 1100 Myr (lower
panels; 1009 objects).  }
\label{fig:fig_HRD_W6}
\end{figure*}

Figure \ref{fig:fig_HRD_W6} shows a time sequence of {\HRD}s for model
W6-III.  The youngest {\HRD} (200 Myr) already shows a white dwarf
sequence.  Note the densely populated ``binary sequence'' $\sim$0.75
magnitudes above the zero-age main-sequence.  One of the objects in
the middle panel (close to but just above the turnoff) is a blue
straggler; the other two are binaries (see also
Figure \ref{fig:Bss_W6III}).  The objects immediately to the left of
the main sequence (the two points in the 600 Myr diagram at $B-V \sim
1.18$ and the single point at $B-V \sim 0.5, V = 8$) are binaries
containing a mass-transfer remnant (a helium star) and a main-sequence
star which has accreted part of its companion's envelope.  Farther to
the blue (between $B-V = 0$ and 0.8), but to the right of the white
dwarf sequence, are binaries consisting of a white dwarf and a
low-mass main-sequence star (several are seen in the 600 Myr and 1100
Myr diagrams). In the bottom panel a break and discontinuity in the
zero-age main-sequence is visible near $B-V = 0.4$ and at $V\simeq
4$. This is an artifact of the stellar evolution fitting formulae
given by Eggleton, Fitchet \&Tout (1989) and appears when the envelope
of a main-sequence star becomes convective.

\begin{figure}
\psfig{figure=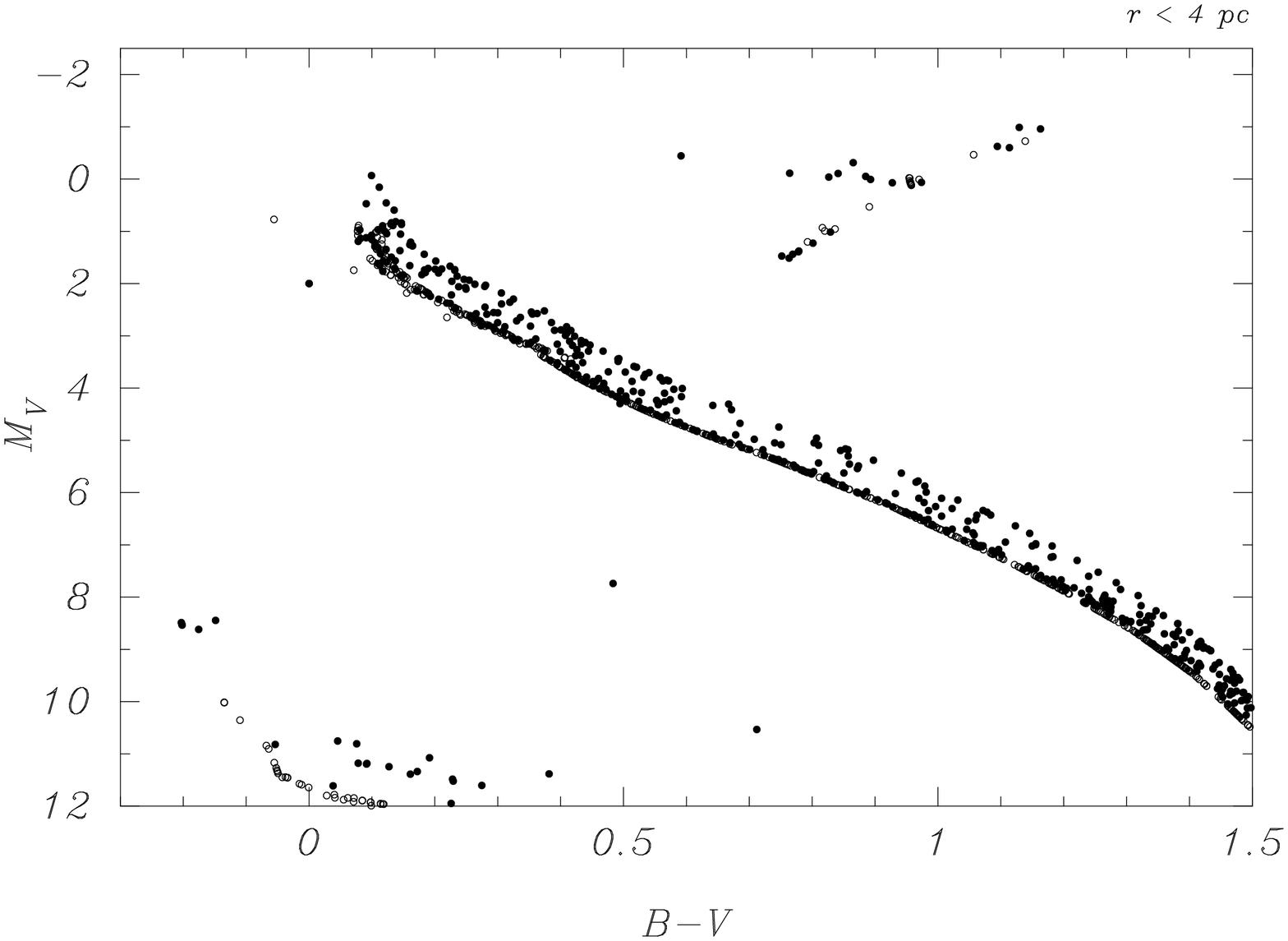,width=7.5cm,angle=-90}
\psfig{figure=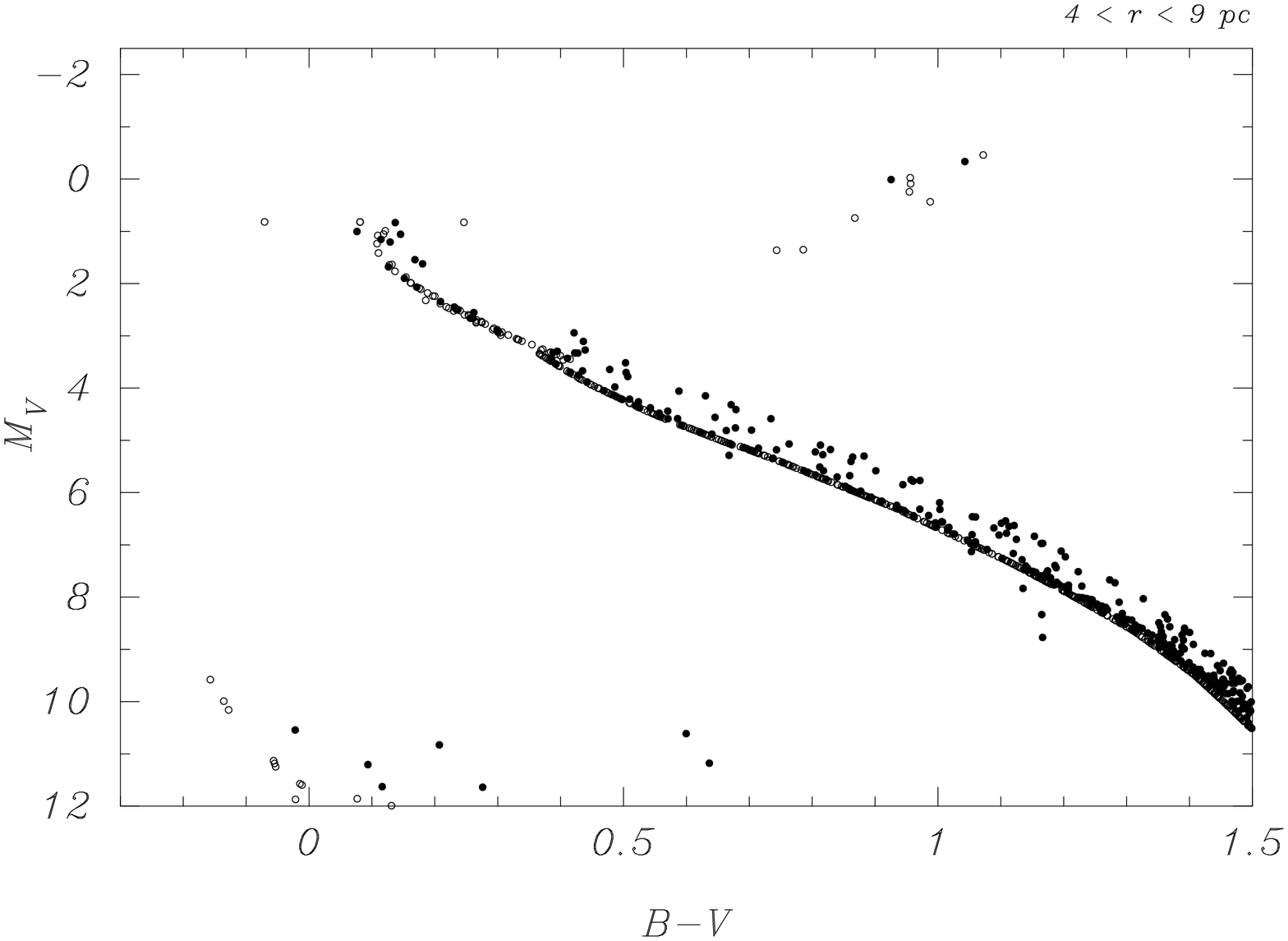,width=7.5cm,angle=-90}
\psfig{figure=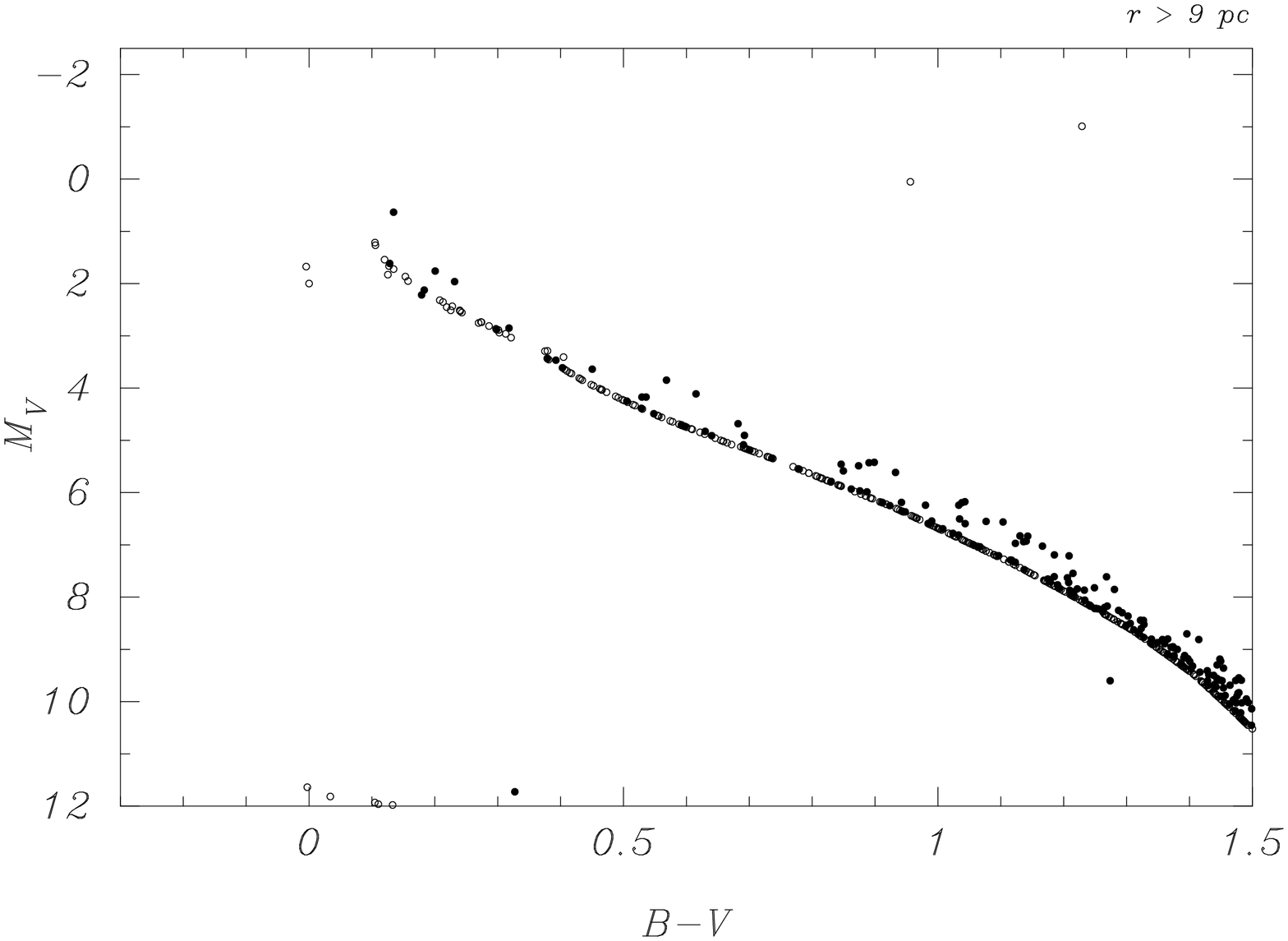,width=7.5cm,angle=-90}
\caption[]{{\HRD}s of the combined W6 models at an age of about 600
Myr. The upper panel shows the innermost (non-projected) 4 pc (2004
objects), the middle panel stars between 4 and 9 pc from the cluster
center (2658 objects), and the bottom panel stars more than 9 pc from
the center (2039 objects).}
\label{fig:fig_BmVt600}
\end{figure}

Figure \ref{fig:fig_BmVt600} shows {\HRD}s for the inner, middle and
outer regions of the combined W6 models at an age of about 600 Myr,
and illustrates the effect of mass segregation.  Each diagram contains
about 2000 objects.  The slight ``fuzziness'' near the main-sequence
turnoff is the result of variations in output times between individual
simulations, which cause the combined diagram to have a small spread
in stellar ages.  About one quarter (one run) of the stars come from a
slightly younger cluster with an age of about 550\,Myr.

%

The {\HRD}s of the inner and outer parts of the cluster show
significant differences due to mass segregation.  Most are a
consequence of the evolving binary population, and will be discussed
in more detail in Paper IVb.  The inner HRD has a clear excess of
(sub)giants and white dwarfs relative to the HRD at the half mass
radius or that in the halo.  Also, the turnoff region is more heavily
populated in the inner HRD than in the others.  Striking also is the
lack of a clear binary sequence in the outer \HRD.  The bottom of the
main-sequence is less clearly affected by mass segregation.

\subsection{Blue Stragglers}

\begin{figure}
\psfig{figure=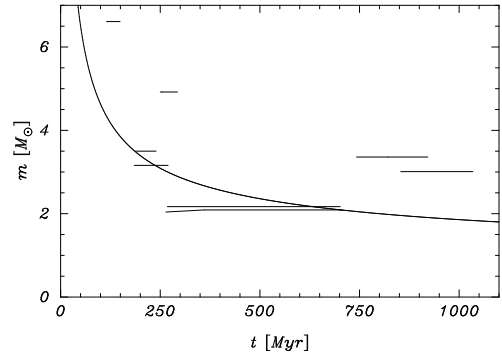,width=7.5cm,angle=-90} 
\caption[]{Blue stragglers in model W6-III.  The solid curve gives the
turnoff mass (in \msun) as a function of time (in Myr).  The
horizontal lines represent the tracks of the blue stragglers in model
W6-III.  The tracks start when the star is rejuvenated (see Appendix
B), and stop when the blue straggler leaves the main sequence.  }
\label{fig:Bss_W6III}
\end{figure}

The (small) numbers of blue stragglers do not depend strongly on the
particular region of the cluster under study.  We count four blue
stragglers in the inner {\HRD}, and one and two in the middle and
outer frames of Figure \ref{fig:fig_BmVt600}, respectively.  These
numbers are fairly typical of our simulated clusters and also quite
typical for the numbers observed (see Tab.\,\ref{Tab:observed}).
Fig\,\ref{fig:Bss_W6III} presents a graphical representation of the
blue stragglers in model W6-III.  (A main-sequence star is identified
as a blue straggler as soon as its mass exceeds the turnoff mass for
that epoch.)

Most blue stragglers are the result of mass transfer in a close
binary.  In about half of all cases (37 out of 76 blue stragglers
formed in all calculations performed), the mass transfer is unstable,
leading to a merger.  Blue stragglers formed from a stable phase of
mass transfer are generally accompanied by a white dwarf or helium
star (the young remnant of mass transfer, see
Figure \ref{fig:Bss_W6III}), causing the blue stragglers to lie
slightly blueward of the turnoff in the {\HRD} (see \S\ref{sec:HRD}).

Three blue stragglers formed via collisions in which a third star
interacted with and became bound to binary, leading to a collision
between the binary components.  The orbit of such a blue-straggler
binary is generally quite eccentric, and the companion to the blue
straggler is most likely to be a main sequence star.  Observing a blue
straggler in an elliptical orbit around a stellar (main-sequence)
companion would provide strong evidence for such dynamical
interactions in star clusters (see Portegies Zwart 1996).

In none of our calculations a blue straggler with more than twice the
turn off mass was formed i.e., there where no collisions between three
or more stars.  A discovery of a blue straggler with a mass more than
twice the turn-off mass would provide strong evidence for effects of
stellar dynamics, though one could imagine a primordial triple to get
into a common-envelope situation in which all three stars spiral in to
a triple merger. Portegies Zwart et al.\, (1999) find runaway
collisions between more than two stars in their simulations of dense
stars clusters without primordial binaries. Thier results, however,
are applicable for a different range of initial conditions, as they
studied the dence and young central star clusters R\,136 in the 30
Doradus region of the Large Machelanic Could.

Many blue stragglers experience mass transfer or a collision long
before actually being classified as blue stragglers by our criterion
(i.e.~exceeding the turnoff mass).  In most of these cases, one or
more phases of mass transfer (stable or unstable, or even accretion
from the stellar wind of a companion) has rejuvenated one of the stars
in a close binary system (see Appendix B).  As the cluster ages, the
star remains behind on the main sequence, and eventually becomes
identifiable as a blue straggler (see also paper II).  This is
illustrated in Figure \ref{fig:Bss_W6III} (the two long tracks with
$m\sim2$ and the track near $m\sim3$).

A blue straggler which was rejuvenated long ago may show no trace of
the event that caused its rejuvenation.  Apart from residing above the
turnoff, the star may appear completely normal; anomalous atmospheric
abundances will have had sufficient time to mix with the stellar
interior.  In addition, if the blue straggler is rejuvenated only a
little, the maximum distance on the {\HRD} between the cluster turnoff
and the blue straggler will be very small; such a star may remain
unidentified as a blue straggler.  This may happen if mass transfer is
unstable but does not lead to a merger, or if a binary is too wide for
Roche-lobe overflow, and the blue straggler is rejuvenated by
accreting a small portion if its companion's stellar wind.

The lifetime of a blue straggler depends on the epoch at which it
formed.  Blue stragglers that form later are generally products of
lower-mass stars, and tend to live longer than blue stragglers that
formed early in the evolution of the stellar system.


\subsection{Isophotes}\label{sect:isophotes}
\begin{figure}
\hspace*{1.cm}
\psfig{figure=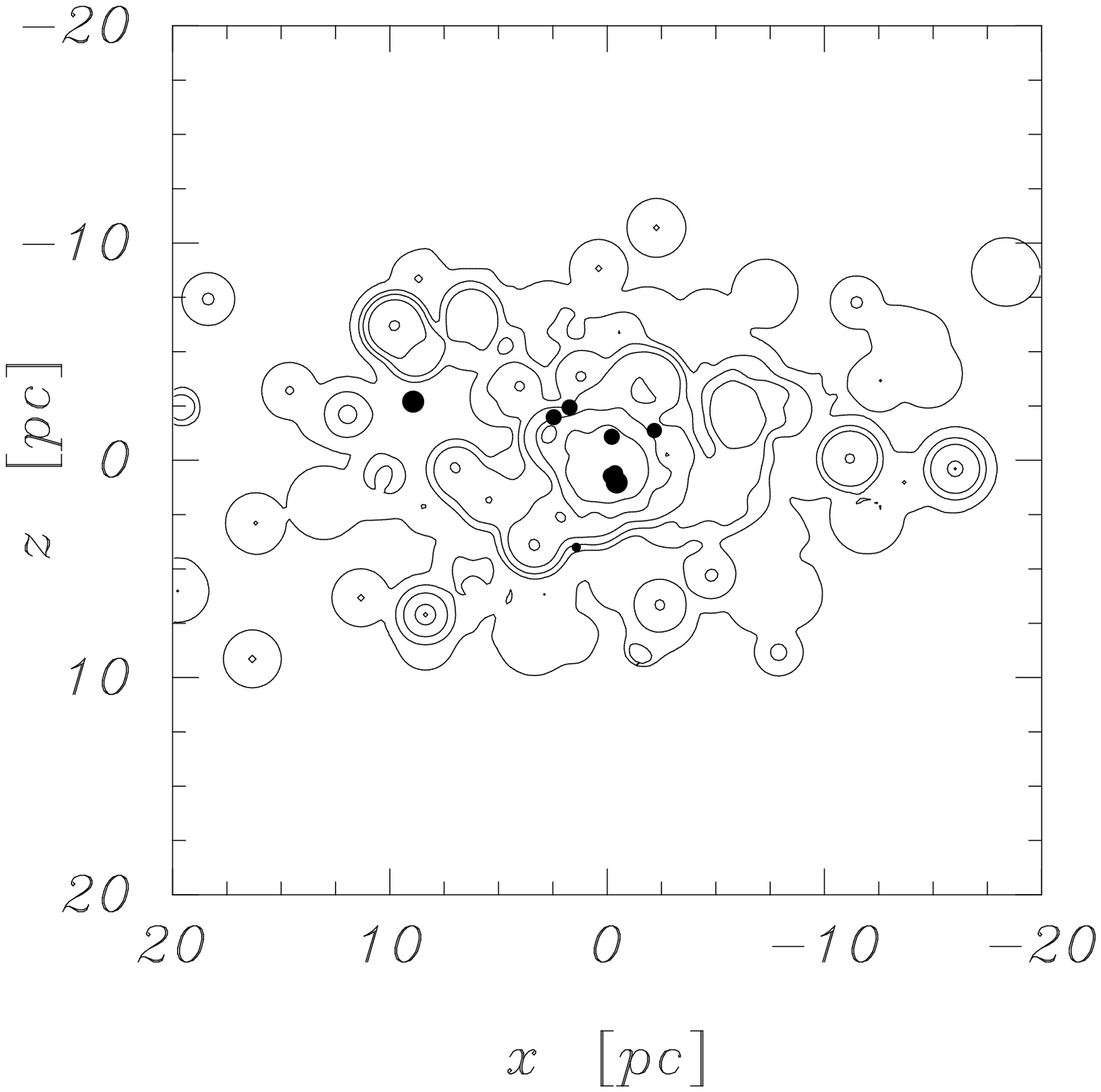,width=7.5cm,angle=-90}
~\psfig{figure=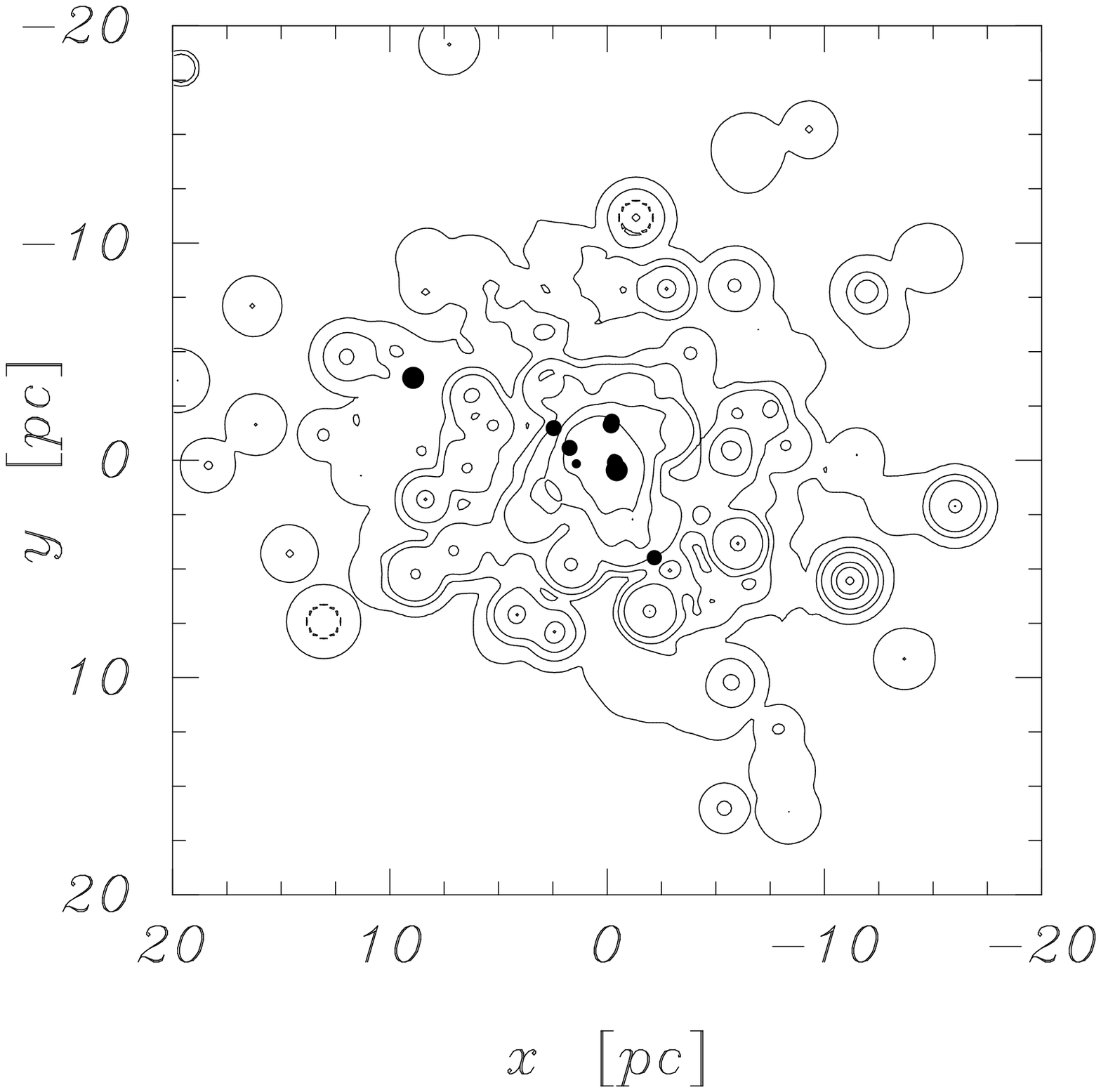,width=7.5cm,angle=-90}
\hspace*{1.cm}
~\psfig{figure=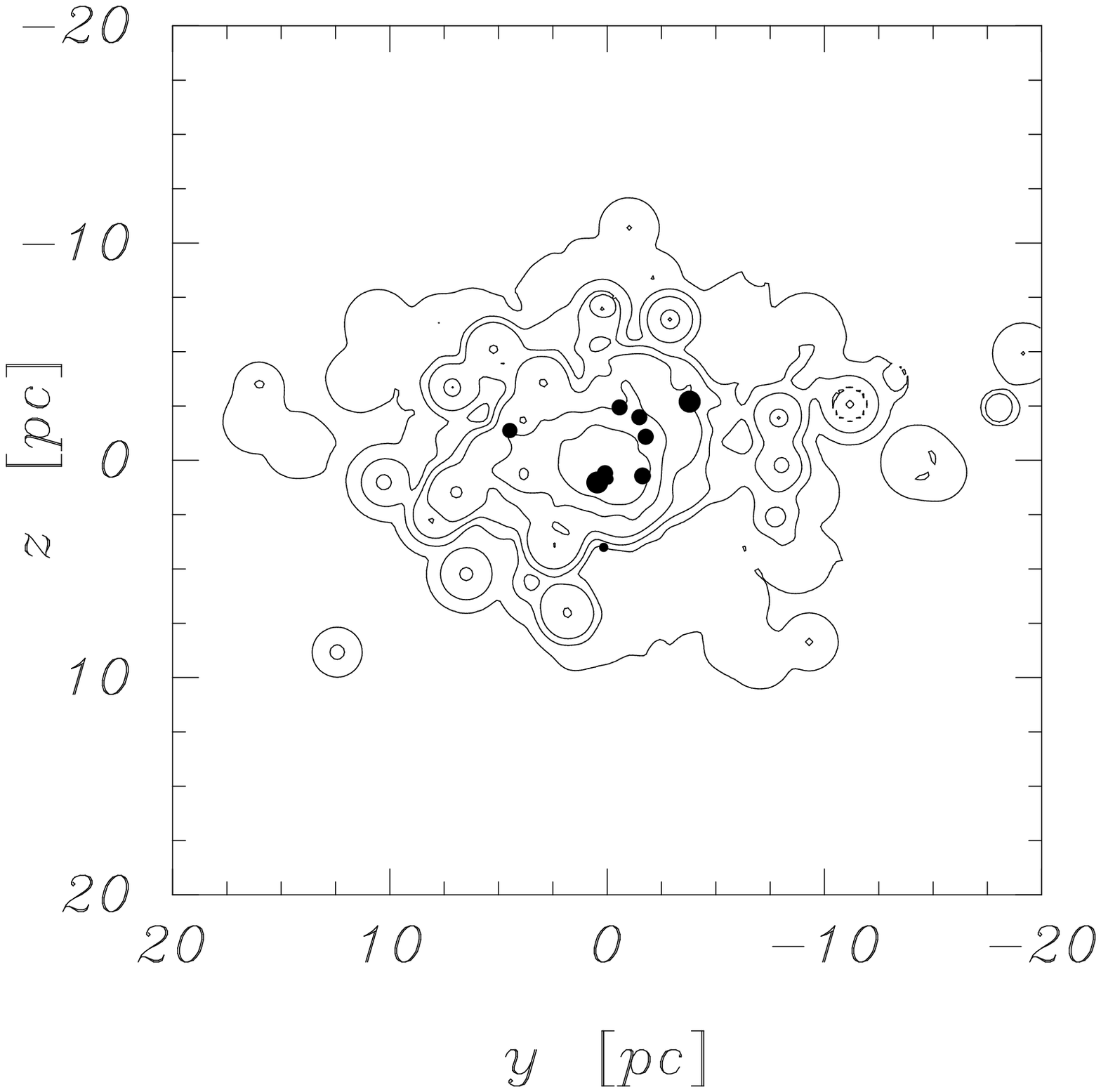,width=7.5cm,angle=-90}
\caption[]{Isophotes in $M_V$ for model W6-III at a cluster age of 600
Myr.  The three panels present views along the three coordinate axes.
The 10 (sub)giants are plotted as dots with size proportional to
magnitude (and are excluded from the isophotes).  The brightest region
in the continuum plot has a surface brightness of -2.2 mag/pc$^{-2}$.
Contours are plotted at -0.86 mag\,pc$^{-2}$, 0.34 mag\,pc$^{-2}$,
1.6\,mag pc$^{-2}$, 2.8 mag\,pc$^{-2}$, 4.1 mag\,pc$^{-2}$, and 7.8
mag\,pc$^{-2}$.  Stars are assigned a Gaussian point spread function
with a dispersion of 0.35 pc.}
\label{fig:isoVt600_W6R2TF}
\end{figure}

Figure \ref{fig:isoVt600_W6R2TF} shows a series of isophotes, as seen
from various directions, for model W6-III at an age of 600 Myr.  The
Galactic center is located to the $-x$ direction (at a distance of
about 12.1\,kpc---see Table \ref{Tab:init}), and $z$ points toward the
Galactic north pole.  While the cluster is barely flattened at
birth,\footnote{This is simply a consequence of the fact that only the
outermost parts of the cluster, near the Jacobi surface, show
significant flattening, and these are initially very sparsely
populated.  Only when cluster evolution drives many stars out to the
Jacobi radius does the flattening become readily apparent.}  by 600
Myr the cluster is significantly flattened by the Galactic tidal
field.  As expected, the flattening is greatest along the $z$ axis,
and also noticeable in the $y$ direction.


\begin{figure}
\framebox[7cm][c]{figure 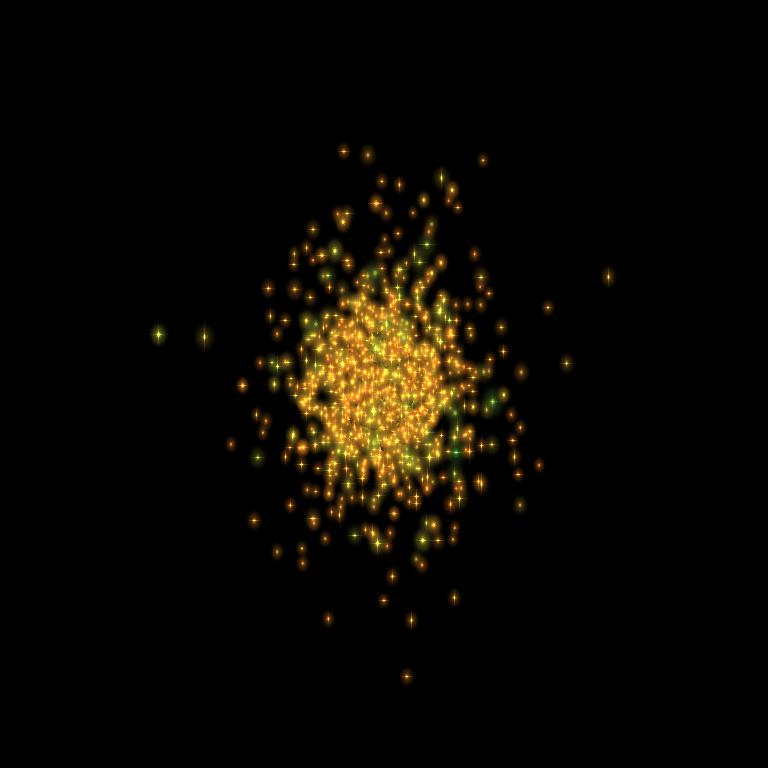}
\framebox[7cm][c]{figure 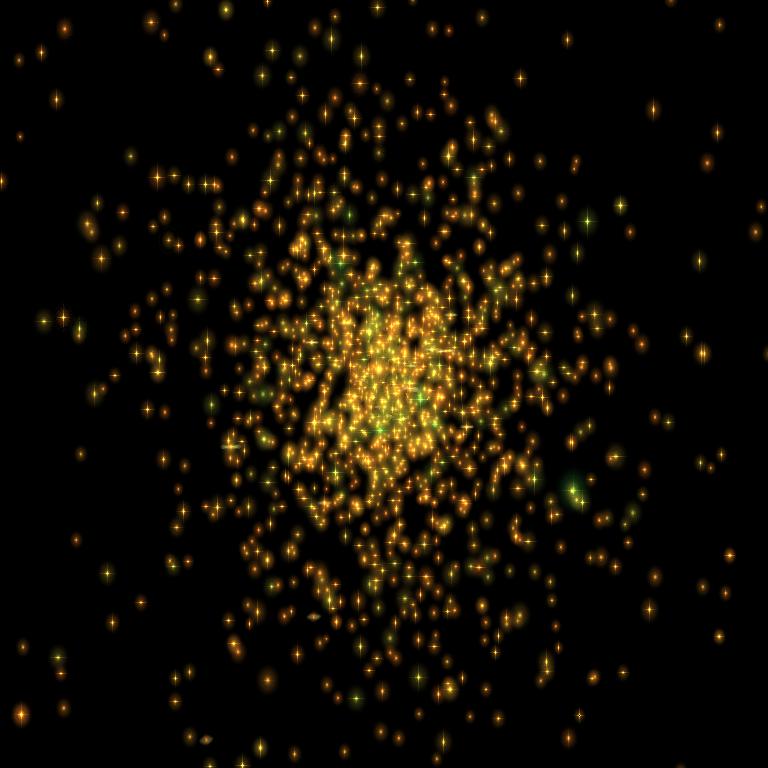}
\framebox[7cm][c]{figure 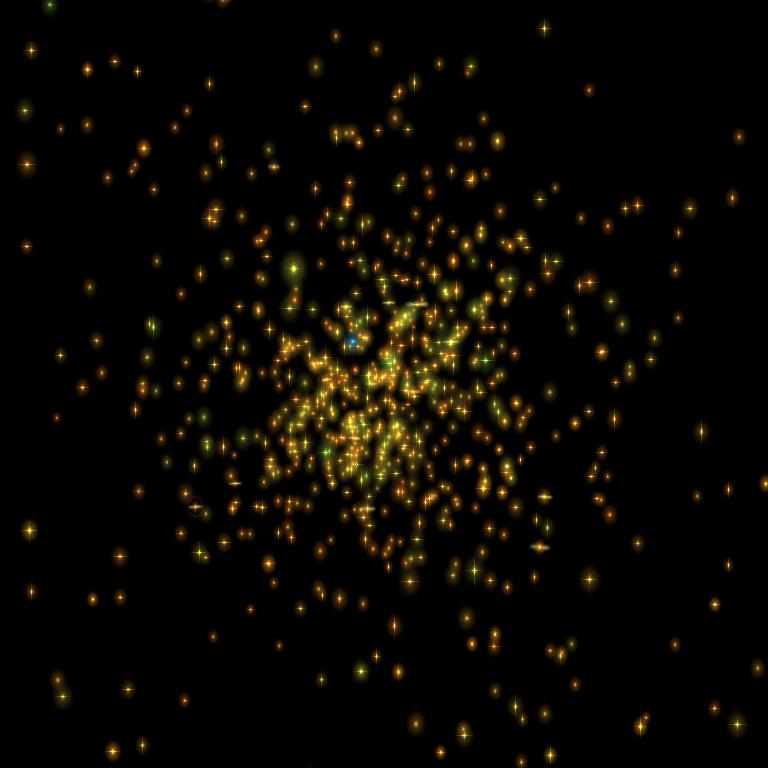}
\caption[]{Visualization of model W6-III at zero age (top image), at
an age of 622\,Myr and at an age of 1512\,Myr.  Images were created
using a ray-tracing technique.  
}
\label{fig:fig_POV_W6}
\end{figure}

Figure\,\ref{fig:fig_POV_W6} shows images of model W6-III at three
different moments in time. The images are created using a ray-tracing
technique.
 

\subsection{Escaping stars}
\begin{figure*}
~\psfig{figure=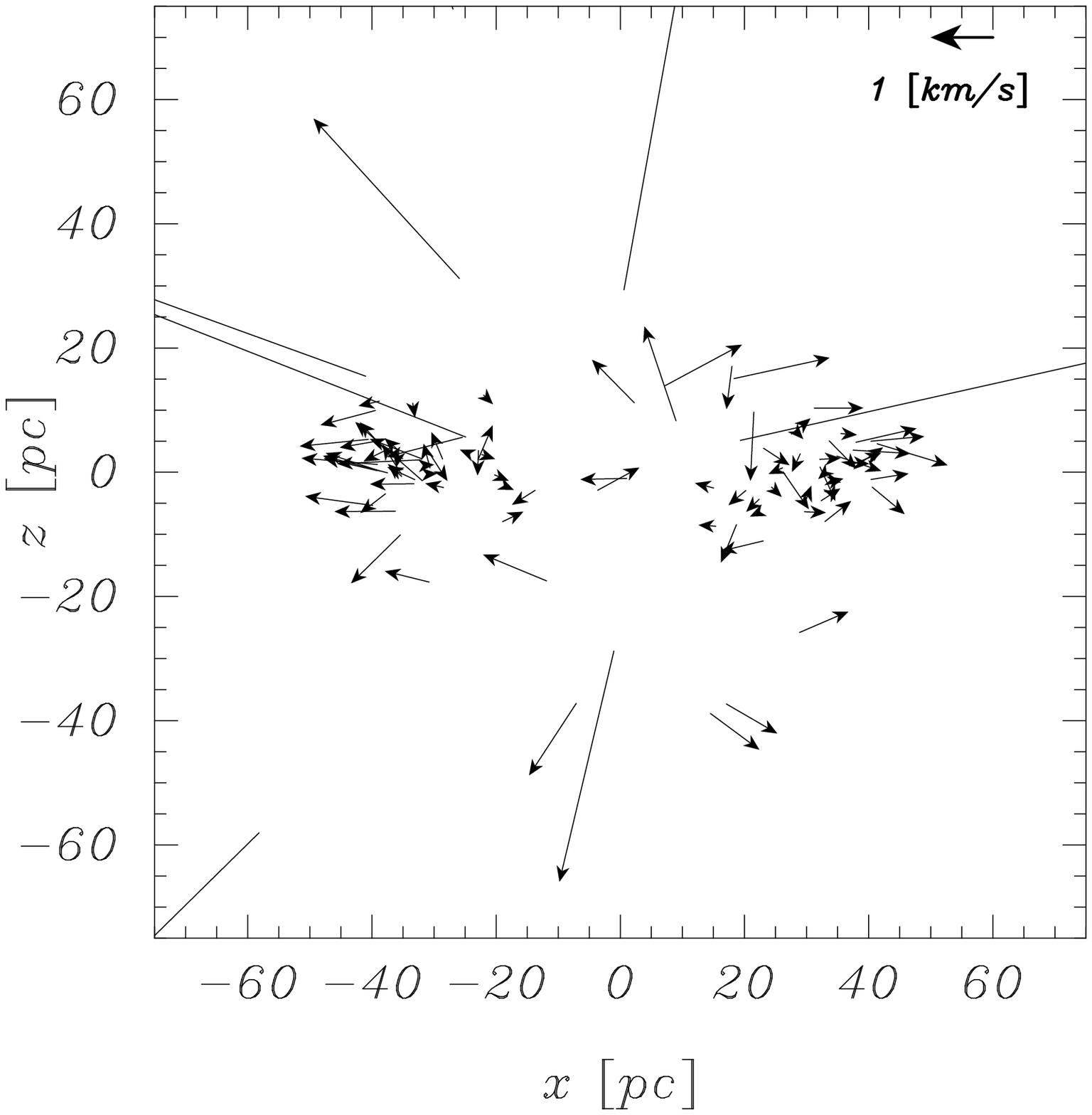,width=7.5cm,angle=-90} 
~\psfig{figure=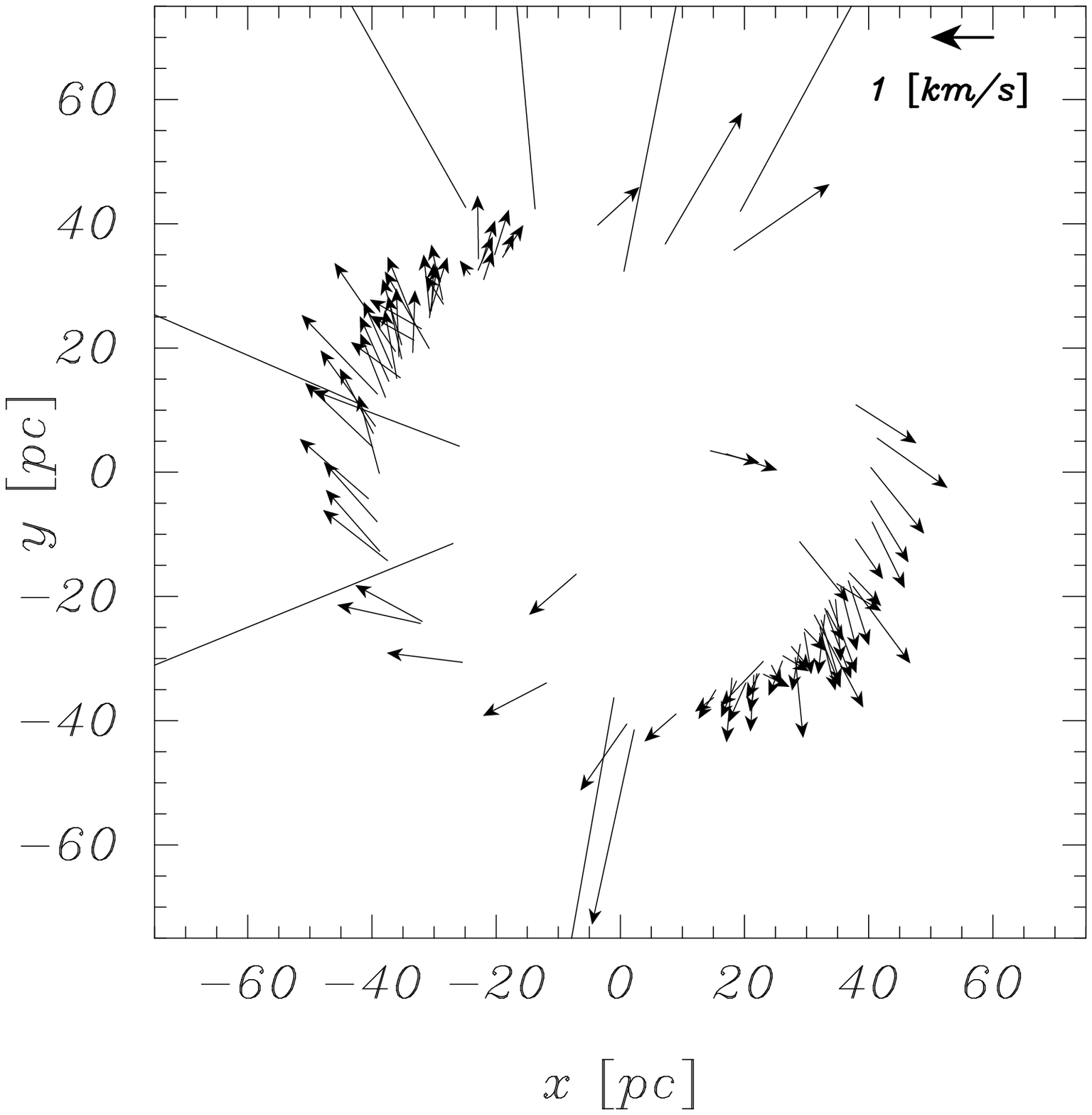,width=7.5cm,angle=-90} 
\caption[]{Vector diagram of the first 100 ($t<218$ Myr) stars
escaping from model W6-III.  Projections onto the $x-z$ plane and the
$x-y$ plane are shown.  A velocity scale is shown in the upper right
corner.}
\label{fig:fig_vW6_first100}
\end{figure*}

\begin{figure*}
~\psfig{figure=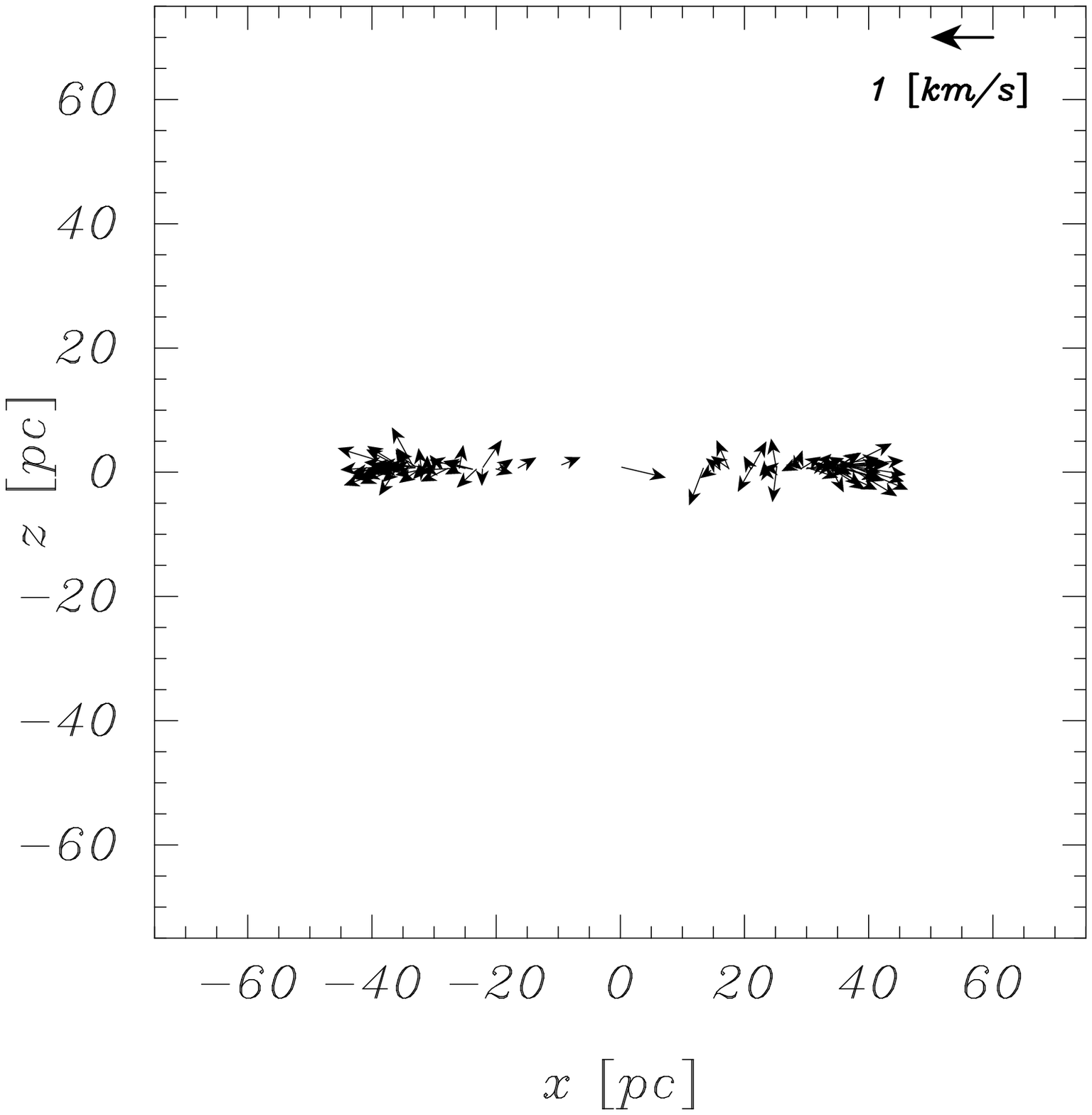,width=7.5cm,angle=-90} 
~\psfig{figure=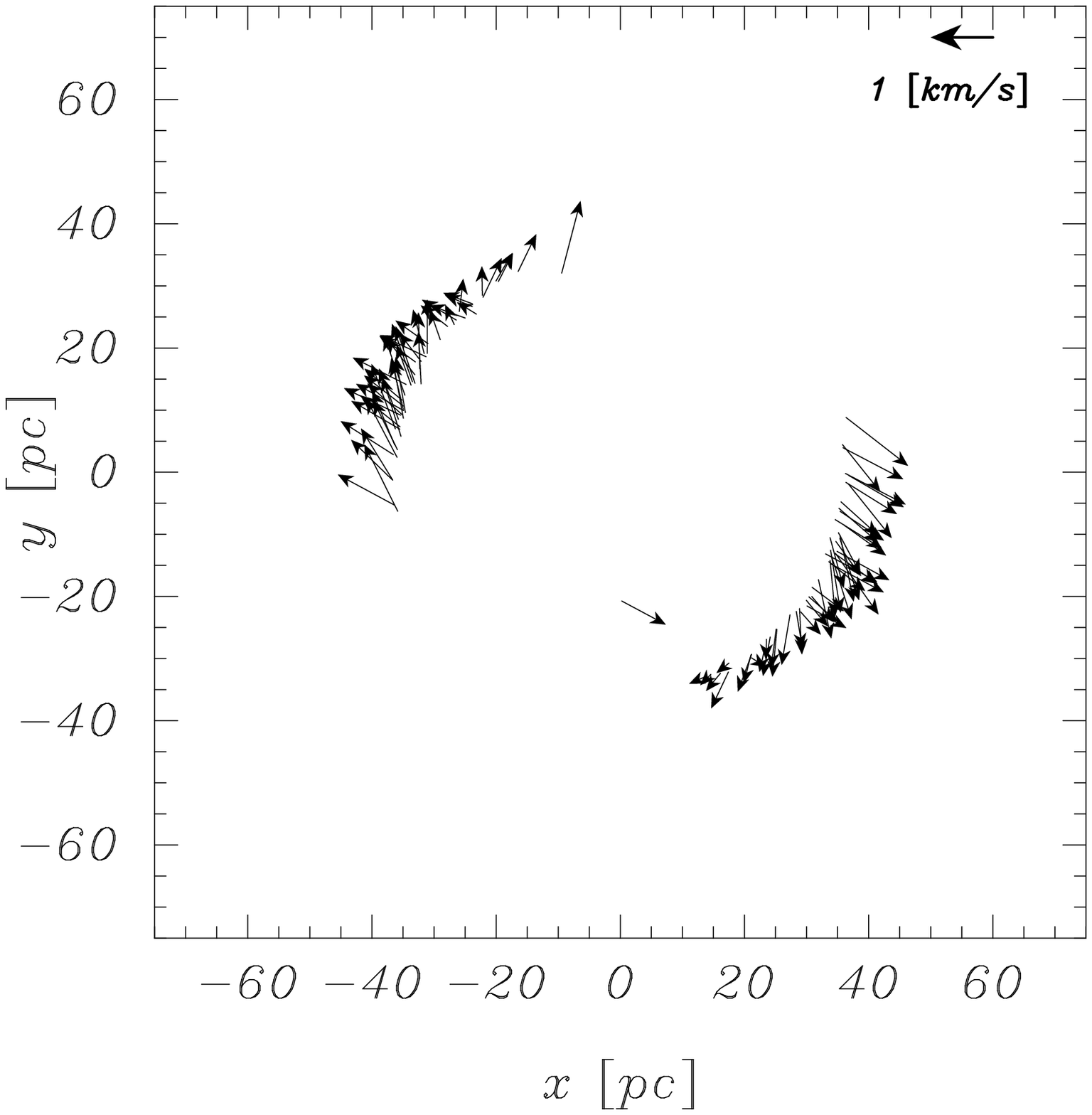,width=7.5cm,angle=-90} 
\caption[]{As for Figure \ref{fig:fig_vW6_first100}, but for the 100
stars escaping from model W6-III between $t=550$ Myr and $t=650$ Myr.}
\label{fig:fig_vW6_last100}
\end{figure*}

Stars escaping from the cluster are lost primarily near the first and
second Lagrange points.  Figures \ref{fig:fig_vW6_first100} and
\ref{fig:fig_vW6_last100} show the positions and projected velocities
of the first 100 escapers from the system, and the 100 stars which
escaped between $t=550$ and $t=650$ Myr.  The left and right panels
show, respectively, projections onto the $x-z$ and $x-y$ planes.  The
first and second Lagrange points lie on the $x$-axis, at distances of
$\sim$20 ($t=0$) and $\sim$17 ($t\sim600$) pc from the cluster center.
The small overall rotations of escapers evident in the $x-y$
projections are consequences of the Coriolis force acting on stars in
the rotating frame of reference in which we perform the simulations.
The high-speed escapers with roughly isotropic velocities in Figure
\ref{fig:fig_vW6_first100} are escaping neutron stars, which receive
high kick velocities on their formation.  They are absent in Figure
\ref{fig:fig_vW6_last100}, as the cluster is by that time too old for
supernovae to occur (except for type Ia supernovae).

The main differences between Figs.\,\ref{fig:fig_vW6_first100} and
\ref{fig:fig_vW6_last100} are (1) the considerably larger spread in
velocities, (2) the larger extent in $z$ of the region over which
stars are lost, and (3) the higher speeds of escaping stars at the
earlier epoch. These differences are readily explained by a
combination of effects; the evolution of the cluster in the Galactic
tidal field, the presence of primordial binaries and the formation of
neutron stars.  As the cluster ages it becomes less massive and the
Galaxy's gravitational pull becomes relatively stronger.  The tidal
radius shrinks and the cluster velocity dispersion decreases, so the
speed of escaping stars and their distances above or below the
Galactic plane also decrease.  The older cluster also lacks massive
stars and no stars are ejected via supernova explosions.  The shallow
core collapse during the first 100 Myr results in increased binary
activity, which also contributes to the higher stellar ejection speeds
at the earlier time.

\subsection{Stellar populations}
Tables \ref{Tab:W6_bin} and \ref{Tab:W4_bin} present, for several
cluster ages, the numbers of single stars and binaries by generic
stellar types.  The numbers are averaged over the calculations
performed for each set of initial conditions, chosen with a different
random seed from the same probability distribution.  Table
\ref{Tab:SeBa_bin} gives the same data for a population of evolving
binaries without dynamics, calculated using {\sf SeBa} (see Appendix
B).

Overall, the evolutionary differences in the populations of single
stars and binaries between models W6 and W4 are quite small.  Clearly,
as already noted, the W4 clusters evaporate more rapidly (Figure
\ref{fig:tm_all}), resulting in a generally more rapid decrease in the
numbers of both stars and binaries.  A more significant difference
between Tables \ref{Tab:W6_bin} and \ref{Tab:W4_bin} is the larger
numbers of white-dwarf binaries in the W4 models compared to other
stellar types.  Table \ref{Tab:SeBa_bin} presents the stellar and
binary properties of an evolving population of isolated binaries.  The
differences between these binaries and the dynamically evolving
population is considerable.  Comparing Table \ref{Tab:SeBa_bin} with
Tables \ref{Tab:W6_bin} and \ref{Tab:W4_bin} reveals that the
dynamically evolving populations are enhanced in both giants and white
dwarfs; the effect is stronger in the W4 models because of the
enhanced escape of the lighter stars.

\begin{table}
\caption{Stellar and binary types in the W6 models at various times.
Binaries which contain two main-sequence stars are identified as (ms,
ms), (ms, gs) contain a main-sequence star and a giant, (gs, gs)
contains two giants, (gs, wd) contains a giant and a white dwarfs and
(wd, wd) contans two white dwarfs. A bracket indicates the binary
component which fills its Roche-lobe and is in a state of mass
transfer to its companion star. binaries with neutron stars or black
holes are omitted.  The bottom row gives the binary fraction.
}\medskip
\begin{tabular}{lr|rrrrrr} \hline
\# runs:      &    5&    5&    5&     5&     4&     4  \\
time [Myr]:   &    0&  100&  200&   400&   600&   800  \\ \hline
ms            & 1024&  939&  915& 830.4& 914.5& 508.5  \\
gs            &    0&  3.2&  3.8&   5.0&   6.0&   4.5  \\
wd            &    0&  4.8&  9.4&  21.2&  33.3&  26.3  \\
(ms, ms)      & 1024&688.6&644.2& 603.0& 520.5& 438.0  \\
$[$ms, ms)    &    0&  4.2&  3.4&   3.2&   3.3&   2.5  \\
(ms, wd)      &    0&  1.0&  2.4&   5.4&   8.8&   9.5  \\
(gs, ms)      &    0&  0.4&  2.4&   1.6&   4.3&   2.8  \\
(gs, gs)      &    0&  0.0&  0.4&   0.2&   0.3&   0.3  \\
(gs, wd)      &    0&  0.4&  0.4&   0.6&   1.5&   2.0  \\
(wd, wd)      &    0&  0.0&  1.2&   3.6&   6.0&   7.5  \\ \hline
$f_{\rm bin}$ &  0.5& 0.42& 0.41&  0.42&  0.36&  0.46  \\ \hline 
\end{tabular}
\label{Tab:W6_bin}
\end{table}

\begin{table}
\caption{Stellar and binary types in the W4 models.}
\medskip
\begin{tabular}{lr|rrrrrr} \hline
\# runs:      &    5&     5&     5&     5&     2&       2 \\
time [Myr]:   &    0&   100&   200&   400&   600&     800 \\ \hline
ms            & 1024& 998.8& 938.8& 771.2& 604.5&   164.0  \\
gs            &    0&   2.8&   4.2&   7.4&   9.8&     7.5  \\
wd            &    0&   4.2&   9.4&  20.6&  33.5&    31.0  \\
(ms, ms)      & 1024& 502.8& 600.6& 453.0& 333.0&   187.5  \\
$[$ms, ms)    &    0&   4.8&   4.1&   2.8&   2.3&     1.1  \\
(ms, wd)      &    0&   1.8&   2.2&   2.4&   5.5&     6.0  \\
(gs, ms)      &    0&   0.6&   2.4&   1.2&   2.5&     3.0  \\
(gs, gs)      &    0&   0.0&   0.0&   0.2&   0.0&     1.0  \\
(gs, wd)      &    0&   0.4&   0.6&   0.4&   1.0&     0.0  \\
(wd, wd)      &    0&   0.2&   1.6&   3.0&   4.0&     5.0  \\ \hline
$f_{\rm bin}$ &  0.5&  0.34&  0.39&  0.37&  0.35&    0.50  \\ \hline 
\end{tabular}
\label{Tab:W4_bin} 
\end{table}

\begin{table}
\caption{Stellar types from population synthesis studies of
$1.5\,\times\,10^5$ binaries.  The numbers of binaries is renormalized
to 1024 because this is the number of primordial binaries in each of
our dynamical calculations. The evolution of population of 1024 single
stars was presented in Tab.\,\ref{Tab:mass}.  Note that the dynamical
models were performed with 1024 primordial binaries and 1024 single
stars.  We added the extra class of binaries which includes a neutron
star or a black hole, such binaries are omitted in the dynamical
models due their small number.  }
\medskip
\begin{tabular}{lr|rrrrrr} \hline
time [Myr]:   &    0&   100&   200&   400&   600&   800  \\ \hline
ms            &    0&   0.79&   0.42&   0.92&   0.34&   0.31 \\
gs            &    0&   0.44&   0.51&   0.73&   1.10&   1.11 \\
wd            &    0&   0.87&   1.96&   3.57&   4.99&   6.55 \\
ns/bh         &    0&   4.37&   4.42&   4.42&   4.43&   4.43 \\
(ms, ms)      & 1024& 985.14& 976.72& 965.18& 956.62& 949.82 \\
(ms, gs)      &    0&   1.81&   3.01&   4.45&   4.82&   4.41 \\
(ms, wd)      &    0&   1.89&   4.35&   8.48&  12.25&  15.61 \\
(ms, ns/bh)   &    0&   0.10&   0.08&   0.05&   0.05&   0.04 \\    
(gs, gs)      &    0&   0.13&   0.25&   0.24&   0.27&   0.21 \\
(gs, wd)      &    0&   0.27&   0.78&   1.33&   1.59&   1.63 \\
(gs, ns/bh)   &    0&   0.01&   0.01&   0.01&   0.00&   0.00 \\
(wd, wd)      &    0&   0.52&   1.98&   4.55&   6.78&   8.88 \\
(wd, ns/bh)   &    0&   0.27&   0.25&   0.23&   0.22&   0.21 \\
(ns/bh, ns/bh)&    0&   0.11&   0.10&   0.10&   0.09&   0.09 \\
\end{tabular}
\label{Tab:SeBa_bin} 
\end{table}

\begin{table}
\caption{Relative numbers of stars and binaries in the dynamical
models, as fractions of the numbers found in the non-dynamical
population synthesis studies.  The normalization is such that the
dynamical and non-dynamical calculations contain equal numbers of
single main-sequence stars and main-sequence binaries.  Numbers
greater than 1 indicate excesses of those stellar type in the
dynamical calculation; numbers less than 1 represent depletion.  }
\medskip
\begin{tabular}{lr|rrrrrr} \hline
              &\multicolumn{6}{l}{Normalized data for the W6 models} \\
time [Myr]:   &    0&  100&  200&   400&   600&   800  \\ \hline
ms            &    1&    1&    1&    1&    1&    1  \\
(ms, ms)      &    1&    1&    1&    1&    1&    1  \\
gs            &    0&  1.0&  0.9&  0.9&  0.9&  1.3  \\
wd            &    0&  1.4&  1.0&  1.3&  1.2&  1.3  \\
(ms, gs)      &    0&  0.3&  1.1&  0.6&  1.2&  1.3  \\
(ms, wd)      &    0&  0.8&  0.8&  1.0&  1.0&  1.3  \\
(gs, gs)      &    0&  0.0&  2.3&  1.3&  1.3&  2.4  \\
(gs, wd)      &    0&  2.1&  0.7&  0.7&  1.3&  2.6  \\
(wd, wd)      &    0&  0.0&  0.9&  1.2&  1.2&  1.8  \\ \hline
              &\multicolumn{6}{l}{Normalized data for the W4 models} \\
time [Myr]:   &    0&  100&  200&   400&   600&   800  \\ \hline
ms            &    1&     1&    1&    1&    1&    1  \\
(ms, ms)      &    1&     1&    1&    1&    1&    1  \\
gs            &    0&   1.0&  1.0&  1.5&  2.3&  6.3  \\
wd            &    0&   1.2&  1.0&  1.3&  1.9&  4.6  \\
(ms, gs)      &    0&   0.6&  1.2&  0.5&  1.4&  3.3  \\
(ms, wd)      &    0&   1.7&  0.8&  0.6&  1.2&  1.8  \\
(gs, gs)      &    0&   0.0&  0.0&  1.6&  0.0&  2.2  \\
(gs, wd)      &    0&   2.6&  0.1&  0.6&  1.7&  0.0   \\
(wd, wd)      &    0&   0.9&  0.5&  1.3&  1.6&  2.7  \\ \hline
\end{tabular}
\label{Tab:renormalized} 
\end{table}

Table\,\ref{Tab:renormalized} gives the fractions of various types of
stars and binaries in the dynamical calculations, relative to the
corresponding numbers from the population synthesis studies.  The
latter are normalized to the same numbers of single main-sequence
stars and main-sequence binaries as in the initial dynamical calculations.
(This normalization is employed here to show trends which are hard to
see in the Tables\,\ref{Tab:W6_bin}, \ref{Tab:W4_bin} and
\ref{Tab:SeBa_bin}.)  Neutron stars and black holes are omitted from
the comparison because the differences are directly obvious: Neutron
stars escape from open star clusters, but they are retained in the
non-dynamical models.  Black holes are omitted because of their small
numbers.

Although the numbers of giants and white dwarfs are also small, a
trend is clearly visible: Single white dwarfs and dwarfs in binaries
are overrepresented at later stages in the dynamical calculations,
especially for the W4 models. The basic reason for this overabundance
of white dwarfs and giants is their larger mass, which makes them more
likely to be retained by the cluster.  White dwarfs have a very
complicated evolution within the stellar system, their progenitors
being among the most massive objects while on the main sequence and
the giant branch, but the white dwarfs having masses comparable to the
mean once their envelopes are lost.  White dwarfs are therefore
preferentially formed in the cores of star clusters.  Once the white
dwarf is formed, it is hard to extract it from the core.  These
effects are more pronounced in smaller clusters ($\trlx\aplt1$\,Gyr).
The W4 models retain more white dwarfs than the W6 models because the
latter relax on a longer time scale and evolve dynamically less
rapidly than the former.


\section{Comparison with observations}\label{sect:discussion}
\subsection{The Pleiades}
\begin{figure}
\hspace*{1.cm}
\psfig{figure=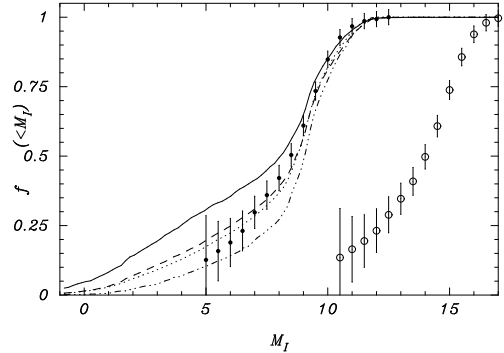,width=7.5cm,angle=-90}
\caption[]{Cumulative luminosity function in $M_I$-magnitudes.  The
open circles ($\circ$) with (Poissonian) error bars show the apparent
luminosity function for the Pleiades (Hambly \& Jameson 1991).  The
corrected absolute luminosity function, assuming distance modulus
$m-M=5.5$ (Gatewood et al. 1990)\nocite{1990ApJ...364..114G} is
indicated by filled circles ($\bullet$). Both luminosity functions are
corrected for the 150 stars brigter than $m_I=10.5$ (Pinfield et al
1998).  The dotted line is the initial luminosity function for all
stars in the models (assuming that binaries are unresolved).  The
solid and dashed lines shows the liuminosity function for all models
at $t=100$\,Myear within a projected (onto the $y$--$z$ plane) 25\%
Lagrangian and the half mass radius of the cluster. The dash-3dotted
line give the luminosity function for the cluster stars in the outer
90\% Lagrangian radius.}
\label{fig:Pleades_LF}
\end{figure}

Figure \ref{fig:Pleades_LF} shows the $M_I$-magnitude luminosity
function for the inner part of the Pleiades cluster (Hambly \& Jameson
1991)\nocite{1991MNRAS.249..137H} and compares it with our model
luminosity functions at 100 Myr.

The best fit between the observed and model luminosity functions is
obtained for the stars within the half mass radius. This suggests that
some mass segregation has already occured in this clusters.  Raboud \&
Mermilliod (1998) also find evidence for mass segregation in this
cluster.  Our luminosity function has too many bright stars and to
make a reasonable fit we have to exclude stars with $M_I<4.5$ from the
sample.  We are not sure why this is the case, but argue that the
brightest stars were possibly overexposed in the observations, and may
therefore have been omitted from the observed luminosity function.

The Pleiades is flattened, with an observed ellipticity $\epsilon
\equiv (1-b/a)$ of 0.17.  Taking into account the orientation of the
cluster in the tidal field of the Galaxy, Raboud \& Mermilliod (1998)
derive an intrinsic ellipticity of almost 0.3 (see also van Leeuwen et
al. 1986),\nocite{1986A&AS...65..309V} comparable to what we find in
our models (see Figure \ref{fig:isoVt600_W6R2TF}) by comparing the
distance to the Jacobi surface along the $z$ axis with the distance to
$L_1$: $\epsilon \simeq 1-r_z/r_J$.


\subsection{Praesepe}\label{sect:praesepe}
\begin{figure}
\hspace*{1.cm}
\psfig{figure=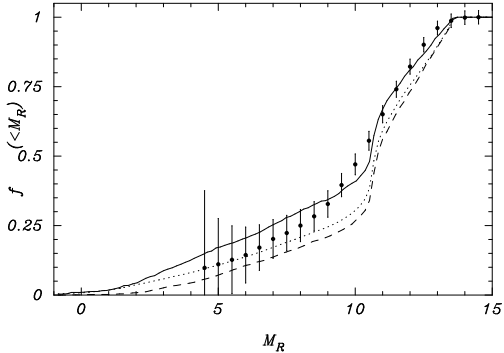,width=7.5cm,angle=-90}
\caption[]{Cumulative luminosity function ($M_R$-magnitude) of the
Praesepe cluster.  The filled circles with error bars give the
observed luminosity function within the half-mass radius (Hambly at
al. 1995a, 1995b), to which 170 stars with $M_R<4.5$ were added. 
The model W6 luminosity function for all stars at
zero age is shown as a dotted line.  The solid and dahsed lines give
the luminosity functions for models W6 at 800 Myear within and outside
the projected (on the $y$--$z$ plane) half mass radii.}
\label{fig:Rt1000_W6}
\end{figure}

Figure \ref{fig:Rt1000_W6} shows the $M_R$-magnitude global luminosity
function for the W6 clusters at birth and at 800 Myr and compares that
with the observed luminosity function for Praesepe reported by Hambly
et al.  (1995a, 1995b),\nocite{1995A&AS..109...29H}
\nocite{1995MNRAS.273..505H} which is shown as filled circles with
error bars.

The two 800\,Myear old luminosity functions are taken from the stars
within the projected half mass radius and stars farther away.  The
outer luminosity function (dashes) fits better to the bright stars
where the inner luminosity function fits better to the dimmer stars.
Again we can argue that omitting the brightest stars from the sample
provides a better fit to the observations, in which case the
luminosity function of the inner half of the cluster provides a better
comparison.

\subsection{The Hyades}
\begin{figure}
\hspace*{1.cm}
\psfig{figure=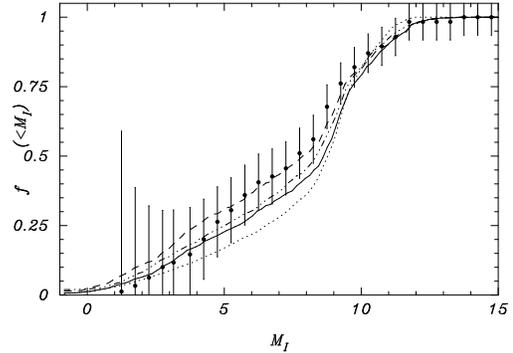,width=7.5cm,angle=-90}
\caption[]{Cumulative $M_I$-magnitude luminosity function (filled
circles with error bars) of the Hyades star cluster (Reid \& Hawley
1999).  The dotted line is the initial luminosity function for all
stars in the model calculations.  The solid line and dashed lines give
the luminosity function for model W4 at an age of 600\,Mears for stars
within the projected (on the $y$--$z$ plane) tidal radius (solid) and
within the half-mass radius (dahes).  The dash-3dotted line gives the
same data as the dashed line but then for model W6.
}
\label{fig:It1000_W6}
\end{figure}

Figure \ref{fig:It1000_W6} compares the $M_I$-band luminosity
functions for the stars and binaries of several models, at birth
(dotted line) and at 600 Myr (other lines), with the observed
luminosity function the Hyades (Reid \& Hawley
1999).\nocite{1999AJ....117..343R} Reid \& Hawley observed the entire
cluster and their luminosity function reportedly extends down to the
hydrogen-burning limit.  The \Wo=4 data for the entire cluster (solid)
and the \Wo=6 model for the stars within the half mass radius
(dash-3dot) does not fit as well as the data for model \Wo=4 for stars
within the inner half mass radius (dashes).  Our models W6 are
somewhat farther from the Galactic center than is the Hyades, which
would tend to suppress mass segregation somewhat by increasing the
cluster tidal radius. The \Wo=4 models do excatly the opposite. Based
on these arguments we conclude that Hyades is somewhat more mass
segregated that our models predict and some degree of promiridal mass
segregation seems to be required.

Oort (1979)\nocite{1979A&A....78..312O} compared observations of
Hyades with Aarseth's (1973,
1975)\nocite{1973VA.....15...13A}\nocite{1975IAUS...69...57A} {\nbody}
calculations and concluded that the outer 4\,pc of the Hyades cluster
are more strongly flattened, with $\epsilon \simeq 0.5$, than the
N-body models implied.  The orientation of the Hyades in the Galaxy
relative to the position of the sun then implies that the intrinsic
flattening is even greater.  Our calculations do not support Oort's
conclusion, and a flattening of $\epsilon=0.5$ is quite consistent
with our {\nbody} models.  The reason for the discrepancy between our
results and the conclusion of Oort is based on Aarseths' models which
were computed with a very small number of stars. The flattening of the
cluster in the tidal field of the Galaxy, however, becomes more
apparent towards the clusters' tidal radius which has smallest stellar
density. Calculations which are performed with a limited number of
stars $\aplt 500$ will hardly show the falttening in the tidal field.


\subsection{NGC\,3680}
\begin{figure}
\hspace*{1.cm}
\psfig{figure=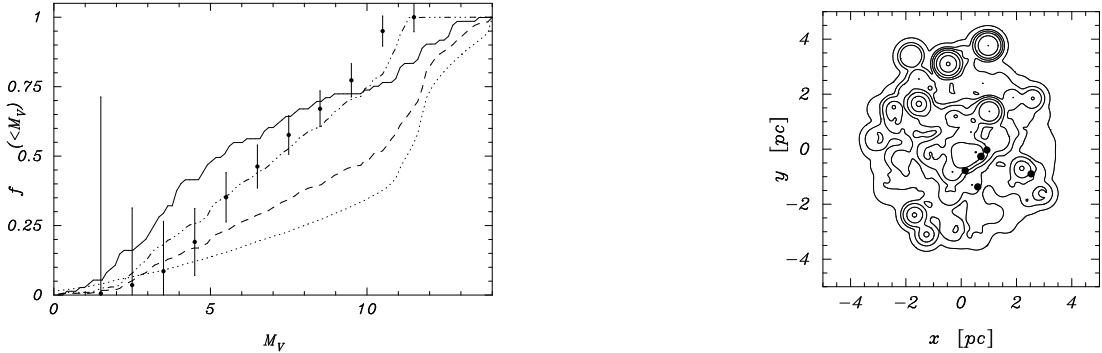,width=7.5cm,angle=-90}
\caption[]{The filled circles with error bars show the observed
cumulative luminosity function, corrected for field stars, of the star
cluster NGC\,3680 (Hawley et al.\, 1999).\nocite{1999AJ....117.1341H}
The dotted line gives the initial luminosity function for all models,
all other lines show luminosity functions at 1400\,Myear.  The solid
and dashed lines give the luminosity function for all stars within the
half mass radius of model W4 and W6, respectively.  The dash-3dotted
line shows only the stars with $M_V<11.5$ from the dashed line (within
the half-mass radius of model W6 at 1400\,Myear).  }
\label{fig:Vt1000_W6}
\end{figure}

Figure \ref{fig:Vt1000_W6} compares the observed $M_V$-band luminosity
function of the old open cluster NGC\,3680 with our model luminosity
functions.  The observed luminosity function is poorly reproduced by
our cluster models.  However, if we remove the least luminous stars
(those with $V>11.5$), the 1.4 Gyr model fits the observed luminosity
function fairly well. The imposed lower limit is rather arbitrary, it
suggests that the observations may not properly correct for the
faintest stars, or they may simply be absent from the data.  Mass
segregation generally causes the lightest stars to escape from the
cluster which, in time, leads to an over abundancy of massive
stars. The observed clusters, however, seem to have too few high mass
stars and also too few low mass stars, which is hard to understand
from a dynamically point of view. We therefore argue that in this case
the lack of low mass stars is an observeational selection effect.

\subsection{Isophotes}
\begin{figure}
~\psfig{figure=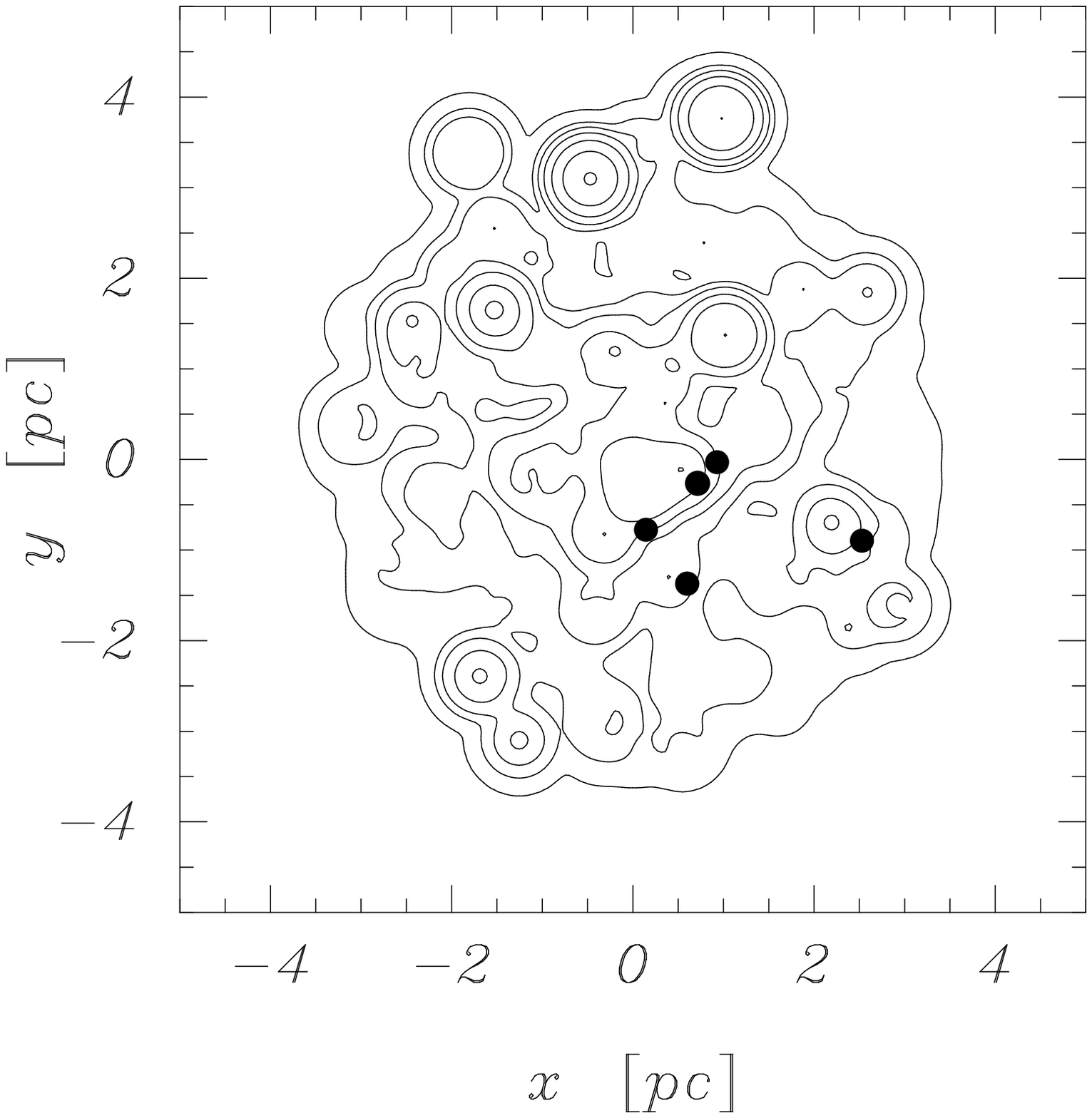,width=7.5cm,angle=-90}
\psfig{figure=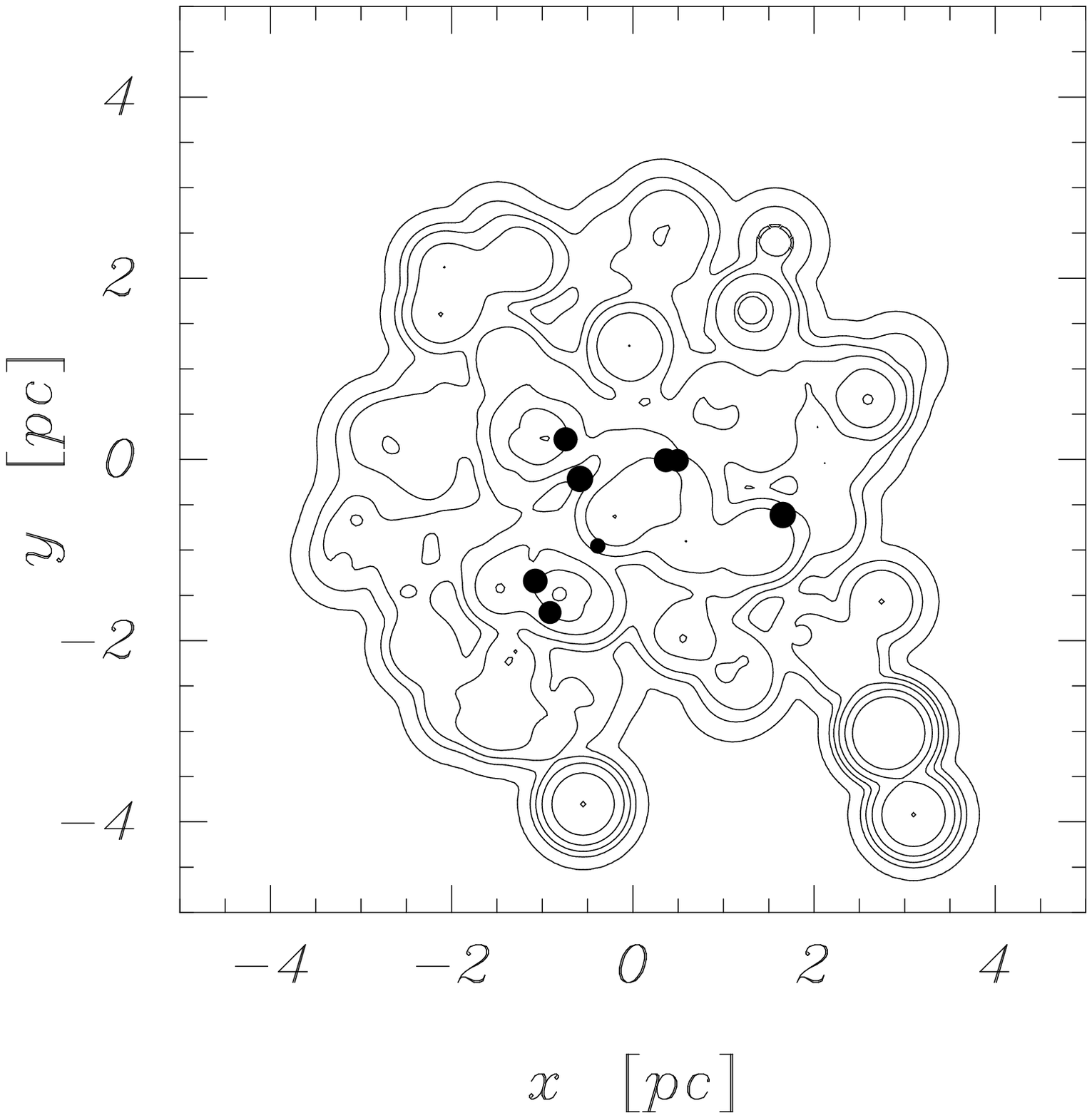,width=7.5cm,angle=-90}
~\psfig{figure=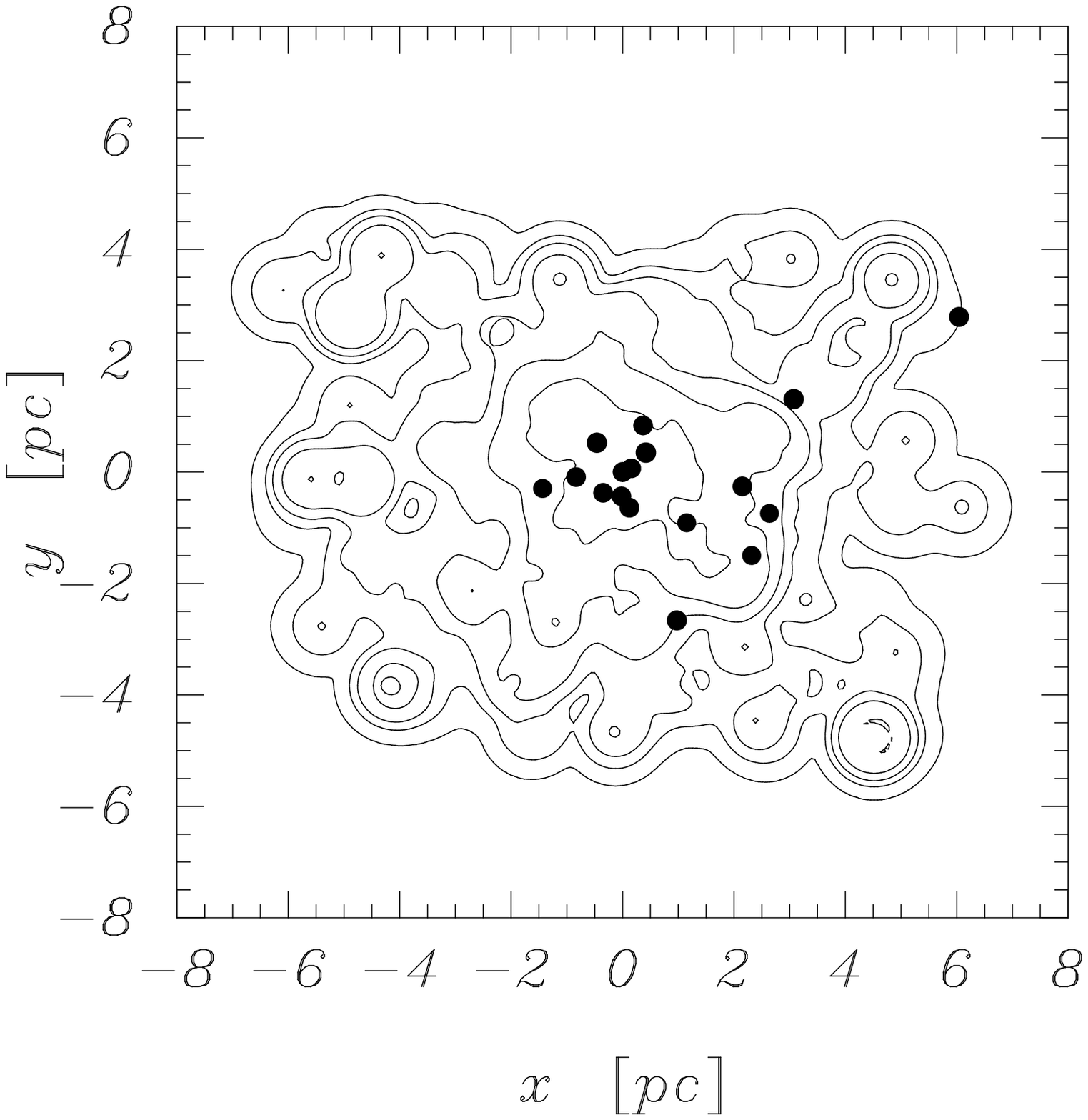,width=7.5cm,angle=-90}
\caption[]{Isophotes in $M_V$ of the cluster NGC\,2516 (top panel)
NGC\,2287 (bottom left), and NGC\,3680 (bottom right).  The adopted
distance moduli are 9.1, 8.2, and 9.95, corresponding to distances of
655\,pc, 373\,pc and 735\,pc for the top and bottom left and right
panels, respectively.  The (sub)giants are plotted as filled circles;
the remaining stars are plotted as isophotes.  The central luminosity
densities are -2.0\,mag\,pc$^{-2}$, -2.2\,mag\,pc$^{-2}$ and
-0.54\,mag\,pc$^{-2}$ for the top, bottom left and right panels,
respectively.  Contours are plotted at constant intervals of
1.25\,mag.  Stars are assigned a Gaussian point spread function with a
dispersion of 0.06\,pc (0.1\,pc for the bottom figure), the figures
contain 594, 178 and 738 stars for the top and bottom left and right
panels, respectively.  }
\label{fig:NGC2287}
\end{figure}

%
%
%
Figure \ref{fig:NGC2287} shows isophotes of the clusters NGC\,2287,
NGC\,2516 and NGC\,3680.  NGC\,3680 is most strongly flattened
($\epsilon \sim 0.23$), the other two are more circular in appearance;
NGC\,2287 has $\epsilon \sim 0.05$ and NGC\,2516 has $\epsilon \sim
0.14$.  We measured these ellipticities for the inner 4\,pc for each
cluster.

The faintest stars in Figure\,\ref{fig:NGC2287} are 13th magnitude for
NGC\,2287 and 15th magnitude for NGC\,2516 and NGC\,3680.  Taking the
distances to these clusters into account, it is clear that only the
top end of the main sequences are included in the isophotes.  Since
mass segregation causes these high-mass stars to be more centrally
concentrated than lower-mass stars, it is likely that we see only the
inner, roughly spherical, regions of NGC\, 2287 and NGC\,2516 (see
Figure \ref{fig:isoVt600_W6R2TF}).  NGC\,3680 appears much somewhat
rectabgular on the image and we are tempted to explain this shape as a
result of the limited imaging; is the cluster is truncated by the edge
of the field of view?. In this case it is not surprising that this
cluster appears somewhat flattened.  If one looks at the inner density
contours the cluster looks much more circular.  Note also that the
tidal radii for these clusters listed in Tab.\,\ref{Tab:observed}, 13,
6.3 and 4.3 pc for NGC\,2516, NGC\,2287 and NGC\,3680 respectively,
seem inconsistent with the above pictures. The isophotes of NGC\,2516
and NGC\,2287 represent only the central portion, where for NGC\,3680
appears much bigger than the tidal radius we derived in
Tab.\,\ref{Tab:observed}. The table possibly underestimates
the tidal radius of NGC\,3680.


\section{Comparison with other work}\label{sect:other_work}

Table\,\ref{Tab:compare} compares the evaporation times of our model
calculations with previously reported results.  We discuss each in
turn.  Table\,\ref{Tab:compare_init} lists the initial conditions of
the models with which we compare our own.


\begin{table*}
\caption{Overview of the model features, initial conditions and
lifetimes of our model calculations and those reported by other
workers.  The Galactic tidal field may be modeled as that of a
self-consistent disc (Disc) or a point mass (PM); initial density
profiles are an anisotropic King model (AKing), an Isothermal sphere
(Iso), or a Plummer sphere (Plummer) the latter two with a cut off.
Stellar and binary evolution are indicated by + (included) or - (for
neglected). Note: Terlevich and de la Fuente Marcos work in terms of the
initial virial radius, which is typically about 20\% bigger than the
half-mass radius.  }
\medskip
\begin{tabular}{lccccrrrrr|rrr} \hline
&Galaxy&Density&Stellar&Binary 
				& $N$  & \mm &$f_{\rm bin}$& \rtide 
		                &\rhm& \trlx& $t_{\rm end}$
			               & $t_{\rm end}/\trlx$ \\
&model &profile&\multicolumn{2}{c}{evolution} 
		               &      &[\msun]&             
				      &\multicolumn{2}{c}{[pc]}
				       &\multicolumn{2}{c}{[Myr]} \\ \hline
W4                 &  Disc&  AKing&      +&      + 
		   & 3072 & 0.56&  50&  6.3& 2.14&    109 & 1200&  11.0 \\
W6                 &  Disc&  AKing&      +&      + 
		   & 3072 & 0.54&  50& 12.1& 2.00&    102 & $\sim$1600& 15.7 \\
Terlevich          &  Disc&    Iso&      +&      - 
		   & 1000 & 0.50&   0& 12.1& 2.00&     71 & $\sim$1400& 19.7 \\
McMillan \& Hut    &    PM&Plummer&      -&      -   
		   & 2048 & 1   &  10& 16.5& 2.06&     86 & 2500& 29.0 \\
Kroupa             &    PM&Plummer&      +&      - 
		   &  400 & 0.32& 100&  6.5& 2.53&     77 &  693& 9.0  \\
de la Fuente Marcos&  Disc&    Iso&      +&      -  
		   &  750 & 0.60&  33&  9.9& 2.47&     67 & 1061& 15.8  \\
\hline
\end{tabular}
\label{Tab:compare}
\label{Tab:compare_init}
\end{table*}


Terlevich (1987)\nocite{1987MNRAS.224..193T} performed direct {\nbody}
calculations with up to 1000 stars, including a power-law stellar mass
function, mass loss from single-star evolution, and the tidal field of
the Galaxy.  The implementation of the Galactic tidal field and the
evolution of single stars were somewhat similar to those presented in
this paper.  Her model XII had initial conditions most similar to our
own, although some significant differences exist.  Terlevich's models
started with spherical distribution of stars with density proportional
to $1/r^2$, virial radius 2 pc (half-mass radius of about 1.6\,pc),
and located at a distance of 10 kpc from the Galactic center.  The
resulting half-mass crossing time was 5.2\,Myr, comparable to the
$\sim 4.1$\,Myr in our models.  Terlevich's model XII initially
consisted of 1000 stars drawn from a Salpeter mass function, with a
mean mass of 0.5\,\msun.  The initial virial radius was $Q \equiv
E_{\rm kin}/|E_{\rm pot}| = 0.25$, less than equilibrium value of 0.5.
Even though the model started with fewer stars, the cool initial
conditions mean that the initial relaxation time was comparable to
that in our own models.  The half-life of model XII was 770 Myr.  The
run was terminated at 1 Gyr, by which time about 70\% of the mass had
been lost.  We estimate that the cluster would have dissolved in about
1.4\,Gyr.

The half-mass lifetimes of our W6 models are all around 800 Myr,
similar to that of Terlevich's model XII.  This is somewhat
surprising, as we might have expected that our W6 models, with more
stars, would live longer than the comparable models of Terlevich.  The
reason for this discrepancy cannot be attributed to the large fraction
of primordial binaries in our calculations, as binaries do not
dramatically affect the evaporation rate of the cluster (Figure
\ref{fig:tm_all}).  The difference may possibly be due to the lower
high-mass cutoff in Terlevich's initial models.  Although her model
XII was initially less concentrated than our W6 models, it did reach
core collapse.  In general, more concentrated models tend to live
considerably longer than shallower models (Takahashi \& Portegies
Zwart 2000).\nocite{2000ApJ_TPZ_inpress}

The evolution to core collapse in Terlevich's model XII was aided by
the absence of stars with masses greater than 10\,{\msun}.  By the
time the turnoff mass has dropped below 10\,{\msun} (after about
22\,Myr) our W6 models have lost between 4\% and 9\% of their mass due
to stellar evolution alone.  The loss of even such a small mass
fraction may have dramatic consequences for the further evolution of
the stellar system, as this mass is lost from the most massive stars,
which reside deep inside the cluster potential well.  The shorter
lifetime of clusters with large populations of massive stars is
demonstrated by Terlevich's model XV, where the initial mean mass was
7.4\,\msun and the cluster did indeed dissolve much more rapidly.

The Galactic tidal field produced a similar flattening effect in
Terlevich's clusters as in our own (compare her Figure 7 with our
Figure \ref{fig:isoVt600_W6R2TF}).


The simulations reported by McMillan and Hut (1994) included up to
2048 equal mass stars, including up to 20\% rather soft primordial
binaries, and incorporated the Galactic tidal field, modeled as the
field of a distant point mass.  However, they excluded stellar
evolution and hence any possibility of stellar mass loss.  In the
absence of a physical time scale associated with stellar evolution,
they presented their results in units of the initial half-mass
relaxation time.  Our W6 models have half-lives of about 6 initial
relaxation times, much shorter than the $\sim 29$ initial relaxation
times for the most comparable McMillan \& Hut models.  Also, as
discussed previously, all the McMillan \& Hut models experienced core
collapse, which is absent in our simulations.

The main reasons for these differences are the effects of stellar mass
loss and the presence of a stellar mass function in the present
studies.  In our models, core collapse is arrested by mass loss.  In
addition, the McMillan \& Hut models all started off well inside their
tidal radii, significantly increasing their lifetimes.



%
In a continuing effort to understand the evolution of young open star
clusters, Kroupa (1995a; 1995b; 1995c)\nocite{1995MNRAS.277.1491K}
\nocite{1995MNRAS.277.1507K}\nocite{1995MNRAS.277.1522K} performed
{\nbody} calculations with up to 400 stars, all of them members of
primordial binaries.  He adopted a distribution in orbital separation
flat in $\log a$, but selected $a$ between $\sim 360$ and $\sim 3.6\,
10^5$\,\rsun; the binaries in his calculations were thus on average
much wider than in our calculations.  His models included the Galactic
tidal field and stellar mass loss, but neglected binary evolution.

Kroupa's model with the highest initial relaxation time started from a
Plummer sphere with a half-mass radius of 2.5\,pc and a crossing time
of $\sim 20$\,Myr.  It dissolved in about 700\,Myr.  Scaled to the
initial relaxation time, this is somewhat faster than our W4 models.
The main reasons for this more rapid evaporation are most probably the
shallower initial density profile in his models and the cluster's
smaller distance from the Galactic center.

We compare the binary and triple properties of our models with those
of McMillan \& Hut and Kroupa in more detail in Paper IVb.


De la Fuente Marcos (1997) studied the effect of the initial mass
function on the dynamical evolution of open star clusters, including
both stellar mass loss and the tidal field of the Galaxy.  His
calculations were limited to 750 stars and included 33\% rather wide
primordial binaries, all with a mass ratio of 0.5, in which (binary)
evolutionary effects were neglected.  His model XX used Scalo's (1986)
mass function and had an initial virial radius of 2.47\,pc.  The
initial relaxation time for this model was about 67\,Myr; the crossing
time was 7.5\,Myr.  This model dissolved in about 1\,Gyr, slightly
faster than our models.  However, scaled to the initial relaxation
time, this result is consistent with the dissolution time of our W6
models (see Table \ref{Tab:compare}).


\section{Summary and Discussion}\label{sect:summary}
The aim of our simulations is to study the evolution of open star
clusters such as the Pleiades, Praesepe and the Hyades.  These
clusters differ significantly in age, but have comparable physical
characteristics, stellar membership, total magnitude and internal
velocity dispersions.  Our {\nbody} calculations incorporate, in a
fully self-consistent fashion, mass loss from single stars, binary
evolution, dynamical encounters among single stars and binaries and
the effect of the Galactic tidal field.

We have compared the luminosity functions, isochrones and projected
luminosity profiles of our models with observations.  For some
clusters, it is hard to find a good match between model and observed
luminosity functions.  Mass segregation and observational limitations
significantly reduce our ability to find a match, and restrict our
understanding of the differences we see.  However, we find that for
models for which we compared the luminosity functions with Praesepe
and the Hyades,
the observations show evidence for significantly more mass segregation
than is seen in the models.  We conclude that these clusters may have
been born somewhat mass segregated.  Alternatively these clusters may
have started out somewhat more massive than assumed here, but with a
shallower density profile.  The selective evaporation of lower mass
stars then results in a ``dynamically old'' appearance (see also
Takahashi \& Portegies Zwart 2000).\nocite{2000ApJ_TPZ_inpress}

The global luminosity function of Praesepe and its degree of mass
segregation suggest an age greater than 800\,Myr, which is at the high
end of the observed range.  The luminosity function of NGC\,2287 is
consistent with the observed age of 150--200\,Myr.  Our age estimates
for the Pleiades and Hyades, based on the structure and dynamical
state of these clustusters, are consistent with ages derived from
isochrone fitting.

\subsection{Mass segregation}
The first effects of mass segregation in our models are discernible in
the cluster core after only a few million years, a small fraction of an
initial half-mass relaxation time.  After about a relaxation time,
mass segregation becomes measurable in the cluster outskirts.  The
luminosity function of the inner parts of the cluster provides a
useful tool for studying mass segregation, although one has to select
regions which are well separated in radius in order to make the effect
visible.  As expected, giants and binary stars are most strongly
affected by mass segregation and it is easiest to identify the effect
by comparing the radial distribution of giants with that of the lower
mass main-sequence stars.

One important effect of mass segregation is that older clusters become
rich in white dwarfs and giants relative to the Galactic field.  These
stars may be single or members of binary systems.  The main reason for
the overabundance of giants and white dwarfs in clusters is the the
depletion of low mass main-sequence stars by evaporation.  This
flattening of the mass function due to mass segregation has also been
studied by Takahashi \& Portegies Zwart
(2000)\nocite{2000ApJ_TPZ_inpress} for more massive clusters.  They
also find that clusters which are close to disruption are rich in
compact objects and giant stars.

The W4 models have shallower initial potentials, lie closer to the
Galactic center, and are more strongly affected by mass segregation
because of their more rapid evaporation.  We suggest that the best
place to look for evidence of mass segregation is in star clusters
with larger half-light radii, which have shallower potentials.  Since
mass segregation manifests itself more clearly in dynamically evolved
systems, it is also better to look at older clusters.  A relatively
old cluster (age $\apgt500$ Myr) with a relatively small mass
($\mtot\aplt500$\,\msun) would be ideal.

\subsection{Age estimates}
The ``dynamical ages''\footnote{We define the dynamical age of a
cluster as its lifetime expressed in units of the initial relaxation
time.}  of our model clusters are often inconsistent with ages
determined by isochrone fitting.  The instantaneous relaxation time is
a poor estimator of a cluster's dynamical age.  The cores of star
clusters lose their memory of the initial relaxation time within a few
million years, well within an initial half mass relaxation time, but
the time scale for global cluster amnesia to set in is far larger than
the initial relaxation time.  The half-mass relaxation time tends to
increase by a factor of two or three, reaching a maximum near the
cluster's half-life and decreasing thereafter.  The increase is caused
by the internal heating of the cluster; the decrease mainly by loss of
stars.

\subsection{Escaping stars}
Stars tend to escape in the direction of the $L_1$ and $L_2$ Lagrange
points of the Galaxy--cluster system.  The velocities of the stars at
these points are highly anisotropic and, as expected, pointing mostly
radially away from the cluster center.  The velocities of the escaping
stars are comparable to the cluster velocity dispersion; very few
stars are ejected with high velocities following a strong dynamical
encounter.

Neutron stars are ejected isotropically and with much higher
velocities, owing to the asymmetric kicks they receive during the
supernovae that create them.  Binaries containing the progenitors of
neutron stars are generally disrupted by the first supernova
explosion; in all the calculations presented in this paper, only one
binary survived the first supernova.  However, the existence of that
single binary does suggest that it may be possible to form X-ray
binaries in open clusters.  Such binaries are only expected to exist
in star clusters which are younger than $\sim 45$\,Myr (the turnoff
age of a 7\,{\msun} star), because mass loss and the velocity kick
imparted to the binary causes it to escape (see paper IVb).

Black holes are more easily retained by the cluster, but are very rare
due to their high progenitor mass and the steepness of the Scalo
initial mass function.

\subsection{Tidal flattening}
The tidal field of the Galaxy flattens the cluster significantly in
the $z$ direction, and to a lesser extent along the $y$ axis.  A
cluster which is spherical at birth develops this flattening in its
outer regions within a few crossing times; the inner parts remain
fairly spherical.  All our models show this flattening, but its
observability from Earth depends on the orientation of the cluster in
the Galactic plane.

Ellipticities reported for the Pleiades ($\epsilon \sim 0.3$; van
Leeuwen et al.\, 1986) and Hyades ($\epsilon \sim 0.5$ in the outer
regions; Oort 1979) are consistent with our model calculations.  The
available data for NGC\,2287 and NGC\,2516 do not show significant
flattening.  However, these data contain only stars from the innermost
3\,pc, which are affected least by the Galactic tidal field.
NGC\,3680 appears more square than elliptical.  This may be caused by
the small field of view of the telescope, which could cause a star
cluster to take on the shape of the CCD frame (see
Fig.\,\ref{fig:NGC2287}).

The flattening persists during mass segregation, in the sense that the
mass-segregated Lagrangian radii are ellipsoids. Projection of the
cluster onto the background may therefore decrease the observed mass
segregation.

\subsection{Core collapse}
Our models experience rather shallow core collapse during their early
evolution.  The clusters then expand more or less homologously,
preserving their initial density profiles.  The expansion is driven by
stellar mass loss and, to a lesser extent, by binary heating;
dynamical models without stellar mass loss but which include a tidal
field and primordial binaries do show core collapse (McMillan \& Hut
1994).  Once the cluster has lost a considerable fraction of its stars
the system shrinks again.  The remnant with a few remaining stars may
become quite compact before it dissolves, but we find no evidence for
late core collapse before complete disruption.

\subsection{Giants and white dwarfs in open clusters}\label{sect:fwdgs}
A cluster's single-star population is not noticeably affected by
cluster dynamics until the age of the system exceeds $\sim2$ initial
half-mass relaxation times.  However, the binary populations are
measurably influenced by dynamics even at early times (see Paper IVb).
At later times ($t\geq 400$\,Myr), our model clusters tend to become
rich in giants and white dwarfs.  Small-number statistics on the
giants limit the degree to which we can quantify this statement, but
the white-dwarf populations in our models increased by factors of 1.3
to 4.6 (for model W6 and W4, resepctively) relative to what one would
expect for a non-dynamically evolving population of single stars and
binaries.  Most of the excess white dwarfs are members of binary
systems.

Table\,\ref{Tab:observed} presents an overview of the numbers of white
dwarfs observed in the clusters studied in this paper.  Explaining the
number of white dwarfs in the Hyades is a long-standing problem,
starting with discussions in the mid-1970s by Tinsley
(1974)\nocite{1974PASP...86..554T} and van den Heuvel
1975),\nocite{1975ApJ...196L.121V} and continuing into the 1990s
(Eggen 1993; Weidemann 1993).
\nocite{1993AJ....106..642E}\nocite{1993A&A...275..158W} All these
papers conclude that the observed number of white dwarfs is too small,
by about a factor of three, for the inferred cluster mass and age.
The three leading explanations were: (1) the upper mass limit for the
production of white dwarfs may be as low as $\sim 4$\,{\msun} (Tinsley
1974), \nocite{1974PASP...86..554T} white dwarfs are born with a
velocity kick as neutron stars do (Weidemann et al.\, 1992) and (3)
mass segregation selectively ejects white dwarfs (van den Heuvel
1975).  \nocite{1992AJ....104.1876W} By studying the white dwarfs in
NGC\,2516, Koester \& Reimers (1996)\nocite{1996A&A...313..810K}
firmly conclude stars up to 8\,\msun\, can form white dwarfs, removing
the first solution to this conundrum.  However, our calculations are
inconsistent with the idea that white dwarfs are preferentially
ejected from star clusters.  On the contrary, the white dwarfs in our
simulations are more easily retained than main-sequence stars of the
same mass, causing the older clusters to become relatively white-dwarf
{\em rich} for their mass.

We can study this problem further by comparing the ratio of the number
of white dwarfs to the number of giants: $\fwdgs \equiv N_{\rm
wd}/N_{\rm gs}$.  For the first five open clusters in
Tab.\,\ref{Tab:observed} this fraction ranges from $\fwdgs=1$ for NGC
2516 and the Pleiades to $\fwdgs=2.5$ for the Hyades.  For an evolving
population of single stars without dynamics (see
Tab.\,\ref{Tab:mass}), we find that the ratio ranges from 1.1 at
100\,Myr (the age of NGC 2516) to 4.2 at 600\,Myr (comparable to the
age of Hyades).

Binary evolution complicates the comparison due to an obvious
selection effect---the giants can probably all be seen, but white
dwarfs are easily hidden near a main-sequence or giant companion.  We
obtain estimates for {\fwdgs} in a field population with 50\%
primordial binaries by combining the single stars from
Tab.\,\ref{Tab:mass} with the binaries from Tab.\,\ref{Tab:SeBa_bin}.
We calculate an upper limit to {\fwdgs} by accounting for all white
dwarfs [counting (wd, wd) binaries as two objects and including (ms,
wd) and (gs, wd) binaries].  A lower limit is obtained by counting
(wd, wd) binaries as one object and excluding white-dwarf binaries
containing a main-sequence or giant companion.  For the field (no
dynamics) population, we find $\fwdgs = 0.80$--1.3 at 100\,Myr, and
$\fwdgs = 2.7$--4.1 at 600\,Myr.  Combining models W4
(Tab.\,\ref{Tab:W4_bin}) and W6 (Tab.\,\ref{Tab:W6_bin}) we obtain
$\fwdgs = 1.2$--1.7 at 100\,Myr and $\fwdgs = 3.0 $--4.0 at 600\,Myr.


The observed value of $\fwdgs\sim1$ at around 100\,Myr seems somewhat
low, but probably not inconsistent with our models.  However, the
value of $\fwdgs\sim2.5$ of the Hyades cluster (at about 600\,Myr) is
smaller than our models predict, suggesting that a considerable
fraction of the white dwarfs are hidden in binaries.

The value of the fraction \fwdgs\, does not seem to pose a serious
problem to understand any of the clusters discussed here.  But
according to the evolution of the field stars and binaries (combining
Tabs.\,\ref{Tab:mass} and \ref{Tab:SeBa_bin}), at an age of 600\,Myr
we expect a total of about 15 giants and 59 white dwarfs.  The
comparable models W4 and W6 contain, respectively, 13 and 12 giants,
and 48 and 57 white dwarfs at that age.  Model W6 has lost about
40\%\, of its mass and model W4 has lost about 60\%, yet the numbers
of giants and white dwarfs have decreased by only 4\% -- 20\%.
Averaging over time, we find a decrease in the number of giants and
white dwarfs of about 2\% per 100\,Myr.  Apparently, the dynamical
evolution of these clusters has little effect on the number of giants
and white dwarfs.  (In fact, this was assumed by von Hippel [1998] in
estimating the mass fractions of white dwarfs in open clusters.)
Number counts of giants and/or white dwarfs may therefore provide a
reasonable estimate of a cluster's initial mass.

For a fifty-fifty mixture of single stars and binaries, meaning that
2/3 of the stars are binary components, the mean mass of a star is
$\mmean=0.46$ at $t=100$\,Myr, 0.41 at 600\,Myr and almost constant
($\mmean\sim0.40$) thereafter.  We use the numbers of giants to
estimate the initial masses of the star clusters in
Tab.\,\ref{Tab:observed}, because the giants are least plagued by
selection effects (although their small numbers significantly limit
the accuracy of our estimates).  The number of giants per 1k stars is
2.7 at 100\,Myr (see Tabs.\,\ref{Tab:mass} and \ref{Tab:SeBa_bin}),
and rises rapidly to about 6.5 at 400\,Myr, after which the specific
number of giants remains roughly constant.  For an open cluster older
than $\sim 400$\,Myr, we thus estimate its intial mass via
\begin{equation}
	M_0 = 65 \msun N_{\rm gs} 
	      \left( 1 + {0.02 t \over [100 {\rm Myr}]} \right).
\end{equation}
For younger clusters, the factor 65 ($= 1024\times0.41/6.5$) is
larger---170 at 100\,Myr and $\sim 100$ at 200\,Myr.

We apply this method to the clusters from Tab.\,\ref{Tab:observed}
with more than 5 giants and obtain the following birth masses:
830\,{\msun} for NGC 2287, 3100\,{\msun} for NGC 2660, and
1400\,{\msun} for NGC 3680.  These mass estimates seem reasonable.
For NGC 2534, Praesepe and the Hyades, the mass estimates are
690\,\msun, 370\,{\msun} and 290\,\msun, respectively, considerably
smaller than the observed masses of these clusters (see
Tab.\,\ref{Tab:observed}).  

The initial mass estimate for these clusters increases proportional to
the number of giants. The ratio \fwdgs\, does not pose a serious
problem and therefore, instead of too few white dwarfs, Hyades may
have too few giants. Where white dwarfs can be hidden easily in
binaries, giants are not that easy to hide.  One way to decrease the
number of giants is by binary activity.  The number of giants can be
decreased when subgiants are stripped in a phase of mass transfer
before they reach the horizontal branch, where they spend most of
their time. In order to reduce the number of giants in the fashion we
require most binaries to be born with short orbital periods ($\aplt
100$\,days). Alternatively the giant lifetimes adopted in our models
may be too long.

\subsection{Notes on individual clusters}

{\bf NGC\,2516} deserves much more study, as its dynamical parameters
(total mass, half-mass and core radii) are poorly known.

{\bf The Pleiades} cluster has been quite thoroughly studied in
searches for brown dwarfs and planets.  Its mass function fits nice
with the luminosity function from our models, with the exception that
our models containt oo many bright stars. 

{\bf NGC\,2287} is quite poorly studied, and with the quoted values for
the tidal radius it is has an extremely low mass for its age (see
Table \ref{Tab:observed}). The presence of 8 giants suggests that its
mass must have been comparable to that of NGC\,2516.

{\bf Praesepe} has been studied recently in great detail, and appears
to fit well with our dynamical models.  However, the cluster is rather
shallow and may have been born somewhat more massive and less
concentrated ($\Wo \aplt 4$) than our models.  This conclusion is
based on the observed shallow density profile and the high degree of
mass segregation.  There is some excess of stars with $V \sim 9$--11
is unexplained by our models. The observed cluster seems to be very
deficient in gaints. Based on the number of white dwarfs and the
dynamical state of the cluster we sould expect at least 20 giants in
this cluster, where only 5 are known.

{\bf The Hyades} cluster fits well with our models, indicating that it
is possible to estimate the initial conditions for an observed star
cluster rather accurately.  The cluster does not appear to be deficient
in white dwarfs if we compare them to the number of giants.  However,
if the mass of Hyades quoted in Table \ref{Tab:observed} is correct,
the observed number of white dwarfs and giants seems too small by a
factor of about three.

{\bf NGC\,3680} Fits well with our W6 models expect for the luminosity
function, which is deficient of low mass stars. A possible solution
may be that these stars were initially absent in the cluster or that
the observations do not go faint enough to reveal the low mass stars.

\subsection{Comparison with other work}
Our models evaporate on time scales generally consistent with
dissolution times reported in previous calculations.  The models of
Terlevich (1987) and de la Fuente Marcos (1997) compare well with the
evaporation times of our W6 models.  Terlevich's models evolve
somewhat more slowly due to the lack of massive stars and the rather
cool initial conditions, which drive the cluster to core collapse and
therefore extend its lifetime somewhat.  The models of McMillan \& Hut
(1994) dissolve more slowly than ours.  This discrepancy can be
completely explained by the absence of stellar evolution mass loss and
binary evolution in their models, along with the small size of
those models relative to the clusters' tidal radii in the Galactic
potential.

The only real discrepancy is in the work of Kroupa (1995a), whose
models dissolve somewhat more rapidly than ours.  Possibly the small
numbers of stars and the large binary fractions drive a more rapid
evaporation than one might naively expect. The evaporation rate of
star clusters is known to depend on total the number of stars (Heggie
et al.\, 1997; Portegies Zwart et al.\,
1998).\nocite{astro-ph/9711191}\nocite{1998A&A...337..363P} In
Section\,\ref{fig:tm_all} we argue that the presence of primordial
binaries has little effect on the cluster lifetime.  It is, however,
not clear how this trend propagates in the scaling of cluster
lifetimes with respect to the number of stars.




\newpage
\begin{center}
	\bf APPENDICES
\end{center}

\appendix\bigskip

\newcommand{\R}{{\bf r}}
\newcommand{\X}{{\bf x}}
\newcommand{\V}{{\bf v}}
\newcommand{\A}{{\bf a}}
\newcommand{\J}{{\bf j}}
\newcommand{\KK}{{\bf k}}
\newcommand{\LL}{{\bf l}}
\newcommand{\half}{{\textstyle\frac12}}
\newcommand{\third}{{\textstyle\frac13}}
\newcommand{\quarter}{{\textstyle\frac14}}
\newcommand{\sixth}{{\textstyle\frac16}}
\newcommand{\twelfth}{{\textstyle\frac1{12}}}
\newcommand{\twentieth}{{\textstyle\frac1{20}}}
\newcommand{\dt}{{\delta t}}

\section{Terminology}\label{sect:terminology}

Throughout the paper (and in future papers in this series) we will use
consistent nomenclature.  Some of these terms are rather confusing and
have been used by different authors in the past with slightly
different meanings.  For clarity we present here a short glossary of
terms. 

\begin{itemize}
\item[] {\bf Binary fraction}: thoughout this paper we define the
binary fraction as:
\begin{equation}
	f_{\rm bin} = { N_{\rm bin} \over N_{\rm sing} + N_{\rm bin}}.
\end{equation}
Here $N_{\rm sing}$ and $N_{\rm bin}$ are the number of single stars
and binaries. 

\item[] {\bf Cluster center}: The density center of the cluster, as
defined below.  Alternative definitions may use the number density or
the luminosity density, or the point where the density is greatest,
and may include projection effects.

\item[] {\bf Collision}: A collision occurs when the distance between
two stars $i$ and $j$ becomes smaller than the sum of their effective
radii: $d < d_{\rm coll} (r_i + r_j)$, with $d_{\rm coll} = 1$.  The
effective radii are determined from detailed fluid-dynamical
calculations.

\item[] {\bf Core radius}: The weighted average distance of all stars
from the density center.  Casertano \& Hut
(1985):\nocite{1985ApJ...298...80C} originally used a weighting
proportional to the local density.  However, in practice this
definition is unsuitable for clusters with near-isothermal density
profiles ($\rho\sim r^{-2}$).  Following Aarseth (1986), we adopt the
modified definition
\begin{equation}
	\rcore \equiv { \sum_i |r_i-r_j| {\rho_j^{(i)}}^2
		          \over \sum_i {\rho_j^{(i)}}^2 }\,,
\end{equation}
where $\rho_j$ is given by Eq.\,\ref{Eq:rhoj}.  Note that this
definition of the core does not {\em necessarily} have any simple
relation to the ``core radius'' normally quoted by observers, nor to
the ``dynamical'' core radius $r_c = \sqrt{\frac{3\langle
v^2\rangle_c}{4\pi G\rho_c}}$, where $\rho_c$ and $\langle
v^2\rangle_c$ are, respectively, the cluster's central density and
velocity dispersion (Binney \& Tremaine 1987).

\item[] {\bf Crossing time}: The time taken by a star with velocity 
equal to the velocity dispersion $v$ to cross the virial radius $r$ of the
stellar system $t_{\rm c} =  r/v$ which for practical reasons is
written as 
\begin{equation}
	t_{\rm c} \equiv \left( {\rvir^3 \over GM} \right)^{1/2}.
\end{equation}
In more convenient units, we write the half-mass crossing time as
\begin{equation}
	\tcrss = 57  \left( {\unit{\msun} \over \mtot} \right)^{1/2}
	             \left( {\rhm \over \unit{\pc}} \right)^{3/2}
		     \unit{Myr}.
\end{equation}

\item[] {\bf Density center} The density weighted average of the
positions of all stars (von Hoerner 1960,
1963):\nocite{1960ZA.....50..184V}\nocite{1963ZA.....57...47V}
\begin{equation}
	\R_{\rm dens}^{\rm VH} \equiv {\sum_i \R_i \rho_j^{(i)} 
					\over \sum_i \rho_j^{(i)} }
\end{equation}
In these expressions, $\rho_j^{(i)}$ is the density estimator of order
$j$ around the $i$-th particle, with position vector $\R_i$.  For
any star $i$ we define the local density within the volume $V_j$ of
the sphere containing the $j$ nearest neighbors ($i_1, i_2,\ldots,
i_j$) of $i$ as
\begin{equation}
	\rho_j^{(i)} \equiv {\sum_{k=1}^{j-1} m_{i_k} \over V_j}\,,
\label{Eq:rhoj}\end{equation}
where $V_j = \frac{4\pi}3 |r_{i_j} - r_i|^3$, and the sum over masses
excludes the masses of both stars $i$ and $i_j$ (see Casertano \& Hut
1985).\nocite{1985ApJ...298...80C} We take $j=12$.

\item[] {\bf Escaper}: A star which is not bound to the cluster,
i.e. whose energy exceeds the energy at the cluster's Jacobi surface.
An escaper may lie within the Jacobi surface.

\item[] {\bf Half mass radius}: The radius of the sphere, centered on
the cluster density center, that contains half of the total cluster
mass (as defined in the text).  It is not always clear which stars to
include in determining this radius, as the cluster is generally
flattened in the Galactic tidal field.

\item[] {\bf Hard binary}: A binary whose binding energy exceeds the
mean stellar energy in the cluster (Heggie
1975).\nocite{1975MNRAS.173..729H} A binary is hard if its semi major
axis exceeds
\begin{equation}
	a =  {GMm (M+m+m_3) \over (M+m)m_3v^2}. 
\end{equation}

\item[] {\bf Jacobi radius}: The distance from the cluster center to
the $L_1$ and $L_2$ Lagrange points---the maximum distance from the
cluster center to the cluster Jacobi surface.

\item[] {\bf Jacobi surface}: The cluster's ``Roche lobe'' in the
tidal field of the Galaxy.  Consider a star moving with Jacobi
integral $E_J \equiv \frac12v^2 + \phi_{eff}(\R)$ in the rotating
frame of reference in which our simulations are performed, where $v$
is velocity and the effective potential $\phi_{eff}$ includes the
cluster's self-gravity, the tidal field of the Galaxy, and the
centrifugal force in the rotating frame.  The {\em zero-velocity
surface} for that value of $E_J$ is defined by $v=0$, so
$\phi_{eff}(\R) = E_J$ (Binney \& Tremaine 1987).  The Jacobi
surface for the cluster is defined to be the last closed zero-velocity
surface that contains the cluster---that is, the surface passing
through the cluster's $L_1$ and $L_2$ Lagrange points.  With the
conventions adopted in {\kira}, the Lagrange points are located along
the $x$-axis, the cluster orbits in the $x$-$y$ plane, and the Jacobi
surface is elongated in the $x$-direction.  The three coordinate axes
intersect the Jacobi surface at distances $r_x = r_J$ (the Jacobi
radius), $r_y$, and $r_z$ from the cluster center.

\item[] {\bf Member}: A star (or multiple system) which is bound to
the cluster.  The energy of such a star is less than the energy at the
cluster Jacobi surface.  A member may lie outside the Jacobi surface.

\item[] {\bf Primary}: The more massive of the two stars in a binary
system. Denoted with $M$.

\item[] {\bf Relaxation time}:
We use Spitzer's (1987) definition of the half mass relaxation time:
\begin{equation}
	\trlx = \left( {\rhm^3 \over G M} \right)^{1/2} 
		{N \over 8 \log\Lambda}.
\end{equation}
Here $\Lambda \simeq 0.4N$ is the Coulomb logarithm.  (In the presence
of a realistic mass function, $\Lambda \aplt 0.1 N$ may be more
appropriate; Farouki \& Salpeter 1982; 1994; Smith, 1992; Fukushige \&
Heggie 1999).
\nocite{1982ApJ...253..512F}\nocite{1994ApJ...427..676F}
\nocite{1992ApJ...398..519S}\nocite{FH_astro-ph/9910468} In convenient
units this may be written
\begin{equation}
	\trlx = 2.05 \left( {\unit{\msun} \over \mtot} \right)^{1/2}
	             \left( {\rhm \over \unit{\pc}} \right)^{3/2}
		            {N \over \log\Lambda}\,\,\,
		     \unit{Myr}.
\label{Eq:trlx}\end{equation}
Note that ``$N$'' here is the number of bound objects in the star
cluster, and is smaller than the total number of stars if the cluster
contains binaries.

\item[] {\bf Secondary}: The less massive of the two stars in a binary
system. Denoted with $m$.

\item[] {\bf Tidal radius}: Since clusters are somewhat elongated in
the Galactic tidal field, the tidal ``radius'' is not well defined.
For definiteness, we take the tidal radius to be the Jacobi radius of
the cluster.  For a disk field described by Oort
(1927)\nocite{Oort1927} constants $A$ and $B$, we have
\begin{equation}
	\rt^3  = {G\mtot \over 4A(A-B)} \; \unit{pc}.
\label{Eq:Rtide}\end{equation}
In the solar neighborhood, $A = 15$\,\kmpskpc, and $B=-12$\,\kmpskpc.

\item[] {\bf Unperturbed binary}: A binary for which the dimensionless
perturbation due to its neighbors is less than some critical value
($\sim10^{-6}$, typically).  Dynamically, unperturbed binaries are
treated in the point-mass approximation, as seen by the rest of the
system.  Only unperturbed binaries can be treated using the \SeBa\,
binary evolution module described in Appendix B.  Perturbed binaries
are treated as two single stars; mass transfer and tidal
circularization are currently not handled in perturbed binaries.

\item[] {\bf Virial radius}: A characteristic length scale for the
system, defined by
\begin{equation}
	\rvir  = \frac{G\mtot^2}{-2U}\,,
\end{equation}
where $M_U$ is the total potential energy of the system (including the
tidal potential).  For an isolated equal-mass system, $\rvir$ is the
harmonic mean of the particle separations (H\'enon
1972).\nocite{1972gnbp.coll..406H}


\end{itemize}


\newpage	

\section{Starlab}\label{sect:Starlab}
The simulations described in this series of papers are carried out
within the ``Starlab'' software environment, version 3.5.  Starlab is
a software package for simulating the evolution of dense stellar
systems and analyzing the resultant data.  It consists of a collection
of loosely coupled programs (``tools'') linked at the level of the
UNIX operating system.  The tools share a common data structure and
can be combined in arbitrarily complex ways to study the dynamics of
star clusters and galactic nuclei.  The main components of Starlab
used in this work are \kira, the {\nbody} integrator, and \SeBa, a
stellar and binary evolution package.  The Starlab system is described
in detail in {\tt http://www.sns.ias.edu/$\sim$starlab}.

\subsection{Kira}\label{sect:Kira}
The {\nbody} integrator {\kira} is the largest single program within
Starlab.  Its basic function is to take an input {\nbody} system and
evolve it forward for a specified period of time, producing snapshot
and other diagnostic output at regular intervals.  In addition to
strictly dynamical evolution of stars and multiple stellar systems,
kira also incorporates stellar and binary evolution (via the \SeBa\,
subpackage), and the possible influence of an external (``tidal'')
gravitational field.  The program is designed to take advantage of the
``GRAPE-4'' special-purpose processor (Makino et
al. 1997),\nocite{1997ApJ...480..432M} if available, although GRAPE is
not required for its operation.

\subsubsection{The Integrator}
Particle motion is followed using a fourth-order, block-timestep
(McMillan 1986)\nocite{1986ApJ...307..126M} ``Hermite''
predictor-corrector scheme (Makino and Aarseth
1992).\nocite{1992PASJ...44..141M} Briefly, during a time step $\dt$,
particle positions $\X$ and velocities $\V$ are first predicted using
the known acceleration $\A$ and ``jerk'' $\J$ (the time derivative of
the acceleration):
\begin{eqnarray}
	\X_p &=& \X + \V\dt + \half\A\dt^2 + \sixth\J\dt^3 \,, \\
	\V_p &=& \V + \A\dt + \half\J\dt^2 \,.
\end{eqnarray}
The acceleration $\A_p$ and jerk $\J_p$ are then computed at the
predicted time using $\X_p$ and $\V_p$, and the motion is corrected
using the additional derivative information thereby obtained,
\begin{eqnarray}
	\KK ~~\equiv~~ \half\A^{\prime\prime}\dt^2
			&=& 2(\A-\A_p) + \dt(\J-\J_p)\,, \\
	\LL ~~\equiv~~ \sixth\A^{\prime\prime\prime}\dt^3
			&=& -3(\A-\A_p) - \dt(2\J+\J_p)\,,
\end{eqnarray}
to obtain the corrected position and velocity:
\begin{eqnarray}
	\X_c &=& \X_p + (\twentieth\LL+\twelfth\KK)\dt^2\,, \\
	\V_c &=& \V_p + (\quarter\LL+\third\KK)\dt\,.
\end{eqnarray}

A single integration step in thus proceeds as follows:
\begin{enumerate}

\item
	Determine which stars are to be updated next.  Each star has
	associated with it an individual time $t$, representing the
	time to which it was last advanced, and an individual timestep
	$\dt$. The list of stars to be integrated consists of those
	with the least value of $t+\dt$.  Time steps are constrained to
	be powers of 2, allowing ``blocks'' of many stars to be
	advanced simultaneously.

\item
	Before the step is actually taken, check for
	\begin{itemize}
	    \item[(a)] termination of the run 
	    \item[(b)] escaper removal 
	    \item[(c)] system reinitialization 
	    \item[(d)] diagnostic (``log'') output, which includes
	    \begin{itemize}
	    	\item[-] information on bulk parameters of the system: total
		  mass, energy, momentum, anisotropy, etc.  
	    	\item[-] technical information on CPU time, timestep
		  distribution, etc.
	    	\item[-] detailed information on the cluster mass
		  distribution: core properties, Lagrangian radii, etc.
	    	\item[-] stellar mass distribution and anisotropy by
			Lagrangian zone
	    	\item[-] luminosity profile, and mass and luminosity functions 
	    	\item[-] cluster stellar content (by spectral type and
		  luminosity class)  
	    	\item[-] detailed dynamical and physical data on all
			binary systems.
	    \end{itemize}
	    \item[(e)] snapshot output, for restart and display
	\end{itemize}

\item
	Perform low-order prediction of all particles to the new
	time. This operation may be performed on the GRAPE, if
	present.

\item
	Recompute the acceleration and jerk on all stars in the
	current block (using the GRAPE, if available), and correct
	their positions and velocities for fourth-order accuracy.

\item
	Check for and initiate unperturbed motion.

\item
	Check for collisions and mergers. 

\item
	Check for tree reorganization (see below).

\item
	Check for and apply stellar and/or binary evolution (\S B.2),
	and correct the dynamics as necessary.

\end{enumerate}

\subsubsection{Tree Structure}
An {\nbody} system in Starlab is represented as a linked-list
structure, in the form of a mainly ``flat'' tree having individual
stars as leaves.  The tree is flat in the sense that single stars
(i.e. stars that are not members of any multiple system) are all
represented as top-level nodes, having the root node (the system
center of mass) as parent.  Binary, triple, and more complex multiple
systems are represented as binary trees below their top-level center
of mass nodes.  The tree structure determines both how node dynamics
is implemented and how the long-range gravitational force is computed.

Each parent node contains ``local'' information about its
dynamics---mass, position, velocity, etc.---relative to its parent
node.  The leaves contain additional information about stellar
properties---effective radius, luminosity, temperature, etc.  The
parent node of a unperturbed binary also contains information on the
binary parameters---semi-major axis, eccentricity, mean anomaly, etc.
The motion of every node {\em relative to its parent node} is followed
using the Hermite predictor-corrector scheme just described.  The use
of relative coordinates at every level ensures that high numerical
precision is maintained at all times, even during very close
encounters.

The tree evolves dynamically according to simple heuristic rules:
particles that approach ``too close'' to one another are combined into
a center of mass and binary node; and when a node becomes ``too
large'' it is split into its binary components.  These rules apply at
all levels of the tree structure, allowing arbitrarily complex systems
to be followed.  In practice, the term ``too close'' is taken to mean
that two stars (1 and 2) approach within the ``close-encounter
distance'' $R_{\rm close}\sim\rv (m_1+m_2) / 2\mtot$, the impact parameter
that would lead to a $90^\circ$ deflection if both bodies moved at
typical stellar speeds.  ``Too large'' means that a node's diameter
exceeds $2.5 R_{\rm close}$.

\subsubsection{Binaries}
How the acceleration (and jerk) on a particle or node is computed
depends on its location in the tree.  Top-level nodes feel the force
due to all other top-level nodes in the system.  Forces are computed
using direct summation over all other particles in the system; no tree
or neighbor-list constructs are used.  (This procedure is designed
specifically to allow efficient computation of these forces using
GRAPE hardware, if available.)  Nearby binary and multiple systems are
resolved into their components, as necessary.

The internal motion of a binary component is naturally decomposed into
two parts: (1) the dominant contribution due to its companion, and (2)
the perturbative influence of the rest of the system.  This
decomposition is applied recursively, at all levels in a multiple
system.  Since the perturbation drops off rapidly with distance from
the binary center of mass, usually only a few near neighbors are
significant perturbers of even a moderately hard binary.  These
neighbors are most efficiently handled by maintaining lists of
perturbers for each binary.  Perturber lists are recomputed at time
the center of mass is integrated.

A further efficiency measure is the imposition of {\em unperturbed
motion} for binaries whose perturbation falls below some specified
value for all or part of an orbit.  Unperturbed binaries may be
followed analytically for many orbits as strictly two-body motion;
they are also treated as point masses, from the point of view of their
influence on other stars.  The use of the unperturbed approximation
near the periastron of eccentric orbits was a key element in our
decision not to use cumbersome regularization schemes for the
computation of binary motion.

Because unperturbed binaries are followed in steps that are integer
multiples of the orbit period, we can relax the perturbation threshold
for unperturbed motion relative to that for a perturbed step (since
most of the perturbative effects of nearby stars are periodic).
Perturbed binaries are resolved into their components, both for
purposes of determining their center of mass motion and for
determining their effect on other stars.  Unperturbed treatments of
multiple systems are also used, based on empirical studies of the
stability of their internal motion.  A hierarchical system is regarded
as stable if (a) the external perturbation is less than some
threshold value, and (b) each component is stable (or single), by the
same criterion.

``Lightly perturbed'' binaries, having external perturbations within a
factor of $\sim$10 of the unperturbed threshold, are treated using a
variant of the method described by Mikkola \& Aarseth
(1998),\nocite{1998NewA....3..309M} in which the internal motion of
the binary is artificially slowed and the perturbation is increased by
the same factor.  Briefly, the result is that long-term secular trends
in the binary orbital elements are properly reproduced, while periodic
perturbative terms are amplified; the latter effect is suppressed by
following the ``slow'' motion over an integral number of orbits.  Our
``slow'' binary treatment differs from that of Aarseth mainly in that
it is not coupled to a regularization scheme---it is applied directly
to the unregularized equations of motion.  In addition, we apply
pairwise corrections to forces between perturbers and the binary
center of mass in order to avoid spurious high derivatives caused by
the mismatch between the (slowed) internal motion and the (normal)
external interaction.

\subsubsection{Tidal Field}
The standard form of the external (tidal) potential is
\begin{equation}
	\phi_{\rm ext} = {\textstyle\frac12} (\alpha_1 x^2 + \alpha_3 z^2)\,.
\end{equation}
This expression includes contributions from both the Galactic tidal
field and the centrifugal force in the cluster's rotating frame of
reference.  The Galactic center is assumed to lie along the negative
$x$ axis and the rotation vector $\bf\Omega$ is in the $z$ direction.
(We assume motion in a circular orbit in the $x$--$y$ plane around the
Galactic center.)  The equations of motion also include a Coriolis
acceleration ${\bf a}_c = -2{\bf\Omega}\times{\bf v}$.  Tidal and
coriolis effects are applied to top-level nodes only---that is, we
neglect the tidal effect of the Galaxy on a binary's internal motion.

The values of $\alpha_1$, $\alpha_3$, and $\Omega$ depend on the
details of the field being modeled.  Some common examples are:
\begin{enumerate}
	\item {\em Point-mass field.} If the Galaxy is represented as
	a point mass $M_G$ at distance $R_G$, we have
	\begin{equation}
		\alpha_3 = -{\textstyle\frac13}\alpha_1 = \Omega^2
			 = \frac{GM_G}{R_G^3}\,.
	\end{equation}

	\item {\em Isothermal field.} For motion in a ``halo'' mass
	distribution modeled as an isothermal sphere ($\rho\sim
	r^{-2}$), with $M(<r) = M_G(r/R_G)$, we have
	\begin{equation}
		\alpha_3 = -{\textstyle\frac12}\alpha_1 = \Omega^2
			 = \frac{GM_G}{R_G^3}\,.
	\end{equation}

	\item {\em Disk field.} For motion in a disk described by local
	Oort constants $A$ and $B$, with local density $\rho_D$, we
	have
	\begin{eqnarray}
		\alpha_1 &=& -4A(A-B)\\
		\alpha_3 &=& 4 \pi G \rho_D + 2 (A^2 - B^2)\\
		\Omega   &=& A-B
	\end{eqnarray}
\end{enumerate}

In the (fairly good) approximation that the gravitational potential of
the cluster stars may be represented close to the Jacobi surface
simply as $\phi_C(r) \sim -GM_{\rm tot}/r$, where $M_{\rm tot}$ is the
cluster mass, the Jacobi radius may straightforwardly be shown to be
\begin{equation}
	r_J \approx \left(\frac{-GM_{\rm tot}}
			{\alpha_1}\right)^{-1/3}.
\end{equation}
The ratio $\alpha_3/\alpha_1$ determines the shape of the Jacobi
surface.

\subsubsection{Escaper Removal}
Stars are removed (``stripped'') from the system when they exceed a
specified distance from the cluster center of mass (or density
center).  For systems without an imposed Galactic tidal field, this
stripping radius is arbitrary.  For systems with a tidal field, the
stripping radius is usually tied to the Jacobi radius of the cluster.
For the runs described in this paper, stars were stripped when their
distance from the cluster center exceeded twice the instantaneous
Jacobi radius.


\subsection{\SeBa}\label{sect:SeBa}
The stellar and binary evolution package {\SeBa}\footnote{The name
{\sf SeBa} is taken from the ancient Egyptian word for `to
teach', `the door to knowledge' or `(multiple) star'.  The exact
meaning depends on the hieroglyphic spelling.} is fully integrated
into the {\kira} integrator, although it can also be used as a
stand-alone module for non-dynamical applications.

\subsubsection{Evolution of a single star}\label{sect:singlestar}
Stars are evolved via the time dependent mass-radius relations for
solar metallicities given by Eggleton et al.~(1989,
\nocite{1989ApJ...347..998E}\nocite{1990ApJ...354..387E} with
corrections by Eggleton et al.\, 1990 and Tout et
al.~1997)\nocite{1997MNRAS.291..732T}
\footnote{New equations which include metallicity dependence have
recently been made available by Hurley et
al.~(2000),\nocite{2000astro.ph..1295H} and will be implemented in the
next version of \SeBa.}. These equations give the radius of a star as a
function of time and the star's initial mass (on the zero-age
main-sequence---ZAMS).  Neither the mass of the stellar core nor the
rate of mass loss via a stellar wind are specified in this
prescription.  However, both quantities are important, both to binary
evolution and to cluster dynamics.  We include them using the
prescriptions of Portegies Zwart \& Verbunt (1996).\nocite{pzv96}

%
\newcommand{\mylist}[2]
	   {\par\parbox[t]{1.25in}{{\bf #1}}~~~\parbox[t]{1.8in}{{#2}}}

Stars are subdivided within {\SeBa} into the following types:

\mylist{planet}{Various types, such as gas giants, etc.; also includes moons.}
\mylist{brown dwarf}{Star with mass below the hydrogen-burning limit.}
\mylist{main sequence}{Core hydrogen burning star.}
\mylist{Wolf-Rayet}{Massive ($m>25$\,\msun) star which has lost its
			hydrogen envelope via a stellar wind.}
\mylist{helium star}{Helium core of a stripped giant, the
			result of mass transfer in a binary.
			Subdivided into helium core, carbon core
			and helium giant.}
\mylist{subgiant}{Hydrogen shell burning star.}
\mylist{horizontal branch}{Helium core burning star.}
\mylist{supergiant}{Double shell burning star.}
\mylist{Thorne-Zytkow}{Shell burning hydrogen envelope with neutron 
			star core.}
\mylist{black hole}{Star with radius smaller than the event
			horizon. The result of evolution of massive
			($m>25$\,\msun) star or collapsed neutron
			star.}
\mylist{neutron star}{Subdivided into radio pulsar, X-ray pulsar 
			and inert neutron star ($m<2$\,\msun).}
\mylist{white dwarf}{Subdivided into helium dwarf, carbon dwarf 
			and oxygen dwarf.}
\mylist{disintegrated}{Result of Carbon detonation to
			Type Ia supernova.}

%

Stellar-wind mass loss is neglected for main-sequence stars with
$m<25$ \msun.  Following Langer (1998),\nocite{1998A&A...329..551L}
more massive stars lose mass with ${\dot m} \propto m^{2.5}$ before
becoming a Wolf-Rayet star (see Portegies Zwart et al. 1999, for the
implementation).  These stars eventually collapse into black holes
with mass $m_{\rm bh} = 0.35 m_0 - 12$ {\msun}, where $m_0$ is the
initial mass of the star.  (For a star whose mass increases due to
collisions or other processes, $m_0$ is the highest mass reached by
the star. The black hole radius equals the Schwarzschild radius: $r =
2Gm/c^2$.)

A star with a helium core mass between 2.2 and 5 {\msun} becomes a
neutron star.  (These limits correspond to 8 {\msun} and 25 {\msun}
ZAMS mass stars which evolve as isolated single stars.)  At birth, a
neutron star receives a velocity ``kick'' in a random direction.  The
magnitude of the velocity kick is chosen randomly from the
distribution proposed by Hartman (1997)\nocite{1997A&A...322..127H}
\begin{equation}
	P(u)du = {4\over \pi} \; {du\over(1+u^2)^2},
\label{eq:kick}\end{equation}
with $u=v/\sigma$ and $\sigma = 600 \,\,\mbox{km}\,\mbox{s}^{-1}$.

A star with a core mass less than 2.2 {\msun} sheds its envelope at the
end of its evolution and becomes a white dwarf.  The mass of the white
dwarf equals the core mass of its progenitor at the tip of the
asymptotic giant branch.

\subsubsection{Schematic evolution of a binary}\label{sect:SeBaschema}

The evolution of a single isolated or unperturbed binary is carried
out in the following steps (see sects.\,\ref{sect:binev} and
\ref{sec:masstransfer} for details):

The evolution of any binary stars by Determining the binary evolution
timestep.  This is the smallest timestep allowed by either of the
stars.  A stellar evolution timestep is 1\% of the time taken for the
star to evolve from the start of one evolutionary stage to the
next---for example, from the zero-age main sequence to the
terminal-age main sequence. (The stellar evolution step is not to
exceed 1Gyear.) A list of these mile-posts along a star's evolutionary
is provided in Sect.\,\ref{sect:singlestar}.  A binary is evolved
whenever one of its stars requires an update.

If a binary is in a state of mass transfer (but not in a common
envelope) the timestep is reduced such that $<1$\% of the donor'
envelope is lost per step.

\begin{enumerate}
\item[1.] Apply angular momentum loss by magnetic stellar wind.
\item[2.] Apply angular momentum loss by gravitational
          wave radiation.
\item[3.] Check for coalescence.
\item[4.] Evolve primary star. 
  \begin{enumerate}
	\item[(a)] adjust binary parameters for stellar wind mass loss 
	\item[(b)] resolve supernova. 
  \end{enumerate}
\item[5.] Check if binary still exists.
	   The evolution of the primary (or secondary) star may
	   have resulted in a supernova which may disrupt the
	   binary or resulted in a collision between the two stars.
\item[6.] Evolve secondary star. 
  \begin{enumerate}
	\item[(a)] Adjust binary parameters for stellar wind mass loss. 
	\item[(b)] Resolve supernova. 
  \end{enumerate}
\item[7.] Check if binary still exists (see [5.]).
\item[8.] Check for tidal circualrization and synchronization.
\item[9.] Check if any star is Roche-lobe filling and
	      identify the donor and the accretor. 
              If no star is filling its Roche-lobe leave 
              binary evolution and notify dynamics,
	      otherwise proceed with the following steps
 \begin{enumerate}
   \item[(a)] Find moment mass transfer starts.
   \item[(b)] Check for binary stability 
     \begin{enumerate}
      \item[-] if binary unstable apply commone envelope
      \item[-] if components merge leave
               binary evolution and notify dynamics.
     \end{enumerate}
   \item[(c)] Calculate $\zeta_{\rm ad}$, $\zeta_{\rm Rl}$ and
	              $\zeta_{\rm th}$. 
   \item[(d)] Determine amount of mass loss from donor.
   \item[(e)] Determine amount of mass gained by accretor.
   \item[(f)] Subtrac mass from donor.
   \item[(g)] Add mass to accretor.
              Calculate new evolutionary state of accretor and 
       	      Rejuvenate.
   \item[(h)] Calculate new binary parameters.
   \end{enumerate}
\end{enumerate}

\subsubsection{Evolution of binary parameters without mass transfer}
\label{sect:binev}
The orbital parameters of a binary are affected by the evolution of
its components.  We will not present here all the many details of
binary evolution, but for clarity we summarize those which affect the
dynamics or are important for interpreting our results.  The details
of the binary evolution program \SeBa\, are discussed in more detail
by Portegies Zwart \& Verbunt (1996) and Portegies Zwart \& Yungelson
(1998).\nocite{pzv96}\nocite{pzy98}

Mass lost in a stellar wind is assumed to escape isotropically from
the binary system.  If the companion accretes a fraction $\xi$ of
the other star's wind this implies
\begin{equation}
	{a \over a_0} = f {M_0+m_0 \over M+m}\,,
\end{equation}
Here $M_0$ and $m_0$ are the initial primary and secondary mass,
respectively, $M$ and $m$ are their final masses. 
And  
\begin{equation}
	f = \left( \left[{M \over M_0}\right]^{\xi}\; 
		   {m \over m_0}\right)^{-2}.
\end{equation}
The fraction $\xi$ is calculated via Bondi--Hoyle
(1944)\nocite{bondy_hoyle44} accretion assuming that the thermal
velocity in the wind equals the escape velocity of the mass-losing
star (for details see Portegies Zwart \& Verbunt 1996).

When the radius of one of the stars exceeds 5 times the orbital
periastron separation $a(1-e)$, orbital energy is transformed into
oscillatory modes in the two stars.  This leads to a decrease in the
orbital separation and, due to conservation of angular momentum
[$a(1-e^2) =$ constant], to the eventual circularization of the
binary. 

Mass loss in a supernova is lost impulsively from the binary system.
As a result both the orbital separation and the eccentricity change:
both increase if the pre-supernova orbit was circular.  The velocity
kick (Eq.\,\ref{eq:kick}) received by the neutron star at formation is
added randomly to its orbital velocity.  New orbital parameters are
then calculated assuming that the positions of the two stars are
unchanged, mass is lost isotropically from the exploding star, and the
companion is unaffected by the explosion. If the pre-supernova orbit
is eccentric things get somewhat more complicated (see Portegies Zwart
\& Verbunt 1996).

Low-mass stars may have magnetically coupled winds, and relatively
large changes in angular momentum may occur even though a negligible
amount of mass escapes.  We follow the prescription described by
Rappaport et al.\,(1983)\nocite{1983ApJ...275..713R} for tidally
synchronized binaries in which at least one component is a
main-sequence star or (sub)giant with mass $0.7 \leq m/\msun \leq
1.5$.

Compact stars in short-period binaries and highly eccentric binaries
lose orbital energy and angular momentum via gravitational radiation.
For such binaries, we use the expressions provided by Peters
(1964)\nocite{Peters64} to compute the time dependence of the orbital
semi-major axis and eccentricity.

\subsubsection{Mass transfer in binaries}\label{sec:masstransfer}
When one star in a binary approaches its Roche-limit, we iteratively
determine the moment at which contact occurs.  The size of the Roche
lobe is calculated as (Eggleton 1983)\nocite{1983ApJ...268..368E}
\begin{equation}
	r_{\rm Rl} = {0.49 \over
		      0.6 + q^{2/3}\ln(1 + q^{-1/3})},	
\end{equation}
where $q \equiv m/M$.  The Roche-lobe-filling star is then identified
as the donor and its companion as the accretor.


\begin{center}
	\ref{sec:masstransfer}.1.~~~Unstable Mass transfer
\end{center}
When a star fills its Roche lobe we first check for the possibility of
Darwin-Riemann instability.  This happens if
\begin{equation}
	J_{\rm donor} > {\textstyle\frac13} J_{\rm bin}, 
\end{equation} 
where $J_{\rm donor}$ and $J_{\rm bin}$ are the angular momenta of the
Roche-lobe-filling star and the binary, respectively.

During spiral-in the envelope of the donor is expelled, at the cost of
orbital energy, following the prescription of Webbink
(1984):\nocite{1984ApJ...277..355W}
\begin{equation}
	{a \over a_0} = {M_c \over M}	
			\left({1 + 2a_0 \over \alpha\lambda} 
			{M_e \over m} \right)^{-1}.
\end{equation}
The parameters governing binary evolution are listed in
Table\,\ref{Tab:free_parameters}.

If the Roche-lobe-filling star is a main-sequence star or compact
object, the two stars simply merge because the donor has no core-halo
structure.  If both the donor and the accretor are (sub)giants, we
expel both envelopes at the cost of the binaries binding energy in a
{\em double inspiral}.  A merger occurs when the binary that remains
after the common envelope phase is semi-detached, in which case no
more mass is lost (see sect.\,\ref{sect:collisions}).

%
\begin{center}
	\ref{sec:masstransfer}.2.~~~Stable mass transfer
\end{center}
We calculate the time scale for mass transfer in a dynamically stable
binary by considering the responses of the donor and the binary
parameters to changes in the donor mass.  For this purpose we define
the logarithmic derivative
\begin{equation}
	\zeta_{\rm i} = \left( {d \ln r \over d\ln m} \right)_{\rm i},
\end{equation}
for each of the following processes:

\newcommand{\mylista}[2]
	   {\par~~\parbox[t]{0.5in}{{#1}}~~\parbox[t]{2.1in}{{#2}}}

\mylista{$\zeta_{\rm ad}$}{the change in donor radius due to
	adiabatic adjustment of hydrostatic equilibrium},
\mylista{$\zeta_{\rm Rl}$}{the change in the size of the donor's
	Roche lobe},
\mylista{$\zeta_{\rm th}$}{the change in donor radius as it
	adjusts to a new thermal equilibrium}.

%

The adopted values for $\zeta_{\rm ad}$ are as follows: For main
sequence stars with $m>0.7$\,{\msun} we use $\zeta_{\rm ad} = 4$ and
half this value for lower-mass main sequence stars.  For stars on the
Hertzsprung gap and horizontal branch we use $\zeta_{\rm ad} = 2.25$
and 15, respectively.  For other stars with a core-halo structure
(subgiants, supergiants and {\TZ} objects), we use the following fit
to the composite polytropic models of Hjellming \& Webbink
(1987):\nocite{1987ApJ...318..794H}
\begin{equation}
   \zeta_{\rm ad} = -0.221 -2.847 x +32.03 x^2 -75.69x^4 +57.81 x^5,
\label{Eq:zeta_ad}
\end{equation}
where $x = m_{\rm core}/m$.  

We use $\zeta_{\rm th}$ between 0 and 0.9 for main sequence stars and
$\zeta_{\rm th} = 0$ for all other stars, except those on the
Hertzsprung gap and on the horizontal branch for which we use
$\zeta_{\rm th} = -2$ and 15, respectively.

The response of the Roche lobe to mass transfer $\zeta_{\rm Rl}$ is
calculated by transfering a infenitesimal amount of mass from the
donor to the accreting star and study the response of the binary
parameters.  This test particle is transfered on the same timescale as
was used in the previous mass transfer step. At first Roche-lobe
contact, when there was no previous mass transfer step, we assume that
the test particle is transfered on a thermal timescale, which is a
rather conservative choise.

The time scale on which mass transfer proceeds is determined as
follows:

\newcommand{\mylistb}[2]
	   {\par~~\parbox[t]{1.5in}{{#1}}~~\parbox[t]{1.75in}{{#2}}}

\mylistb{$\zeta_{\rm ad} < \zeta_{\rm Rl}$}{dynamically unstable mass transfer proceeds on time scale $\tau_{\rm dyn}$}
\mylistb{$\zeta_{\rm ad} > \zeta_{\rm Rl}$ and  $\zeta_{\rm th} < \zeta_{\rm Rl}$}{thermally unstable mass transfer proceeds on time scale $\tau_{\rm th}$} 
\mylistb{$\zeta_{\rm ad} > \zeta_{\rm Rl}$ and $\zeta_{\rm th} \ge \zeta_{\rm Rl}$}{nuclear unstable mass transfer proceeds on time scale $\min (\tau_{\rm nuc}, \tau_{\rm J})$,}


where the time scales associated with the various criteria are as
follows:

\begin{center}\
\begin{tabular}{llll}
{\bf dynamic:} & $\tau_{\rm dyn}$ & $\simeq$ & 
			$5.1\, 10^{-11}\sqrt{r^3/m}\, \;\; \unit{Myr}$\\
{\bf thermal:} & $\tau_{\rm th}$ & $\simeq$ & 
			$32 m^2/(rL)\, \;\; \unit{Myr}$ \\
{\bf nuclear:} & $\tau_{\rm nuc}$ & $\simeq$ & 
			$0.1 t_{\rm ms}$ \\
{\bf angular momentum loss:} & $\tau_{\rm J}$ & $\simeq$ & 
			$J_{\rm bin}/({\dot J}_{\rm gr} + {\dot J}_{\rm mb}).$
\end{tabular}
\end{center}
%
%
Here $t_{\rm ms}$ is a star's main-sequence lifetime, and $m$, $r$ and
$L$ are its mass, radius and luminosity, respectively.  The loss of
angular momentum via gravitational radiation and magnetic braking are
denoted by ${\dot J}_{\rm gr}$ and ${\dot J}_{\rm mb}$, respectively.

Table \ref{Tab_mdot} gives a flavor of the various time scales on
which mass transfer generally proceeds.  However, the details depend
critically on the orbital separation and on the mass and evolutionary
state of both the donor and the accreting star.

\begin{table*}
\caption[]{Schematic diagram of the time scales on which stable mass
transfer (donor to accretor) proceeds.}
\begin{flushleft}
\begin{tabular}{l|cccc}
{\em Donor:}  &  main sequence & subgiant & supergiant & compact
							   object \\
\hline
{\em Accretor:}&               &                                  \\
main sequence & nuclear/thermal& thermal         & dynamic        & ---    \\
(sub)giant    & ---            & nuclear/thermal & thermal/dynamic& --     \\ 
compact object& thermal/aml    & dynamic         & dynamic        & aml \\ \hline
\end{tabular}
\end{flushleft}
\label{Tab_mdot}
\end{table*}

Some mass may be lost from the binary system during mass transfer.
The new orbital parameters are calculated assuming that the mass lost
from the binary carries specific angular momentum $\eta_J$ (see
Table\,\ref{Tab:free_parameters}).  We calculate the final orbital
separation using
\begin{equation}
	{a \over a_0} = \left({Mm \over M_0 m_0}\right)^{-2}
			\left({M+m \over M_0+m_0}\right)^{2\eta_J+1}.
\end{equation}
Here $M = M_0 - dM$ and $m = m_0 + dm$ (so $dM$ and $dm$ are defined
as positive quantities).  The binary thus loses mass if $dM - dm \geq
0$.  For the amount of mass accepted by the accretor, see Portegies
Zwart \& Verbunt (1996).

\subsection{Rejuvenation of the accretor}
An accreting star generally becomes more massive, which shortens its
evolutionary time scale.  The method described here is rather ad-hoc.
We assume that a star accreting $\delta m$ of mass remains in the same
evoutionary state (see the list in Sect.\,\ref{sect:singlestar}).  The
age of the star with mass $m$ is $t(m)$ and we want to know what is
the age $t(m+\delta m)$ of the star with mass $m+\delta m$.  At he
moment the mass of the accretor increases from $m$ to $m+ \delta m$
the star is in evolutionary state $i$. It took the star $t_{i}(m)$ to
reach that evolutionary state and this state lasts for $\tau_i(m)
\equiv t_{i+1}(m)- t_{i}(m)$ for a star with mass $m$.  The same stage
for a star with mass $m+\delta m$ lasts for $\tau_i(m+\delta m)$.  The
age of the star after accretion then becomes
\begin{equation}
	t(m+\delta m) = t_{i}(m+\delta m) 
		      + \left( { t(m)-t_{i}(m) \over \tau_i(m)}
			\right)
	                \tau_i(m+\delta m) {\cal R}.
\label{Eq:accretion}\end{equation}
Here ${\cal R}$ is a fraction introduced to mimick the rejuvenation of
the accretor (${\cal R} > 1$).  The mass dumped on its surface may
lead to some internal mixing, refreshing some of the helium core
material with some of the freshly accreted Hydrogen.  This
rejuvenation fraction is calculated with
\begin{equation}
	{\cal R} =  \left({m+\delta m \over m} \right)^\kappa,
\label{Eq:rejuvenation}\end{equation}
and we use ${\cal R} = 1$ if the accreted material is not Hydrogen, in
which case the accretor is not rejuvenated.  Allowing ${\cal R} <1$
would mimick that a star becomes older upon accreting material.  The
adopted value for $\kappa$ is listed in
Table\,\ref{Tab:free_parameters}.

We will give two examples of a $m=2$\,\msun\, star which accretes
$\delta m= 0.2\,\msun$\, from a Helium rich companion (${\cal R} =
1$).  For simplicity we assume here that this amount of mass is
transferred in an infenitesimal timestep. The main-sequence lifetime
for a 2\,\msun\, star is about 801\,Myear and about 608\,Myear for a
2.2\,\msun\, star.  If mass transfer starts at $t=700$\,Myear, the
2\,\msun\, accretor is still on the main sequence. After mass transfer
the accreting star has an age of 531\,Myear and is still on the main
sequence.

If mass transfer started at $t=1$\,Gyear things become somewhat more
complicated. The 2\,\msun\, accretor is then on the horizontal branch.
The time it takes from zero-age to the beginning of the horizontal
branch is about 938\,Myear, for a 2.2\,\msun\, star this is about
712\,Myear. The 2\,\msun\, star spends roughtly 84\,Myear on the
horizontal branch, where a 2.2\,\msun\, star spends only 82\,Myear in
that stage. Substitution of these numbers into Eq.\,\ref{Eq:accretion}
results in an age of the post mass transfer star of 773\,Myear.

\begin{table}
\caption[]{Free parameters in binary evolution}
\medskip
\begin{tabular}{lrl} \hline
term             &value& description  \\ \hline
$\kappa$         & 1   & accretor rejuvenation factor \\
$\eta_J$         & 2   & specific angular momentum loss per unit mass \\
$\lambda$        & 0.5 & envelope binding energy fraction \\
$\alpha_{\rm ce}$& 4   & common envelope constant \\
\hline
\end{tabular}
\label{Tab:free_parameters}
\end{table}

\subsection{Result of a merger or collision}\label{sect:collisions}


We have adopted a set of simple prescriptions to specify the outcome
of stellar collisions.  In the future these prescriptions can be
refined when more accurate calculations become available.  As a rule
of the thumb, the result of a collision is the conservative accretion
of the lower-mass star onto the more massive star.  The accretor will
then be rejuvenated as described in Eq.\,\ref{Eq:rejuvenation}.  This
rule is violated when one component is a giant or a compact object.  A
detailed prescription of how to calculate the evolutionary state of
such a merger is presented in \paperI.

We describe our treatment of the possible outcomes of encounters
between two stars, ordered by the evolutionary state of the more
massive of the two (the primary). Table~\ref{Tab_mprod} summarizes
this treatment.

\begin{table}
\caption[]{Simplified representation of possible merger outcomes.
           The four columns correspond to the four choices given for the
          type of massive star (primary), while the four rows indicate
           the type of less massive star (secondary):
           main-sequence star (ms), (sub)giant (sg), white dwarf (wd) and
          neutron star (ns).
           In this table we do not distinguish between stars
           in the Hertzsprung gap (Hg) or on the first and second ascent
          on the asymptotic-giant branch (AGB).}
\begin{flushleft}
\begin{tabular}{l|cccc}
          & \multicolumn{4}{c}{primary} \\
      star& ms & sg& wd  & ns \\ \hline
          &    &   & wd  & ns \\
       ms & ms & sg& +   & +  \\
          &    &   & disc&disc\\ \hline
          &    &   & wd  & ns \\
       sg & Hg &AGB& +   & +  \\
          &    &   & disc&disc\\ \hline
          &    &   &     &    \\
       wd & sg &AGB& --  & -- \\
          &    &   &     &    \\ \hline
          &    &   &     &    \\
       ns &\TZO&\TZO& -- & -- \\
          &    &   &     &    \\ \hline
\end{tabular}
\end{flushleft}
\label{Tab_mprod}\end{table}

\subsubsection{Main-sequence primary}
If both stars involved in the encounter are main-sequence stars, then
the less massive star is accreted conservatively onto the more massive
star.  The resulting star is a rejuvenated main-sequence star (see Lai
et al.\ 1993, Lombardi et al.\ 1995).
\nocite{1993ApJ...406L..63L}\nocite{1995ApJ...445L.117L} The details
of this procedure are described in Appendix C4 of Portegies Zwart \&
Verbunt (1996).

If the less massive star in the encounter has a well developed core
(giant or subgiant), this core becomes the core of the merger product.
The main-sequence star and the envelope of the giant are combined to
form the new envelope of the merger.  In general, the mass of the core
is relatively small compared to the mass of the envelope, and the star
is assumed to continue its evolution through the Hertzsprung gap.
Note that this type of encounter can only occur when the main-sequence
star is itself a collision product (e.g.~a blue straggler).

When a main-sequence star encounters a less massive white dwarf, we
assume that the merger product is a giant whose core and envelope have
the masses of the white dwarf and the main-sequence star,
respectively.  We then determine its evolutionary state as follows.
We calculate the total time $t_{\rm agb}$ that a single, unperturbed
star with mass equal to that of the merged star would spend on the
asymptotic giant-branch, and the mass $m_{\rm c,agb}$ of its core at
the tip of the giant branch.  The age of the merger product is then
calculated by adding $t_{\rm agb}m_c/m_{\rm c,agb}$ to the age of an
unperturbed star with the same mass at the bottom of the asymptotic
giant branch.  For example, a single, unperturbed 1.4$\msun$ star
leaves the main-sequence after 2.52$\,$Gyr, spends 60$\,$Myr in the
Hertzsprung gap, moves to the horizontal branch at $2.96\,$Gyr, and
reaches the tip of the asymptotic giant branch after $3.06\,$Gyr, with
a core of $0.64\,\msun$.  Thus, if a 0.6$\,\msun$ white dwarf merges
with an $0.8\,\msun$ main-sequence star, the merger product has an age
of 2.87$\,$Gyr, leaving it another 180$\,$Myr before it reaches the
tip of the asymptotic giant-branch.

If the less massive star is a neutron star or black hole a
Thorne--{\Zytkow} object (1977)\nocite{tz77} is formed.

\subsubsection{Evolved primary}
When a (sub)giant or asymptotic branch giant encounters a less massive
main-sequence star, the main-sequence star is combined with the
envelope of the giant, which stays in the same evolutionary state.
Its age within that state is changed, however, according to the
rejuvenation calculation described in Sect.\ C3 of Portegies Zwart \&
Verbunt (1996).  For example, an encounter of a giant of $0.95\,\msun$
and age $11.34\,$Gyr with a 0.45$\,\msun$ main-sequence star produces
a giant of 1.4$\,\msun$ with an age of $2.67\,$Gyr.

When both stars are (sub)giants, the two cores are merged and form the
core of the merger product (see Davies et al.\ 1991 and Rasio \&
Shapiro 1995).\nocite{dbh91}\nocite{rs95} Half the envelope mass of
the less massive star is accreted onto the primary.  The merger
product continues its evolution starting at the next evolutionary
state---a (sub)giant continues its evolution on the horizontal branch,
and a horizontal branch star becomes an asymptotic-giant branch star.
The reasoning behind this assumption is that an increased core mass
corresponds to a later evolutionary stage.

If the less massive star is a white dwarf, then its mass is simply
added to the core mass of the giant, and the envelope is retained.  If
the age of the giant before the encounter exceeds the total lifetime
of a single unperturbed star with the mass of the merger, then the
newly formed giant immediately sheds its envelope and its core turns
into a single white dwarf.  Otherwise the merged giant is assumed to
have the same age (in years) as the giant before the collision, and
continues its evolution as a single unperturbed star.

If the other star is a less massive neutron star, a Thorne--{\Zytkow}
object is formed.

\subsubsection{White-dwarf primary}
In an encounter between a white dwarf and a less massive main-sequence
star, the latter is assumed to be completely disrupted, and forms a
disk around the white dwarf (Rasio \& Shapiro
1991).\nocite{rm90}\nocite{rs91} The white dwarf accretes from this
disk at a rate of 1 percent of the Eddington limit.  If the mass in
the disc exceeds 5\% of the mass of the white dwarf, the excess mass
is expelled from the disc at a rate equal to the Eddington
limit. 

If a white dwarf encounters a less massive (sub)giant, a new white
dwarf is formed with a mass equal to the sum of the pre-encounter core
of the (sub)giant and the white dwarf.  The newly formed white dwarf
is surrounded by a disk formed from half the envelope of the
(sub)giant before the encounter. The factor half is rather arbitrary
and based on the lack of detailed calculations which provide a proper
number.  If the mass of the white dwarf exceeds the Chandrasekhar
limit it explodes in a type Ia supernova, leaving no remnant (Nomoto
\& Kondo 1991; Livio \& Truran 1995).  \nocite{nk91}\nocite{lt95}

A collision between two white dwarfs results in a single white dwarf
with mass equal to the sum of the original masses.  If the total mass
of the collision product exceeds the Chandrasekhar mass, it explodes
in a type Ia supernova.

Collisions between white dwarfs and neutron stars or black holes
result in the formation of an accretion disc around
the compact object; the white dwarf is destroyed.  Following the
accretion, the neutron star may collapse into a black hole.

\subsubsection{Neutron-star or black-hole primary}
All collisions involving a neutron star or black hole primary lead to
the formation of a massive disk around the compact star.  If the
compact star had a disk prior to the collision, this disk is expelled.
This disk accretes onto the compact star. We chose, rather arbitrarily
that the accretion rate is 5\% of the Eddington limit. An accreting
neutron star turns into a millisecond radio pulsar, or---when its mass
exceeds $2\msun$---into a black hole.


\subsection{Communication between {\SeBa} and \Kira}
Due to the interaction between stellar evolution and stellar dynamics,
it is difficult to solve for the evolution of both systems in a
completely self-consistent way.  The trajectories of stars are
computed using a block timestep scheme, as described earlier.  Stellar
and binary evolution is updated at fixed intervals (every 1/64 of a
crossing time, typically a few thousand years).  Any feedback between
the two systems may thus experience a delay of at most one timestep.
Internal evolution time steps may differ for each star and binary, and
depend on binary period, perturbations due to neighbors, and the
evolutionary state of the star.  Time steps in this treatment vary
from several milliseconds up to (at most) a million years.

After each $1/64$ of a crossing time, all stars and binaries are
checked to determine if evolutionary updates are required.  Single
stars are updated every 1/100 of an evolution timestep or when the
mass of the star has changed by more than 1\% since the last update.
A stellar evolution timestep is the time taken for the star to evolve
from the start of one evolutionary stage to the next (see
sect.\,\ref{sect:SeBaschema}).

After each stellar evolution step the dynamics is notified of changes
in stellar radii, but changes in mass are, for reasons of efficiency,
not passed back immediately (mass changes generally entail recomputing
the accelerations of all stars in the system).  Instead, the
``dynamical'' masses are modified only when the mass of any star has
changed by more than 1\%, or if the orbital parameters, semi-major
axis, eccentricity, total mass or mass ratio of any binary has changed
by more than 0.1\%.

\subsection{Mass loss from stars and binaries}
Fast (sudden) and slow (gradual) mass loss affects the dynamics of the
stellar system in different ways.  Mass loss is considered fast when
it takes place within a fraction of an orbital time scale.  For single
stars, this time scale is on the order of the crossing time of the
star cluster.  For binaries, it is much shorter---on the order of the
binary orbital period.  Mass loss during a supernova explosion is
considered fast, stellar winds and mass lost from a binary during mass
transfer are considered slow.

Due to the discretized time steps of the stellar dynamics and the
stellar evolution, from the point of view of the dynamics mass is lost
in ``bursts.''  For example, an asymptotic giant star with a strong
stellar wind may lose its entire envelope in a hundred steps spanning
roughly one crossing time, while a supergiant might lose its entire
envelope instantaneously in a supernova.  Mass loss for single stars
affects the dynamics of the entire stellar system.  For binaries and
multiple systems, mass loss from a member star directly affects the
orbital characteristics of its neighbors.

The rate of mass loss is particularly important for binaries.  Slow
mass loss via a stellar wind will soften a binary system, but will not
affect its eccentricity or its center of mass velocity.  (This is true
if the binary is unperturbed.  In a perturbed binary, the eccentricity
and center of mass velocity are both affected by stellar wind mass
loss.)  Sudden mass loss, on the other hand, can dramatically affect
the binary's internal parameters.  For unperturbed binaries, the
effects of mass loss from both component stars are computed
consistently using \SeBa.  Changes in binary parameters are calculated
and the dynamics is notified, thereby transmitting the information to
the rest of the stellar system via the integrator.

For perturbed binaries and multiples (and also hierarchical systems
where the inner binary is unperturbed), the integrator takes care of
the dynamical effects of stellar mass loss.  By construction, mass
transfer cannot occur in a perturbed binary or multiple system.  If a
supernova occurs in a perturbed binary, any slow mass loss is
accounted for before fast mass loss occurs, since a star which is
about to explode generally loses a significant fraction of its mass in
a stellar wind before the supernova event itself.  Supernova remnants
do not lose mass.  This assumption breaks down when the binary
companion of the exploding star loses a significant fraction of its
mass between the moment of the supernova and the end of the stellar
update timestep.  (This can happen if the binary companion is either a
Wolf-Rayet star or a supergiant.)  The stellar evolution time steps in
these cases are taken sufficiently small (on the order of a hundred
years) to ensure that this causes a negligible error.

\subsection{Collisions and mergers}
We draw a distinction between ``mergers'' and ``collisions.''  A
merger may result from mass transfer or a common-envelope phase during
the evolution of an unperturbed binary.  The binary node is then
replaced by the merger product.  The product of a merger is generally
different from the result of a collision, since a merger is often
preceded by a phase of mass transfer which affects the masses of both
stars.

Collisions may occur between between single stars (which are part of a
binary tree) or between stars in a perturbed binary.  Since the
integrator may miss the precise moment of closest approach, the
orbital elements of each ``close'' pair of stars is calculated after
each integration step.  A collision occurs when the stars are found to
be within $d_{\rm coll}$ times the sum of their radii at periastron:
$p < d_{\rm coll}(r_1+r_2)$.  In this case, the two stars are replaced
by the collision product, which is
placed in the center of mass and with the center of mass velocity of
the original two-body system.  The characteristics of the collision
product are calculated using {\SeBa} (Sect.\,\ref{sect:collisions}).

A collision may also occur when an unperturbed binary in a state of
mass transfer is perturbed by a close encounter with another cluster
member.  Such a induced collision may be triggered by a close flyby,
or in a multiple system with a perturbed outer orbit.  The collision
occurs if the sum of the component radii exceeds the distance between
the two stars at the moment the binary becomes perturbed.


\bigskip\noindent{\bf Acknowledgments} We thank Douglas Heggie, Ken
Janes, Gijs Nelemans, Koji Takahashi and Lev Yungelson for numerous
discussions.  This work was supported by NASA through Hubble
Fellowship grant HF-01112.01-98A awarded (to SPZ) by the Space
Telescope Science Institute, which is operated by the Association of
Universities for Research in Astronomy, Inc., for NASA under contract
NAS\, 5-26555, and by ATP grant NAG5-6964 (to SLWM).  SPZ is grateful
to Drexel University, Tokyo University and the University of Amsterdam
(under Spinoza grant 0-08 to Edward P.J. van den Heuvel) for their
hospitality. Special thanks goes to the University of Tokyo for
providing time on their GRAPE-4 system.

\bibliographystyle{./inc/aabib}

\end{document}